\input harvmac
\input epsf
\baselineskip=24pt


\lref\rsennet{A. Sen, {\it F Theory and Orientifolds,} Nucl. Phys.
{\bf B475} (1996) 562, {\tt hep-th/9605150}.}

\lref\eddie{E. Witten, {\it Anti de Sitter Space and Holography,} Adv.\ Theor.\ Math.\ Phys.\ 
{\bf 2}, 253 (1998), {\tt hep-th/9802150}.}

\lref\AMP{A.M. Polyakov, {\it Quantum Geometry of Bosonic Strings,} Phys.\ Lett.\ {\bf 103B}, 
207 (1981).} 

\lref\gubbie{S.S. Gubser, I.R. Klebanov, A.M. Polyakov, {\it Gauge Theory Correlators from 
Noncritical String Theory,} Phys.\ Lett.\ {\bf B428}, 105 (1998), {\tt hep-th/9802109}.} 

\lref\hooft{G. 't Hooft, {\it A Planar Diagram Theory for Strong Interactions,} Nucl.\ Phys.\ 
{\bf B72}, 461 (1974).}

\lref\lyk{Joseph D. Lykken, {\it Introduction to Supersymmetry,} {\tt hep-th/9612114}.}

\lref\gsw{M.B. Green, J.H. Schwarz, E. Witten, {\it  Superstring Theory. Vol. 1: Introduction,}   
Cambridge, Uk: Univ. Pr. (1987) 469 P. (Cambridge Monographs On Mathematical Physics).}


\lref\mal{ J.~Maldacena, {\it The large $N$ limit of superconformal field theories and supergravity,} Adv.\ Theor.\ Math.\ 
Phys.\ {\bf 2}, 231 (1998) [Int.\ J.\ Theor.\ Phys.\ {\bf 38}, 1113 (1998)], {\tt hep-th/9711200}.  } 

\lref\joep{J.~Polchinski, {\it Dirichlet-Branes and Ramond-Ramond Charges,} Phys.\ Rev.\ Lett.\  {\bf 75}, 4724 (1995), {\tt hep-th/9510017}.}

\lref\tasi{J.~Polchinski,{\it Tasi Lectures on D-Branes,} {\tt hep-th/9611050}.} 

\lref\holo{Leonard Susskind, {\it The World as a Hologram,} J.\ Math.\ Phys.\ {\bf 36}, 6377 
(1995), {\tt hep-th/9409089}.}

\lref\holt{G. 't Hooft,{\it Dimensional Reduction in Quantum Gravity,} {\tt gr-qc/9310026}.}

\lref\bht{J.D. Bekenstein, {\it Black Holes and the Second Law}, Lett.\ Nuovo Cim.\ {\bf 4}, 
737 (1972).} 

\lref\strom{Gary T. Horowitz, Andrew Strominger, {\it Black Strings and p-branes,} 
Nucl.\ Phys.\ {\bf B360}, 197 (1991).}

\lref\klebo{Igor R. Klebanov, {\it World Volume Approach to Absorption by Nondilatonic Branes,} 
Nucl.\ Phys.\ {\bf B496}, 231 (1997), {\tt hep-th/9702076}.}  

\lref\rNojiri{S. Nojiri and S. Odintsov, {\it Two-Boundaries AdS/CFT Correspondence 
in Dilatonic Gravity,} Phys.Lett. {bf B449} (1999) 39-47, {\tt hep-th/9812017}.}
\lref\rAhn{C. Ahn, K. Oh and R. Tatar, {\it The Large N Limit of ${\cal N}=1$ 
Field Theories from F-theory,} Mod.Phys.Lett. {bf A14} (1999) 369-378, {\tt hep-th/9808143}.} 
\lref\rHyun{S. Hyun, Y. Kiem and H. Shin, {\it Effective Action for Membrane Dynamics 
in DLCQ $M$ theory on a Two-torus,} Phys. Rev. {\bf D59} (1999) 021901, {\tt hep-th/9808183}.}
\lref\rSW{N. Seiberg and E. Witten, {\it Monopoles, Duality and Chiral
Symmetry Breaking in N=2 Supersymmetric QCD,} Nucl. Phys. {\bf B431}
(1994) 484, {\tt hep-th/9408099.}}
\lref\rdorey{N. Dorey, V.V. Khoze and M.P. Mattis, {\it On N=2 
Supersymmetric QCD with Four Flavors,} Nucl.Phys. {\bf B492} (1997) 
607, {\tt hep-th/9611016}.}
\lref\rHen{M. Henningson, {\it Extended superspace, higher derivatives and 
SL(2,Z) duality,} Nucl. Phys. {\bf B458} (1996) 445, {\tt hep-th/9507135}.}
\lref\rDS{M. Dine and N. Seiberg, 
{\it Comments on Higher Derivative Operators in
some SUSY Field Theories,} Phys.Lett. {\bf B409} (1997) 239, 
{\tt hep-th/9705057}.}
\lref\rKD{N. Dorey, V.V. Khoze, M.P. Mattis, M.J. Slater and W.A. Weir,
{\it Instantons, Higher Derivative Terms and Non-renormalization Theorems in
Supersymmetric Gauge Theories,} Phys.Lett. {\bf B408} (1997) 213, 
{\tt hep-th/9706007}.}
\lref\rOvrut{I.L. Buchbinder, S.M. Kuzenko and B.A. Ovrut,
{\it On the D=4 N=2 Nonrenormalization Theorem,} Phys. Lett. {\bf B433}
(1998) 335, {\tt hep-th/9710142.}}
\lref\rOLoop{B. de Wit, M.T. Grisaru and M. Rocek, {\it Non-holomorphic
Corrections to the One Loop N=2 Super Yang-Mills Action,}
Phys. Lett. {\bf B374} (1996) 297, {\tt hep-th/9601115}
\semi
F. Gonzalez-Rey, U. Lindstrom, M. Rocek and R. Von Unge, {\it On N=2
Low Energy Efective Actions,} Phys. Lett. {\bf B388} (1996) 581, 
{\tt hep-th/9607089}
\semi
S. Ketov {\it On the next-to-leading-order correction to the effective action
in N=2 gauge theories,} Phys. Rev. {\bf D57} (1998) 1277,
{\tt hep-th/9706079.}}
\lref\rocek{F. Gonzalez-Rey, B. Kulik, I.Y. Park and M. Rocek, {\it Selfdual 
Effective Action of N=4 super Yang-Mills,} {\tt hep-th/9810152}.}
\lref\rOfer{O. Aharony, A. Fayyazuddin and J. Maldacena, {\it The Large N
Limit of N=2,1 Field Theories from Threebranes in F-theory,} JHEP 9807
(1998) 013, {\tt hep-th/9806159}.}
\lref\rGSVY{B.R. Greene, A.Shapere, C.Vafa and S.T.Yau, {\it Stringy Cosmic
Strings and Noncompact Calabi-Yau Manifolds,} Nucl. Phys. {\bf B337}
(1990) 1
\semi
M. Asano, {\it Stringy Cosmic Strings and Compactifications of F-theory,}
Nucl. Phys. {\bf B503} (1997) 177,  {\tt hep-th/9703070}.}
\lref\rKeh{A. Kehagias, {\it New Type IIB vacua and their F-Theory
Interpretation,} Phys. Lett. {\bf B435} (1998) 337, 
{\tt hep-th/9805131}.}
\lref\Sen{A. Sen, {\it F-theory and Orientifolds,} Nucl. Phys.
{\bf B475} (1996) 562, {\tt hep-th/9605150}.}
\lref\rDoug{T.Banks, M.R.Douglas and N.Seiberg, {\it Probing F-theory with branes,}
Phys. Lett. {\bf B387} (1996) 74, {\tt hep-th/9605199}
\semi
M.R.Douglas, D.A.Lowe and J.H.Schwarz, {\it Probing F-theory with multiple
branes,} Phys. Lett. {\bf B394} (1997) 297, {\tt hep-th/9612062}.}
\lref\rJev{A. Jevicki and T. Yoneya, {\it Spacetime Uncertainty Principle
and Conformal Symmetry in D-particle Dynamics,} Nucl. Phys. {\bf B535}
(1998) 335, {\tt hep-th/9805069}
\semi
A. Jevicki, Y. Kazama and T. Yoneya, {\it Quantum Metamorphosis
of Conformal Transformation in D3 Brane Yang-Mills Theory,}
Phys. Rev. Lett. {\bf 81} (1998) 5072, {\tt hep-th/9808039}
\semi
A. Jevicki, Y. Kazama and T. Yoneya, {\it Generalized Conformal Symmetry
in D Brane Matrix Models,} Phys. Rev. {\bf D59} (1999) 066001,
{\tt hep-th/9810146}}
\lref\rHo{V. Periwall and R. von Unge, {\it Accelerating D Branes,} Phys. 
Lett. {\bf B430} (1998) 71, {\tt hep-th/9801121}
\semi
J. de Boer, K. Hori, H. Ooguri and Y. Oz, {\it Kahler Potential and Higher 
Derivative Terms from M Theory Fivebranes,} Nucl. Phys. B518 (1998) 173-211,
{\tt hep-th/9711143}. }
\lref\rKh{N. Dorey, V.V. Khoze and M.P. Mattis, {\it Multi Instanton Calculus 
in N=2 Supersymmetric Gauge Theory. 2. Coupling to Matter} Phys. Rev. 
{\bf D54} (1996) 7832, {\tt hep-th/9607202}.}
\lref\rDaction{A.A. Tseytlin, {\it Self-duality of Born-Infeld action and Dirichlet
 3-brane of type IIB superstring theory,} Nucl. Phys. {\bf B469} (1996) 51, 
{\tt hep-th/9602064}
\semi
M. Aganic, J. Park, C. Popescu and J. Schwarz, {\it Dual D-Brane Actions,}
Nucl. Phys. {\bf B496} (1997) 215, {\tt hep-th/9702133.}}
\lref\rTseyt{I. Chepelev and A. Tseytlin, {\it On Membrane Interaction in
Matrix Theory,} Nucl. Phys. {\bf B524} (1998) 69, {\tt hep-th/9801120.}}
\lref\rBob{R. de Mello Koch and R. Tatar, {\it Higher Derivative Terms
from Threebranes in F-theory,} Phys.Lett. {\bf B450} (1999) 99, 
{\tt hep-th/9811128}.}
\lref\rKD{A. Yung, {\it Instanton Induced Effective Lagrangian in the 
Seiberg-Witten Model,} Nucl. Phys. {\bf B485} (1997) 38, 
{\tt hep-th/9605096}
\semi
N. Dorey, V.V. Khoze, M.P. Mattis, M.J. Slater and W.A. Weir,
{\it Instantons, Higher Derivative Terms and Non-renormalization Theorems in
Supersymmetric Gauge Theories,} Phys.Lett. {\bf B408} (1997) 213, 
{\tt hep-th/9706007}
\semi
D. Bellisai, F. Fucito, M. Matone and G. Travaglini, {\it Nonholomorphic terms
in N=2 Susy Wilsonian Actions and RG Equation,} Phys. Rev. {\bf D56}
(1997) 5218, {\tt hep-th/9706099.}}
\lref\rDas{S.R. Das, {\it Brane Waves, Yang-Mills Theory and Causality,}
{\tt hep-th/9901006.}}
\lref\rAlv{E. Alvarez and C. Gomez, {\it Noncritical Coinfining Strings and 
the Renormalization Group,} {\tt hep-th/9902012.}}
\lref\rDie{I. Antoniadis, {\it A Possible New Dimension at a Few TeV,}
Phys. Lett. {\bf B246} (1990)~377\semi
J.D. Lykken, {\it Weak Scale Superstrings,} Phys. Rev. {\bf D54} (1996)
3693, {\tt hep-th/9603133}\semi
K.R. Dienes, E. Dudas and T. Gherghetta, {\it Extra Spacetime Dimensions and
Unification,} Phys. Lett. {\bf B436} (1998) 55 {\tt hep-th/9803466}
\semi
K.R. Dienes, E. Dudas and T. Gherghetta, {\it Grand Unification at Intermediate
Mass Scales through Extra Dimensions,} Nucl. Phys. {\bf B537} (1999) 47,
{\tt hep-th/9806292}
\semi
Z. Kakushadze, {\it TeV Scale Supersymmetric Standard Model and Brane
World,} {\tt hep-th/9812163}.}
\lref\rNeg{For a recent review see for example: 
J.W. Negele, {\it Instantons, the QCD Vacuum and Hadronic Physics,}
{\tt hep-lat/9810153}.}
\lref\rLoop{S-J. Rey and J. Yee, {\it Macroscopic Strings as heavy quarks
in Large N Gauge Theory,} {\tt hep-th/9803001}
\semi
J. Maldacena, {\it Wilson Loops in Large N Field Theories,}
Phys. Rev. Lett. {\bf 80} (1998) 4859, {\tt hep-th/9803002}
\semi
E. Witten {\it Anti-de Sitter Space, Thermal Phase Transition and Confinement
in Gauge Theories,} Avd. Theor. Math. Phys. {\bf 2} (1998) 505
{\tt hep-th/9803131}.}
\lref\rIgK{I.R. Klebanov and A.A. Tseytlin, {\it D Branes and Dual Gauge 
Theories in Type-0 Strings,} {\tt hep-th/9811035}
\semi
J.A. Minahan, {\it Glueball Mass Spectra and other issues for Supergravity
Duals of QCD Models,} {\tt hep-th/9811156}
\semi
J.A. Minahan, {\it Asymptotic Freedom and Confinement from Type 0 String
Theory} {\tt hep-th/9902074}
\semi
I.R. Klebanov and A.A. Tseytlin, {\it Asymptotic Freedom and Infrared
Behaviour in the Type-0 String Approach to Gauge Theory,}
{\tt hep-th/9812089.}}
\lref\rWK{For nice discussions of this see: E. Witten, 
{\it New Perspectives in the Quest for Unification,} 
{\tt hep-ph/9812208}
\semi
I.R. Klebanov, {\it From Threebranes to Large N Gauge
Theory,} {\tt hep-th/9901018.}}
\lref\rSon{A. Brandhuber, N. Itzhaki, J. Sonnenschein and S. Yankielowicz,
{\it Wilson Loops in the Large N Limit at Finite Temperature,}
Phys. Lett. {\bf B434} (1998) 36, {\tt hep-th/9803137.}}
\lref\rMal{J. Maldacena, {\it The Large N Limit of Superconformal
Field Theories and Supergravity,} Adv.Theor.Math.Phys. {\bf 2} (1998) 231,
{\tt hep-th/9711200}
\semi
E. Witten, {\it Anti-de Sitter Space and Holography,} Adv.Theor.Math.Phys.
{\bf 2} (1998) 253, {\tt hep-th/9802150}
\semi
S.S. Gubser, I.R. Klebanov and A.M. Polyakov, {\it Gauge Theory Correlators
from Noncritical String Theory,} Phys.Lett. {\bf B428} (1998) 105, 
{\tt hep-th/9802109}.}
\lref\rTh{G. 't Hooft, {\it A Planar Diagram Theory for Strong Interactions,}
Nucl. Phys. {\bf B72} (1974) 461.}
\lref\rWS{N. Seiberg and E. Witten, {\it Electric-Magnetic Duality, Monopole
Condensation and Confinement in N=2 Super Yang-Mills Theory,} 
Nucl. Phys. {\bf B426} (1994) 19, {\tt hep-th/9407087.}}
\lref\rPol{A. Polyakov, {\it The Wall of the Cave,} {\tt hep-th/9809057.}}
\lref\rES{E. Witten and L. Susskind, {\it The Holographic Bound in Anti-de 
Sitter Space,} {\tt hep-th/9805114.}}
\lref\rGuby{A. Kehagias and K. Sfetsos, {\it On Running Couplings in Gauge
Theories from IIB supergravity,} {\tt hep-th/9902125}
\semi
S.S. Gubser, {\it Dilaton-driven Confinement,} {\tt hep-th/9902155.}}
\lref\rbuch{I.L. Buchbinder and S.M. Kuzenko, {\it Comments on the Background
Field Method in Harmonic Superspace: Nonholomorphic Corrections in N=4
super Yang-Mills,} Mod. Phys. Lett. {\bf A13} (1998) 1623, 
{\tt hep-th/9804168}
\semi
E.L. Buchbinder, I.L. Buchbinder and S.M. Kuzenko, {\it Nonholomorphic
Effective Potential in N=4 SU(n)
SYM,} Phys. Lett. {\bf B446} (1999) 216, {\tt hep-th/9810239}.}
\lref\rSpa{A. Fayyazuddin and M. Spalinski, {\it Large N Superconformal Gauge
Theories and Supergravity Orientifolds,} Nucl. Phys. {\bf B535} (1998) 219,
{\tt hep-th/9805096.}}


\lref\rWitten{A. Hanany and E. Witten, {\it Type IIB Superstrings, BPS
Monopoles and three dimensional gauge dynamics,} Nucl. Phys. {\bf B492}
(1997) 152-190, {\tt hep-th/9611230}
\semi
E. Witten, {\it Solutions of four-dimensional field theories
via M-theory,} Nucl. Phys. {\bf B500} (1997) 3-42, {\tt hep-th/9703166}.}
\lref\rGukov{S. Gukov, {\it Seiberg-Witten Solution from Matrix Theory,}
{\tt hep-th/9709138}
\semi
A. Kapustin, {\it Solution of N=2 Gauge Theories via Compactification to
three dimensions,} Nucl. Phys. {\bf B534} (1998) 531-545, 
{\tt hep-th/9804069}
\semi
R. de Mello Koch and J.P. Rodrigues, {\it Solving four dimensional field
theories with the Dirichlet fivebrane,} Phys.Rev. {\bf D60} (1999) 027901, 
{\tt hep-th/9811036}.}
\lref\rOz{N.D. Lambert and P.C. West, {\it N=2 Superfields and the 
M-Fivebrane,} Phys. Lett. {\bf B424} (1998) 281, {\tt hep-th/9801104}
\semi
N.D. Lambert and P.C. West, {\it Gauge Fields and M-Fivebrane Dynamics,}
Nucl. Phys. {\bf B524} (1998) 141, {\tt hep-th/9712040},
\semi
P.S. Howe, N.D. Lambert and P.C. West, {\it Classical M-Fivebrane Dynamics
and Quantum N=2 Yang-Mills,} Phys. Lett. {\bf B418} (1998) 85,
{\tt 9710034}
\semi
J. de Boer, K. Hori, H. Ooguri and Y. Oz, {\it Kahler Potential and Higher
Derivative terms from M Theory Fivebrane,} Nucl. Phys. {\bf B518} (1998)
173, {\tt hep-th/9711143}.}
\lref\rOoguri{E. Witten, {\it Branes and the Dynamics of QCD,} Nucl. Phys.
{\bf B507} (1997) 658, {\tt hep-th/9706109}
\semi
H. Ooguri, {\it M Theory Fivebrane and SQCD,} Nucl.Phys.Proc.Suppl. 68 (1998)
84, {\tt hep-th/9709211}.}
\lref\rOfer{O. Aharony, A. Fayyazuddin and J. Maldacena, {\it The Large N
Limit of N=2,1 Field Theories from Threebranes in F-theory,} JHEP 9807
(1998) 013, {\tt hep-th/9806159}.}
\lref\rRadu{R. de Mello Koch and R. Tatar, {\it Higher Derivative Terms
from Threebranes in F-theory,} Phys.Lett. {\bf B450} (1999) 99, 
{\tt hep-th/9811128}.}
\lref\rUs{R. de Mello Koch, A. Paulin-Campbell and J.P. Rodrigues,
{\it Non-holomorphic Corrections from Threebranes in F-theory,} Phys.Rev. D60 (1999) 106008,
{\tt hep-th/9903029}.}

\lref\rUst{R. de Mello Koch, A. Paulin-Campbell and J.P. Rodrigues,
{\it Monopole Dynamics in {\cal N}=2 super Yang Mills from a Threebrane Probe,} Nucl. Phys. {\bf B559} (1999) 143,
{\tt hep-th/9903207}.}

\lref\rSen{A. Sen, {\it BPS States on a Threebrane Probe,} 
{\tt hep-th/9608005}.}
\lref\rFayy{A. Fayyazuddin, {\it Results in supersymmetric field theory
from 3-brane probe in F-theory,} Nucl. Phys. {\bf B497} (1997) 101,
{\tt hep-th/9701185}.}
\lref\rES{E. Witten and L. Susskind, {\it The Holographic Bound in Anti-de 
Sitter Space,} {\tt hep-th/9805114.}}
\lref\rJe{A. Jevicki and T. Yoneya, {\it Spacetime Uncertainty Principle
and Conformal Symmetry in D-particle Dynamics,} Nucl. Phys. {\bf B535}
(1998) 335, {\tt hep-th/9805069}
\semi
A. Jevicki, Y. Kazama and T. Yoneya, {\it Quantum Metamorphosis
of Conformal Transformation in D3 Brane Yang-Mills Theory,}
Phys. Rev. Lett. {\bf 81} (1998) 5072, {\tt hep-th/9808039}
\semi
A. Jevicki, Y. Kazama and T. Yoneya, {\it Generalized Conformal Symmetry
in D Brane Matrix Models,} Phys. Rev. {\bf D59} (1999) 066001,
{\tt hep-th/9810146}
\semi
F. Gonzalez-Rey, B. Kulik, I.Y. Park and M. Rocek, {\it Selfdual 
Effective Action of N=4 super Yang-Mills,} {\tt hep-th/9810152}.}
\lref\rSW{N. Seiberg and E. Witten, {\it Electric-Magnetic Duality, Monopole
Condensation and Confinement in N=2 Super Yang-Mills Theory,} 
Nucl. Phys. {\bf B426} (1994) 19, {\tt hep-th/9407087}
\semi
N. Seiberg and E. Witten, {\it Monopoles, Duality and Chiral
Symmetry Breaking in N=2 Supersymmetric QCD,} Nucl. Phys. {\bf B431}
(1994) 484, {\tt hep-th/9408099.}}
\lref\rFerrari{F. Ferrari, {\it The dyon spectra of finite gauge theories,}
Nucl. Phys. {\bf B501} (1997) 53, {\tt hep-th/9608005}.}
\lref\rLerche{W. Lerche, {\it Introduction to Seiberg-Witten Theory and its
Stringy Origin,} Nucl. Phys. Proc. Suppl. {\bf 55B} (1997) 83,
{\tt hep-th/9611190.}}
\lref\rWit{E. Witten, {\it New Perspectives in the Quest for Unification,}
{\tt hep-th 9812208} 
\semi
I. Klebanov, {\it From Threebranes to Large N Gauge Theories,}
{\tt hep-th 9901018}.}
\lref\rGubser{A. Kehagias and K. Sfetsos, {\it On Running Couplings in Gauge
Theories from IIB supergravity,} {\tt hep-th/9902125}
\semi
S.S. Gubser, {\it Dilaton-driven Confinement,} {\tt hep-th/9902155}
\semi
H. Liu and A.A.Tseytlin, {\it D3-brane-D-Instanton configuration and
N=4 Super YM Theory in constant self-dual background,} 
{\tt hep-th/9903091}
\semi
A. Kehagias and K. Sfetsos, {\it On Asymptotic Freedom and Confinement
from Type IIB Supergravity,} {\tt hep-th/9903109}
\semi
S. Nojiri and S.D. Odinstov, {\it Running gauge coupling and quark-antiquark
from dilatonic gravity,} {\tt hep-th/9904036}
\semi
S. Nojiri and S.D. Odinstov, {\it Two Boundaries AdS/CFT correspondence in 
dilatonic gravity,} Phys.Lett. {\bf B449} (1999) 39 {\tt hep-th/9904036}.}
\lref\rGaber{M.R. Gaberdiel, T. Hauer and B. Zweibach, {\it Open String-String junction transitions,}
Nucl. Phys. {\bf B525} (1998) 117, {\tt hep-th/9801205}.}
\lref\rBerg{O. Bergmann and A. Fayyazuddin, {\it String junctions and BPS states in Seiberg-Witten theory,}
Nucl.Phys. {\bf B531} (1998) 108, {\tt hep-th/9802033}
\semi
A. Mikhailov, N. Nekrasov and S.Sethi, {\it Geometric realisations of BPS states in N=2 theories,} Nucl.Phys. {\bf B531} (1998) 345,
{\tt hep-th/9803142}
\semi
O. De Wolfe and B. Zweibach, {\it String junctions for arbitrary Lie algebra representations,} Nucl. Phys. {\bf B541} (1999) 509,
{\tt hep-th/9804210}.}
\lref\rMal{C.G.Callan and J.M.Maldacena, {\it Brane Dynamics from the
Born-Infeld Action,} Nucl.Phys. {\bf B513} (1998) 198, 
{\tt hep-th/9708147}
\semi
G.W.Gibbons, {\it Born-Infeld Particles and Dirichlet p-branes,}
Nucl.Phys. {\bf B514} (1998) 603, {\tt hep-th/9709027}.}
\lref\rHash{A.Hashimoto, {\it Shape of Branes pulled by strings,}
Phys.Rev. {\bf D57} (1998) 6441, {\tt hep-th/9711097}.}
\lref\rChalm{G Chalmers, M. Rocek, and R. von Unge, {\it Monopoles in quantum corrected N=2 super Yang Mills theory},
{\tt hep-th/9612195}.}
\lref\rDenef{F. Denef, {\it Attractors at weak gravity,} {\tt hep-th/9812049}.}
\lref\rSel{A.Fayyazuddin, {\it Some Comments on N=2 Supersymmetric Yang-Mills,}
Mod. Phys. Lett. A10 (1995) 2703, {\tt hep-th/9504120}
\semi
S.Sethi, M.Stern and E.Zaslow, {\it Monopole and Dyon Bound States in N=2
Supersymmetric Yang-Mills Theories,} Nucl. Phys. {\bf B457} (1995) 484,
{\tt hep-th/9508117}
\semi
J.P. Gauntlett and J.A. Harvey, {\it S-Duality and Dyon Spectrum in N=2 Super
Yang-Mills Theory,} Nucl.Phys. {\bf B463} (1996) 287, {\tt hep-th 9508156}
\semi
F.Ferrari and A.Bilal, {\it The Strong-Coupling Spectrum of the Seiberg-Witten
Theory,} Nucl. Phys. {\bf B469} (1996) 387, {\tt hep-th/9602082}
\semi
A.Klemm, W.Lerche, P.Mayr, C.Vafa and N.Warner, {\it Self-dual
Strings and N=2 Supersymmetric Field Theory,} Nucl. Phys. {\bf B477}
(1996) 746, {\tt hep-th/9604034}
\semi
F.Ferrari and A.Bilal, {\it Curves of Marginal Stability and Weak and Strong
Coupling BPS Spectra in N=2 Supersymmetric QCD,} Nucl.Phys. {\bf B480} (1996)
589, {\tt hep-th/9605101}
\semi
A.Brandhuber and S.Steiberger, {\it Self Dual Strings and Stability of BPS
States in N=2 SU(2) Gauge Theories,} Nucl.Phys. {\bf B488} (1997) 199, {\tt
hep-th/9610053}.}
\lref\rWst{N.D.Lambert and P.C.West, {\it Monopole Dynamics from the 
M-theory fivebrane,} {\tt hep-th/9811025}.}

\lref\rPS{M.K.Prasad and C.M.Sommerfeld, {\it An Exact Classical Solution
for the `t Hooft monopole and the Julia-Zee dyon,} Phys.Rev.Lett. {\bf 35}
(1975) 760.}
\lref\rGSVY{B.R. Greene, A.Shapere, C.Vafa and S.T.Yau, {\it Stringy Cosmic
Strings and Noncompact Calabi-Yau Manifolds,} Nucl. Phys. {\bf B337}
(1990) 1
\semi
M. Asano, {\it Stringy Cosmic Strings and Compactifications of F-theory,}
Nucl. Phys. {\bf B503} (1997) 177,  {\tt hep-th/9703070}.}
\lref\rnh{See for example, J.D. Bekenstein, {\it Black Hole Hair : twenty-five
years after,} {\tt gr-qc/9605059}.}
\lref\rDaction{A.A. Tseytlin, {\it Self-duality of Born-Infeld action and Dirichlet
3-brane of type IIB superstring theory,} Nucl. Phys. {\bf B469} (1996) 51, {\tt hep-th/9602064}
\semi
M. Aganic, J. Park, C. Popescu and J. Schwarz, {\it Dual D-Brane Actions,}
Nucl. Phys. {\bf B496} (1997) 215, {\tt hep-th/9702133.}}
\lref\rHarv{For a nice review see J.A. Harvey, {\it Magnetic Monopoles, Duality
and Supersymmetry,} {\tt hep-th/9603086}.}
\lref\rSols{P.Forgacs, Z.Horvath and L.Palla, {\it Nonlinear Superposition
of Monopoles,} Nucl.Phys. {\bf B192} (1981) 141
\semi
M.K.Prasad, {\it Exact Yang-Mills Higgs Monopole Solutions of arbitrary
topological charge,} Commun.Math.Phys. {\bf 80} (1981) 137.}
\lref\rSut{H.W.Braden and P.M.Sutcliffe, {\it A Monopole Metric,} Phys.Lett.
{\bf B391} (1997) 366, {\tt hep-th/9610141}
\semi
P. Sutcliffe, {\it BPS Monopoles,} Int.J.Mod.Phys. {\bf A12}
(1997) 4663, {\tt hep-th/9707009}.}
\lref\rBiel{R. Bielawski, {\it Monopoles and the Gibbons-Manton Metric,}
Commun.Math.Phys. {\bf 194} (1998) 297, {\tt hep-th/9801091}.}
\lref\rManton{N.S.Manton, {\it A Remark on the Scattering of BPS Monopoles,}
Phys.Lett. {\bf B110} (1982) 54
\semi
N.S.Manton, {\it Monopole Interactions at Long Range,} 
Phys.Lett. {\bf 154B} (1985) 395, Erratum-ibid. {\bf 157B} (1985) 475.}
\lref\rGM{G.W.Gibbons and N.S.Manton, {\it Classical and Quantum Dynamics
of BPS Monopoles,} Nucl.Phys. {\bf B274} (1986) 183
\semi
G.W.Gibbons and N.S.Manton, {\it The Moduli Space Metric for well
Separated BPS Monopoles,} Phys.Lett. {\bf B356} (1995) 32, 
{\tt hep-th/9506052}.}
\lref\rAH{M.F.Atiyah and N.J.Hitchin, {\it The Geometry and Dynamics of 
Magnetic Monopoles,} Princeton University Press, 1988.}
\lref\rSusyO{G.Gibbons, G.Papadopoulos and K.Stelle, {\it HKT and OKT 
Geometries on Soliton Black Hole Moduli Spaces,} Nucl.Phys. {\bf B508}
(1997) 623, {\tt hep-th/9706207}.}
\lref\rMSD{E. Weinberg, {\it Parameter Counting for Multi-Monopole
Solutions,} Phys.Rev. {\bf D20} (1979) 936
\semi
E. Corrigan and P. Goddard, {\it An N Monopole Solution with 4N-1
Degrees of Freedom,} Commun.Math.Phys. {\bf 80} (1981) 575.}
\lref\rGaunt{J.P. Gauntlett, {\it Low Energy Dynamics of N=2 Supersymmetric
Monopoles,} Nucl. Phys. {\bf B411} (1994) 443, {\tt hep-th/9305068}.}
\lref\rTown{J.P.Gauntlett, C.Koehl, D.Mateos, P.K.Townsend and M.Zamaklar,
{\it Finite Energy Dirac-Born-Infeld Monopoles and String Junctions,}
{\tt hep-th/9903156}.}
\lref\rDia{M.B.Green and M.Gutperle, {\it Comments on Threebranes,}
Phys.Lett. {\bf B377} (1996) 28, {\tt hep-th/9602077}
\semi
D.E.Diaconescu, {\it D-branes, Monopoles and Nahm's Equations,}
Nucl.Phys. {\bf B503} (1997) 220, {\tt hep-th/9608163}.}


\Title{ \vbox {\baselineskip 12pt\hbox{}
\hbox{}  \hbox{}  }}
{\vbox {\centerline{Threebranes in F-theory}
}}

\centerline{Alastair Paulin-Campbell}
\smallskip

\centerline{\it Department of Physics and Centre of Theoretical Physics,}
\centerline{\it University of the Witwatersrand,}
\centerline{\it Wits, 2050, South Africa}

\bigskip
\medskip

A thesis submitted to the Faculty of Science, University of the Witwatersrand,
Johannesburg, South Africa, in fulfillment of the degree of Doctor of Philosophy.

\Date{October 2001}

\vfill
\eject

{\bf Declaration}
\bigskip
I declare that this thesis is my own unaided work. It is being submitted for the degree of Doctor of Philosophy in the University of the Witwatersrand, Johannesburg, South Africa. It has not been submitted before for any degree or examination at any other University.

\bigskip\bigskip\bigskip
\hrule width 2.5 in

\smallskip
\noindent
(Alastair Paulin-Campbell)

\vfill\eject

\vglue 3 in

\centerline{\it{This thesis is dedicated to my mother, and the memory of Helen.}}

\vfill\eject

\noindent
{\bf{Acknowledgments}}

The hardest part of writing a thesis is undoubtedly the acknowledgements. In writing a thesis, there are usually two types of support, the tangible and the intangible. The tangible focuses on the nuts and bolts of obtaining a thesis, which in itself is a subtle fusion of imparting facts while teaching the special way of thinking so intrinsic to the discipline. The intangible, while necessarily much more difficult to characterise, encompasses moral support, love, and all those more subtle things required to make a complete individual. 

Firstly, let me give thanks to those individuals who were able to provide both. In this regard I extend my thanks to my supervisor, Jo\~ao Rodrigues, who made this journey possible by accepting a fresh faced student into the fold of theoretical physics. Under his guidance I took my first steps in this exciting field, an opportunity for which I will always be grateful. He has always been unstintingly generous in his financial support, a quality greatly appreciated by monetarily impoverished graduate students. I thank him for all he has done for me during the gestation of this thesis. 

In addition, I would like to mention my colleague and friend, Robert de Mello Koch. Robert has a deep love for physics and a great ability to communicate complex ideas simply and elegantly, which has the effect of both instructing and inspiring the listener at the same time. Robert will always go that extra mile for a friend, and I am indebted to him for the many kindnesses he has shown me when times become tough, as they inevitably do during the course of a PhD. Our friendship has inspired and enriched me in every sphere of my life.

While not involved directly in my thesis, special mention must also be made of Dieter Heiss, who was instrumental in inspiring me to pursue further my studies in physics. The contribution of his enthusiasm and energy cannot be overemphasized, and have played a central r\^ole in making the Physics department what it is today. 

In the intangibles category I would like to thank my friends Richard Lynch and Jonathan Crowhurst for their support, both moral and otherwise. Richard and I have enjoyed a special friendship over the years, attending both school and university together. Jonathan and I met as undergraduates. We surprised ourselves by becoming friends. As fellow physicists, I have had many inspiring and enjoyable discussions with both of them about the discipline we share; however, by far the most enduring result of our friendships is the myriad ways in which they have enriched my life and broadened my perspectives. I thank the both of you for all of the fun (as well as the angst) we shared in navigating that single river that ultimately split into the divergent paths we now follow. I am fortunate enough to be secure in the fact that, wherever life may take us, our friendships will remain firm.

I would also like to thank my friend Helen. Although she passed away under tragic circumstances many years ago, her memory is with me still. Sometimes, when my life is infused with chaos, I like to think that I hear her voice in the calm, gently admonishing my fears and banishing my indecision. I owe a great deal of my achievements to her gentle prodding. 

In the office, I must thank my colleagues Johan, Thuto, Mike and Phil who have always provided light hearted relief and a sense of camaraderie. It has been a pleasure to share an office with them, and I wish them the very best in the conclusion of their degrees.  

Lastly, and perhaps most significantly, I would like to thank my family, Rosamund, Catriona, Annemarie, Patricia and Sophie. They, more than most, have had to endure the brunt of my oscillating moods during the writeup of this thesis. They have provided support in innumerable ways, creating an environment that is pleasant and uplifting. They are the heartbeat of my day to day life, always there when I have needed them, always able to finesse my darker moods with an unerring intuition. For dealing with my unconventional and often difficult personality, I am indebted to them always. I thank my grandparents, Hugh and Margaret, for all of the things they have done for me, practical or otherwise. A more wonderful set of grandparents could not be found in all the length and breadth of Scotland. I treasure you both.

My cousin Mike and I have always been especially close; I wish to thank him for the innumerable times I have bent (if not permanently warped) his ear with discussions of string theory. He has always been an attentive and enthusiastic audience. I thank him for the many summers we spent together as children, and for the thousands of hours we dispensed in the pursuit of mischief. My younger years would have been all the poorer without his companionship.

I would like to especially mention my uncle, John Bruyns-Haylett, who understands better than most the rigours of research, for his constant encouragement and the amusing anecdotes of his days as a research student. I no longer take the internet for granted. 

This acknowledgements page has taken on a life of its own--however, part of the pleasure of concluding this thesis derives not only from enjoying the fruits of a job well done, but from savouring the special relationships which have placed me, whether directly or indirectly, in the happy position in which I now find myself. I thank you all again for bringing me here.

\vfill\eject

\noindent
{\bf{Preface}}
\medskip
\noindent
This thesis is based on published material. The work concerning non-holomorphic corrections to the effective action of ${\cal{N}}=2$ super Yang-Mills theory has appeared in Phys.Rev. D60 (1999) 106008. The section devoted to BPS states of ${\cal{N}}=2$ super Yang-Mills theory from the worldvolume theory of a threebrane probe (section (11) onward) appeared in Nucl.Phys. B559 (1999) 143-164.

\vfill\eject

\newsec{Introduction}

In this thesis we will study the effective action of ${\cal{N}}=2$ super Yang-Mills theory by means of a threebrane probe in F theory. The thesis is structured to place this work in context. We begin in section (2) by giving an introduction to bosonic string theory--this provides the necessary framework for a general understanding of the central concepts of string theory. In section (3), we discuss the notion of duality, specifically electromagnetic duality, as a preparation for introducing $T$-duality in string theory. The discussion of $T$-duality is required to motivate the introduction of Dirichlet branes, which play a central r\^ole in our present understanding of string theory, and in the work of this thesis. In section (4), we review the argument of Polchinski which states that Dirichlet branes carry $R$-$R$ charge and consequently source forms in the $R$-$R$ sector of $IIA$ and $IIB$ string theories, which are perturbatively theories of closed strings.
In section (5), the notion of open$\backslash$closed string duality is discussed, first in the context of the Veneziano amplitude, and then in relation to the conjecture of Maldacena, which is a duality between supergravity (a closed string theory) and a super Yang-Mills theory (an open string theory). In this section a link between Born-Infeld and the effective action of a gauge theory is motivated, in which the effective action of the gauge theory is realised as the worldvolume theory of a threebrane probing a supergravity background specific to the field theory of interest. In section (6) some relevant aspects of ${\cal{N}}=2$ supersymmetric Yang-Mills theory are reviewed. In the following section, we discuss the observation of Sen that certain exact backgrounds of $IIB$ correspond to solutions of the Seiberg-Witten monodromy problem, which has relevance for the construction of the supergravity backgrounds probed by the threebrane in the description of ${\cal{N}}=2$ supersymmetric Yang-Mills theory. In section (8), and the remainder of the thesis thereafter until section(11), relevant field theory results are reviewed followed by the results of the probe analysis. The bulk of the results presented in this section concern the structure of the four derivative terms as predicted by the probe analysis of ${\cal{N}}=2$ supersymmetric Yang-Mills theory with $N_F=4$ massless flavors, $N_F=4$ massive flavors and, lastly, the case of the pure gauge theory. From section (11) onwards, BPS states of the field theory are investigated as non-trival finite energy solutions on the probe worldvolume. From the point of view of the probe worldvolume, these solutions correspond to magnetic monopoles. The metric on moduli space of these solutions is studied. Finally, in section (15), we provide a summary of results and conclusion.

\vfill\eject

\newsec{Introduction to String Theory}

String theory as it exists today began as an attempt to construct the quantum theory of a one dimensional extended object, the string. We will attempt to give a brief introduction to this subject to give a flavour for some important concepts which will be needed throughout. In order to introduce the formalism used in string theory, a natural starting point is to first introduce and motivate the formalism underlying the point particle, and then proceed to develop the formalism of string theory guided by the point particle analogy{\foot{An excellent reference for this section is given by \gsw\ and references therein.}}.

We will begin by trying to construct a suitable relativistic action for a massive point particle. One of the simplest possible choices is given by

\eqn\act
{S=-m\int ds}

\noindent
This action is proportional to the length of the worldline of the particle, a quantity which has a general coordinate invariance. One may rewrite this action as \foot{Our metric has signature $(1,-1,-1,-1)$, and we have set $\hbar =c=1$.}

\eqn\rewr
{S=-m\int d\tau \sqrt{g_{\mu\nu}{\partial x^\mu\over\partial \tau}{\partial x^\nu\over\partial \tau}}}

\noindent
The indices $\mu ,\nu $ are spacetime indices, and the quantity $\tau $ is a
parameter which, when varied, moves one along the wordline. One can verify
that the action has an invariance under reparametrizations of the form\foot{Strictly speaking, one should also require that ${d\tau^{\prime}\over d\tau}$ does not change sign, to prevent `zigzagging'.}
\eqn\tot
{\tau \to \tau^{\prime}(\tau)}

\noindent
One might view this heuristically as arising from the fact that the length
of the worldline is geometrical in character, and cannot depend on how it is
parametrized. Let us examine the low energy limit of this action in order to
gather more evidence for its suitability as the action describing a point
particle. Due to the reparametrization symmetry of the action, we are free
to choose $\tau =x^0,$ in which case

\eqn\gauge
{S=-m\int dx^0 \sqrt{1-{\partial x^i\over\partial x^0}{\partial x^i\over\partial x^0}}}

\noindent
At low velocities $\dot{x}^i<<1$ we may expand the square root to obtain
\eqn\expand
{\eqalign{S&=-m\int dt \left( 1-{1 \over 2}\vec{v}\cdot \vec{v}+O(v^4)\right)
\cr &=\int dt \left(-m+{1\over2}mv^2+O(v^4)\right)}}

\noindent
which has the form
\eqn\action
{S=\int dt \ (T-V)}

\noindent
where the potential energy is interpreted as the energy due to the mass of
the particle. This action reduces to the correct low energy result, and has
the expected symmetries, which ensures its suitability for describing a
point particle. In order to guess a suitable candidate for the action of a
string theory, let us proceed by analogy. For the point particle, we chose
an action such that $S\sim L,$ where $L$ is the length of the particle
worldline. Consequently, it seems reasonable to choose an action for a
string where $S\sim A,$ where $A$ is the area of the string ``worldsheet''.

The action must be of the form

\eqn\stringact
{S=-{1\over2\pi \alpha ^{\prime }}\int d\sigma d\tau {\cal{L}}_{NG}}

\noindent
where $\tau $ and $\sigma $ are coordinates parametrizing the worldsheet. The new quantity $\alpha^{\prime}$ is the string tension. The explicit form of the Lagrangian density is

\eqn\stlagra
{{\cal{L}}_{NG}=\sqrt{|\det \partial _\alpha X^\mu \partial _\beta X_\mu }|}

\noindent
The index $\mu $ is a spacetime index, while $\alpha ,\beta $ are worldsheet indices. This action was first proposed by Nambu and Goto \gsw. It is easy to verify that the above action has a reparametrization invariance under the replacements{\foot{As in the single particle case, we require that the Jacobian of the transformation does not change sign.}} 
\eqn\reparam
{\eqalign{&\sigma \to \sigma^{\prime }(\tau ,\sigma) \cr &\tau  \to \tau ^{\prime }(\tau ,\sigma )}}

\noindent
which one may interpret heuristically as arising from the fact that the area of the string worldsheet must be independent of its parametrization.

In the case of the string, our intuition about a low energy limit is not as clear cut as it is for the point particle. A reasonable starting point might be to assume that, at low energies, the extended nature of the string is no longer evident and so it appears point like. Accordingly, we should expect that at low energies the action should reduce to a point particle action. In order make the low energy limit more precise, we make use of the reparametrization invariance of the Nambu--Goto action to identify $X^0=\tau$, $X^1=\sigma$ which corresponds to an open string lying along the $X^1$ direction. It is easily seen that the Nambu-Goto action becomes 

\eqn\lequoo{S \sim -{1 \over \alpha ^{\prime }}\int d\sigma dt \sqrt{|(1-\partial _t X^j \partial _t X_j)(1+\partial _\sigma X^j \partial _\sigma X_j)+ (\partial _\sigma X^j \partial _\tau X_j)^2}|}

\noindent
where $j \neq 0,1$. In taking the low energy limit, we assume that the fields $X^j$ do not depend on the `small' coordinate $\sigma$, and hence one obtains 

\eqn\lee{{S \sim - {1 \over \alpha ^{\prime }}\int dt\sqrt{(1-\partial _t X^j \partial _t X_j)} \int d\sigma}=- {l \over \alpha ^{\prime }}\int dt\sqrt{(1-\partial _t X^j \partial _t X_j)} }

\noindent
where we take $l$ to be the string `length'. One may now identify $m \equiv{l \over \alpha ^{\prime }}$ as a quantity with units of mass. Consequently, the action may be written as 

\eqn\lee{{S \sim -m\int dt\sqrt{(1-\partial _t X^j \partial _t X_j)}}}

\noindent
which is identical to the point particle action in the gauge $\tau=x^0$.

The low energy limit and symmetries of this action suggest that it is indeed a suitable candidate to describe an extended one dimensional object; however, it is not clear how to handle this action as it stands due to the appearance of the square root. In the formalism of quantum field theory, a starting point is given by the terms quadratic in the fields. These terms have an additive energy spectrum, a property one would expect any theory of free particles to exhibit. The significance of this is that it naturally implies a particle interpretation which is inherited by the interacting theory. The central idea of perturbation theory is to add small corrections to the exactly soluble free part, which has the advantage that it provides a heuristic physical picture of particles and their interactions. The Nambu-Goto action does not have such a quadratic structure, so in the context of quantum field theory it is not clear how to make progress.

It was the work of Polyakov \AMP\ who showed that the Nambu-Goto action can be written in the form

\eqn\polyakov
{S=-{1\over{{4\pi \alpha ^{\prime }}}} \int d\tau d\sigma \sqrt{|\det \gamma |}\gamma ^{ab}\partial _aX^\mu \partial _bX_\mu}

\noindent
where the worldsheet metric $\gamma ^{ab}$ is a new feature. The indices $a,b$ are worldsheet indices, and $\mu $ is a spacetime index. It is easily verified that eliminating $\gamma _{ab}$ using its equation of motion yields the Nambu-Goto action.

 At first sight it appears that we have traded one difficulty for another, since although the action is quadratic in the fields $X^\mu $ the worldsheet metric enters in a rather nontrivial way. One possible way to deal with this might be to use the equation of motion to obtain a classical solution for $\gamma _{ab}$ which may be expanded about a vacuum configuration, yielding a term quadratic in the worldsheet metric. Alternatively, the action possesses a number of classical symmetries which we might be able to use to our advantage. 

In two dimensions, the worldsheet metric has only three independent components. At the classical level, we have two diffeomorphism invariances, and an invariance under Weyl rescalings. One might hope that these three symmetries could be used to eliminate the worldsheet metric from the theory. However, this relies crucially on the requirement that these classical symmetries still hold at the quantum level. It can be shown that requiring that the classical symmetries do not develop anomalies at the quantum level restricts us to a certain dimensionality of spacetime--the so called critical dimension. In the case of the bosonic string, the critical dimension is $d=26,$ however, for the superstring it is $d=10.$ In what follows, it will be assumed that we work in the critical dimension. 

A natural question to ask is whether our specification of the action is as general as it could be, that is to say, are there any additional terms which could be added to the action? This question is important since, in general, any terms which are consistent with the symmetries of the theory will have a non-zero coefficient after renormalization. One possibility is the introduction of a cosmological constant term,

\eqn\cosmo
{S=\int d\tau d\sigma \sqrt{-\det \gamma }\,\Lambda}
\noindent
However, this term explicitly breaks Weyl symmetry. It cannot be included on the grounds that we would be unable to use the Weyl symmetry to eliminate the worldsheet metric, which was the strategy employed to get the Polyakov action into a manageable form. {\foot{This term does not appear in the classical action. However, at the quantum level issues arise in defining a Weyl invariant measure. Consequently, this term may make an appearance in the form of a loop correction in such a fashion as to ensure the invariance of the path integral under Weyl transformations in the critical dimension.}} 

Another possibility is given by the term
\eqn\hilbert
{S=\int d\sigma d\tau {\lambda\over4\pi}R}
\noindent
which proves to be unnecessary, since in two dimensions this can be written as a total derivative which does not affect the equations of motion. However, when considering string interactions, this term will play a r\^ole since it is sensitive to the topology of the string worldsheet.

We now work in the conformal gauge, where $\gamma _{ab}=\delta _{ab}$, mindful of the fact that this is only possible at the quantum level in the critical dimension. The action is written
\eqn\gauact
{S=-{1\over4\pi \alpha ^{\prime }}\int d\tau d\sigma \partial _aX^\mu \partial _aX_\mu}

\noindent
At first sight, this looks like the theory of a massless scalar in two dimensions. However, this is not so since one has still to impose

\eqn\vira
{{\delta S\over\delta \gamma _{ab}}=T^{ab}=0}

\noindent
as a constraint. Bearing in mind that the coordinate of the worldsheet $\sigma $ is compact, the mode expansion of the $T^{ab}$ will be discrete. The modes corresponding to the $T^{ab}$ are denoted $L_n,$ and the constraint may be simply expressed in terms of them as

\eqn\stconst
{L_n=0}

\noindent
The comment is often made that the gauge fixed action describes a massless scalar, in which case one would just have the statement

\eqn\sccond
{p_ap^a=0}

\noindent
where $a$ is a worldsheet index. This statement amounts to the fact that the spacetime coordinate of the string $X^{\mu}$ is made up of the sum of holomorphic and antiholomorphic terms, {\it{i.e.}} it satisfies   

\eqn\sccond
{\partial_z \partial_{\bar{z}}X^{\mu}=0}

\noindent
where we have assembled the worldsheet coordinates into a single complex coordinate $z$. However, in this case we must also enforce the additional constraint Eq. \stconst. In the quantum theory, we satisfy the constraint Eq. \stconst\ by requiring that physical states are annihalated by $L_n$, that is,

\eqn\sccond
{L_n|phys\rangle=0, \; \; \; n \geq 0}

\noindent
It turns out that the mass spectrum of the theory arises from the constraint

\eqn\stmass
{(L_0-a)|phys\rangle=0}
\noindent
where $a$ is a normal ordering constant, which implies

\eqn\massfom
{m^2=-p^\mu p_\mu = {1\over\alpha ^{\prime }}\left( \sum_{m=1}^\infty mN_m-a\right)}
\noindent
where $N_m$ is the number of excitations in the $m$'th mode.

\vfill
\eject

\newsec{Duality}

In order to gain greater insight into duality it will be instructive to examine the case of classical electromagnetism\rHarv. Our ultimate goal will be to understand $T$-duality in the context of string theory. Electromagnetism has the advantage that it exhibits many of the important features of duality in the relatively simple setting of a point particle theory. We will identify some generic features of duality in this setting, which will be used to guide us in our analysis of string theory.

\subsec{Duality in Classical Electromagnetism}
\noindent
The sourceless equations of motion of classical electromagnetism are

\eqn\emequ
{\partial _\mu F^{\mu \nu }=0}
\noindent
where the field strength $F^{\mu \nu }$ is defined in the usual way,

\eqn\fdefn
{F^{\mu \nu }=\partial ^\mu A^\nu -\partial ^\nu A^\mu}
\noindent
The field strength is expressed in terms of the electric and magnetic fields as
\eqn\deef
{F^{0i}=-E^i}
\noindent
and

\eqn\mdef
{F_{ij}=-\varepsilon _{ijk}B^k}
\noindent
Setting $\nu =0$ in \emequ\ yields

\eqn\ingauss
{\partial _iF^{i0}=0}
\noindent
which is Gauss' law,

\eqn\gauss
{\vec{\nabla}\cdot \vec{E}=0}
\noindent
Similarly, setting $\nu =i$ in \emequ\ one obtains

\eqn\max
{\vec{\nabla}\times \vec{B}-{\partial \vec{E}\over\partial t}=0}
\noindent
These are two of the four Maxwell equations, yet we have already exhausted the content of the equations of motion. The key to recovering the two remaining Maxwell equations lies in realising that the field strength $F^{\mu \nu }$ must obey a constraint. The reasoning behind this has its origin in the fact that one can formulate electromagnetism entirely in terms of the four potential $A^\mu$, which has only four components, while the antisymmetric tensor $F^{\mu \nu }$ has six. Accordingly, it is clear that repackaging the four component $A^\mu$ in terms of the field strength tensor implies that $F^{\mu \nu }$ must satisfy a constraint. To uncover the constraint, define the dual field

\eqn\fstar
{^{\ast }F^{\alpha \beta }={1\over2}\varepsilon ^{\alpha \beta \mu \nu }F_{\mu \nu}}

\noindent
where $\varepsilon ^{\alpha \beta \mu \nu }$ is the maximally antisymmetric tensor density in $3+1$ dimensions. One may easily verify that the following relation is satisfied,

\eqn\const
{\partial _\alpha {}^{*}F^{\alpha \beta }=0}
\noindent
The above contraction vanishes since we contract a symmetric tensor with an antisymmetric tensor\foot{We have assumed that the partial derivatives commute, {\it i.e.} that the vector potential $A^\mu $ is smooth.}; it must be stressed that this is in stark contrast to \emequ\ which is a statement about the dynamics of the system. The constraint equation has a topological origin; this can be seen by defining the one form

\eqn\defna
{A\equiv A_\mu dx^\mu}
\noindent
The action of the exterior derivative is

\eqn\extder
{\eqalign{dA &=\partial _\nu A_\mu \,dx^\mu \wedge dx^\nu \cr &={1\over2}F_{\mu \nu }dx^\mu \wedge dx^\nu \cr &\equiv F}}

\noindent
The above equation indicates that the components $F^{\mu \nu }$ are the components of an exact form. It is known that all exact forms are closed, that is, the action of the exterior derivative on $F$ is
\eqn\ext
{dF=0=\partial _\rho F_{\mu \nu }dx^\rho \wedge dx^\mu \wedge dx^\nu}
\noindent
which is equivalent to
\eqn\formcons
{\varepsilon ^{\alpha \beta \mu \nu }\partial _\beta F_{\mu \nu }=0}
\noindent
since the wedge products require us to consider the antisymmetric part of $\partial _\rho F_{\mu \nu }$.
\noindent
The topological nature of the constraint is now manifest, since it essentially arises from the statement
\eqn\poincare
{d^2=0}
\noindent
which can be mapped into the statement ``the boundary of a boundary is zero" by Poincare duality,
\eqn\poincare
{\partial^2=0}
\noindent
where $\partial$ is the boundary operator.

The field strength $F^{\mu \nu }$ may be thought of as describing curvature in ``charge space''--for this reason, the constraint equation is often called the Bianchi identity, since it is the analogue of the relation obeyed by the Riemann curvature tensor in general relativity. Thus, if we wish to formulate the theory in terms of a field strength, we must supplement the equation of motion by the Bianchi identity.

\noindent
The definition \fstar\ straightforwardly gives

\eqn\dedefn
{^{\ast }F^{ij}=\varepsilon ^{ijk}E^k}
\noindent
and

\eqn\defdmfn
{^{\ast }F^{0i}=-B^i}

\noindent
The equations implied for the electric and magnetic fields by the constraint are \eqn\vconso
{\vec{\nabla}\cdot\vec{B}=0}
\noindent
and

\eqn\vcont
{{{\partial \vec{B}} \over {\partial t}}+\vec{\nabla}\times \vec{E}=0}
\noindent
We note that the Maxwell equations are left invariant under the following discrete transformation,

\eqn\distran
{\eqalign{\vec{E} &\to \vec{B} \cr \vec{B} &\to -\vec{E}}}
\noindent
The transformation acts on the field strength and its dual as
\eqn\ftrans
{\eqalign{F^{\mu \nu } &\to ^{*}F^{\mu \nu } \cr ^{\ast }F^{\mu \nu } &\to -F^{\mu \nu }}}

\noindent
In the presence of electrical charges, the equation of motion for the field strength is modified to

\eqn\sources
{\partial _\mu F^{\mu \nu }=j^\nu}
\noindent
with the constraint \eqn\cnt
{\partial _\mu ^{*}F^{\mu \nu }=0}
\noindent
One sees that the duality transformation given above no longer holds. Classically, it is easy to remedy this by simply adding magnetic charges to the system, such that the above equations are modified to

\eqn\emcharge
{\eqalign{\partial _\mu F^{\mu \nu}&=j^\nu \cr \partial _\mu ^{*}F^{\mu \nu }&=k^\nu}}
\noindent
One can arrange this by choosing an appropriate vector potential $A_\mu $ which requires more than one chart for its description.
\noindent
It is clear that the original symmetry

\eqn\restore
{\eqalign{F^{\mu \nu } &\to ^{*}F^{\mu \nu } \cr ^{\ast }F^{\mu \nu } &\to -F^{\mu \nu }}}
\noindent
is restored when supplemented by a transformation law for the four currents, \eqn\transk{\eqalign{j^\nu &\to k^\nu \cr k^\nu &\to -j^\nu}}
\noindent
A striking feature is that the restoration of the duality symmetry forces us to have two charges; one with a dynamical origin, and one of a topological nature. What is the significance of the above symmetry? It is telling us that the theory has two equivalent descriptions--the electric and magnetic descriptions. For example, instead of writing an action in terms of the field strength, one might imagine writing down an action in terms of the dual field,

\eqn\dualact
{S=-{1\over 4}\int d^4x^{*}F^{\mu \nu }\,^{*}F_{\mu \nu }}
\noindent
where the dual field strength is defined in terms of a dual vector potential $A_D^\mu $ as follows
 \eqn\dualfdef
{^{\ast }F^{\mu \nu }=\partial ^\mu A_D^\nu -\partial ^\nu A_D^\mu}

\noindent
The equation of motion is found by varying the action with respect to the dual potential,
\eqn\demequ
{\partial _\mu \,^{*}F^{\mu \nu }=0}
\noindent
We can now define the field strength by inverting our previous definition,

\eqn\fdefd{F^{\mu \nu }={1\over 2}\varepsilon ^{\mu \nu \alpha \beta }\,^{*}F_{\alpha \beta }}
\noindent
In this case, one can easily verify that we obtain the constraint
\eqn\dualcons{\partial _\mu \,F^{\mu \nu }=0}
\noindent
In the presence of magnetically charged matter, the equation of motion is modified to include a magnetic four-current by coupling the current in the usual way to the dual potential, \eqn\dualcurr{\partial _\mu \,^{*}F^{\mu \nu }=k^\nu}
\noindent
If we wish to engineer an electric four current, the appropriate dual potential cannot be described globally by a single chart,

\eqn\dualemmot{\partial _\mu \,F^{\mu \nu }=j^\nu}
\noindent
Hence writing the theory in terms of the dual field (and thus the dual potential $A^\mu )$ gives rise to the description we obtained by performing a duality transformation on the electric description. One sees that there are two equivalent descriptions of the same theory. Switching between these descriptions is achieved by exchanging the field strength and dual field, and their respective four currents as prescribed above. The choice of which variables to use is entirely arbitrary, although one may exploit the advantages of the electric or magnetic description depending on the physics one wishes to describe. For example, in the electric description, the magnetic monopole is realised via a vector potential which cannot be described globally by a single chart, in contrast to the dual vector potential, which may have a single globally valid chart. However, perhaps one of the most striking insights afforded us is that duality symmetry requires two conserved charges of differing origins--in the electric description, the electric charge follows from the equations of motion, while the magnetic charge derives from the constraint. Performing the duality symmetry swaps the origins of the charges.

\vfill\eject

\subsec{$T$-Duality}
How can we apply the insights we have gained to the case of the bosonic string? The key feature of electromagnetic duality was the appearance of the two conserved charges which are dynamical and topological in origin, respectively. Applying this insight to string theory one might wish to look for two such charges in the hope that exchanging these will lead to dual descriptions of the same physics. In the case of the bosonic string one obvious conserved Noether charge is the centre of mass momentum. Extending the parallel with electromagnetism, one would hope to identify a topological charge. One way to generate such a charge is to compactify one of the directions, admitting the possibility that a string may wind around this compact direction. If the string is closed, one would expect the associated winding number to be conserved. In addition, the non-trivial cycles of the space we consider implies the topological nature of the winding number. The duality we review in this section is covered comprehensively in \tasi. 

Consider the solution to the equation of motion for the closed bosonic string\gsw,

\eqn\lr{X^\mu (z,\bar{z})=X_L^{\mu} (z)+X_R^{\mu} (\bar{z})}
\noindent
with

\eqn\xzee
{X_L^\mu (z)={1\over 2}x^\mu +\sqrt{{{\alpha ^{^{\prime }}}\over 2}}\left( -\alpha _0^\mu \ln z+\sum_{n\neq 0}{{\alpha _n^\mu }\over {nz^n}}\right) }

\noindent
and

\eqn\xzb
{X_R^\mu (\bar{z})={1\over 2}x^\mu +\sqrt{{{\alpha ^{^{\prime }}}\over 2}}\left( -\tilde{\alpha}_0^\mu \ln \bar{z}+\sum_{n\neq 0}{{\tilde{\alpha}_n^\mu }\over {n\bar{z}^n}}\right) }

\noindent
where $z=e^{\tau -i\sigma }.$

A closed string in a non-compact space must satisfy the boundary condition
\eqn\bc{X^\mu (\tau ,\sigma )=X^\mu (\tau ,\sigma +2\pi )}
\noindent
which ensures that the string closes on itself. One sees that the oscillator terms satisfy this requirement; however, the zero modes contain a term linear in $\sigma ,$

\eqn\zero{X^\mu (\tau ,\sigma )=x^\mu -i\sqrt{{{\alpha ^{\prime }}\over 2}}\left( \alpha _0^\mu -\tilde{\alpha}_0^\mu \right) \sigma -\sqrt{{{\alpha ^{\prime }}\over 2}}\left( \alpha _0^\mu +\tilde{\alpha}_0^\mu \right) \tau +oscillators}

\noindent
Consistency with the boundary condition requires

\eqn\bcnoncom{\alpha _0^\mu=\tilde{\alpha}_0^\mu }
\noindent
since this removes the linear $\sigma$ dependence. We may now write
\eqn\clbc{X^\mu (\tau ,\sigma )=x^\mu -\alpha ^{\prime }p^\mu \tau +\sum_{n\neq 0}{1\over n}\left({{\alpha _n^\mu }\over {z^n}}+{{\tilde{\alpha}_n^\mu }\over {\bar{z}^n}}\right) }

\noindent
where we have identified\foot{It is not too difficult to convince oneself of this. From the action, the momentum ``density'' ${\cal{P}}\sim \dot{X}.$ Only the zero modes can contribute to the charge $Q\sim \int d\sigma \,\cal{P}$ since the oscillating terms integrate to zero.} 

\eqn\mom{p^\mu =\sqrt{2\over {\alpha ^{\prime }}}\alpha _0^\mu}

\noindent
Note that the boundary conditions have halved the number of conserved charges.

One may contrast this solution with the open string, which requires that we enforce Neumann boundary conditions

\eqn\neu{\partial_{\sigma} X^{\mu}(\tau,0)=0=\partial_{\sigma} X^{\mu}(\tau,2\pi)}

\noindent
which leads to the following solution,

\eqn\open
{X^\mu (\tau ,\sigma )=x^\mu -\sqrt{2\alpha ^{\prime }}\alpha^{\mu}_0 \tau +\sqrt{2\alpha ^{\prime }} \sum_{n\neq 0}{1\over n}\alpha _n^\mu e^{-n\tau}\cos(n\sigma)} 

\noindent
where we now have only {\it{one}} set of oscillators $\alpha^{\mu}_n$.

Returning to the closed string solution,the two sets of oscillators ${\tilde{\alpha}_n^\mu }$, ${{\alpha}_n^\mu }$ create left and right moving modes which propagate independently about the string. The quantities $x^\mu ,$ $p^\mu $ are to be interpreted as the position and momentum of the centre of mass of the string in spacetime, while the oscillator terms describe internal excitations of the string. Some insight can be gained by realising that the solution has the following form,

\eqn\newton
{X\sim x_{cm}+v_{cm}\tau +oscillators}

\noindent
which implies {\it(i)} that it describes an object whose centre of mass is not subject to any external forces, (implying in turn the conservation of the centre of mass momentum $p^\mu $), and {\it (ii)} has internal degrees of freedom described by the oscillator terms.

The boundary condition given above assumed that the space was non-compact. In the case of a compact direction another possibility arises, namely that the string may wind around this direction. So, under $\sigma \to \sigma +2\pi $ we can have

\eqn\compactbc
{X^{25}(z,\bar{z})\to X^{25}(z,\bar{z})+2\pi wR}

\noindent
where the integer $w$ is the number of times the string wraps the compact direction. Studying the zero modes of the general solution and setting $\sigma \to \sigma +2\pi$ gives

\eqn\windcomp
{X^{25}(\tau ,\sigma+2\pi )=X^{25}(\tau ,\sigma )+2\pi \sqrt{{{\alpha ^{\prime }}\over 2}}\left( \alpha _0^{25}-\tilde{\alpha}_0^{25}\right)}

\noindent
Imposing the boundary conditions requires us to set
\eqn\walpha
{\sqrt{{{\alpha ^{\prime }}\over 2}}(\alpha _0^{25}-\tilde{\alpha}_0^{25})=wR}
\noindent
This gives rise to our second conserved charge; we note that it has its origins in the boundary conditions, not the dynamics. The centre of mass momentum is given by the zero modes as
\eqn\momalpha
{p^\mu ={1\over{\sqrt{2\alpha ^{\prime }}}}\left( \alpha _0^\mu +\tilde{\alpha}_0^\mu \right)}

\noindent
However, the momentum of a particle on a ring is quantized as $n\over R,$ where $n$ is an integer, and hence the momentum of the centre of mass must satisfy \eqn\test{{1\over{\sqrt{2\alpha ^{\prime }}}}\left( \alpha _0^{25}+\tilde{\alpha}_0^{25}\right) ={n\over R}}

\noindent
The left and right moving momenta conjugate to $z,\bar{z}$ may be written as

\eqn\conjmomz
{\alpha _0^{25}={{wR}\over {\sqrt{2\alpha ^{\prime }}}}+\sqrt{{\alpha ^{\prime }\over 2}}{n\over R}}

\noindent
and

\eqn\conjmomzb
{\tilde{\alpha}_0^{25}=\sqrt{{\alpha ^{\prime } \over 2}}{n\over R}-{{wR}\over{\sqrt{2\alpha ^{\prime }}}}}

\noindent
The zero mode expansion may be rewritten in terms of the winding and momentum as \eqn\zmpw{X^{25}(\tau ,\sigma )=x^{25}-\sigma wR-i\alpha ^{\prime }{n\over R}\tau +oscillators}

\noindent
Now that we have found two conserved charges the next step is to swap the roles of momentum and winding. Exchanging $n\leftrightarrow w,$ and requiring that the zero mode spectrum should remain unchanged, we must set

\eqn\dualradius{R\rightarrow {{\alpha ^{\prime }}\over R}=\tilde{R}}
where we call $\tilde{R}$ the dual radius. Thus we see that performing this set of transformations takes us to a theory with the same zero mode spectrum and a compact direction of radius $\tilde{R}$. In the literature of string theory this duality is known as `T-duality'\tasi.

The above transformation requires the zero mode oscillators to transform as

\eqn\zeroosc{\tilde{\alpha}_0^{25}\rightarrow -\tilde{\alpha}_0^{25}}
\noindent
while ${\alpha}_0^{25}$ remains unchanged. Recall that the zero mode oscillators are proportional to the left and right moving momenta, 

\eqn\zero{\tilde{\alpha}_0^{25} \sim p_R^{25}, \ \alpha_0^{25} \sim p_L^{25}}
\noindent
Consequently, performing a $T$-duality takes $p_R^{25} \to -p_R^{25}$ and $p_L^{25} \to p_L^{25}$, which sets 

\eqn\su{X^{25}(z,\bar{z})=X_L^{25}(z)-X_R^{25}(\bar{z})}
\noindent
Hence, performing a $T$-duality acts with spacetime parity on one half of the theory.

Thus far, the notion of $T$-duality has been explored in the context of the closed string. If we are to consider swapping momentum and winding in the case of the open string, things differ somewhat in comparison to the closed string case. Firstly, the momentum for an open string is conserved, however the concept of winding is not as well defined. We might try to define a notion of winding by walking along the open string and counting the number of times it wraps the compact direction. However, it is clear that since the ends of the open string are free to move independently, the dynamics of the string will not, in general, preserve this quantity. If we were to naively swap the momentum with this quantity, it would seem that we would be moving to a system with a conserved winding and a non-conserved momentum. The appearance of a non-conserved momentum implies that translational invariance in the compact direction has been broken. This signals the appearance of an object in the compact direction which provides a natural origin in the space, and hence breaks translational invariance. However, in the directions transverse to the compact space, the momenta are all conserved. This suggests that the object, while appearing as a point in the compact direction, is space filling in the transverse directions. We have managed to explain away the non conservation of momentum by realising that the exchange of momentum and winding has led to the appearance of a twenty five dimensional object. In this context, would one be able to define a conserved winding number? As noted above, if the ends of the string are free to move, we can easily unwrap the string and change the winding number. However, we could imagine wrapping the string about the compact direction and affixing its endpoints to a plane--the winding number is now conserved. The natural conclusion is that the conservation of winding arises since the endpoints of the open string are constrained to end on the object which exists in the twenty five transverse directions. The requirement that the string endpoints are held fixed is simply the Dirichlet boundary condition--hence, the object is called a `$D$-brane', or Dirichlet brane. In what follows, we will employ the notation $Dp$-brane, to indicate a Dirichlet brane of spacetime dimension $p+1$.

Assuming that there are additional compact directions, further $T$-dualities may be performed on this system, along directions either parallel or transverse to the brane. It is clear that the directions transverse to the brane have non-conserved momenta associated with them since the $Dp$-brane appears as a point in these directions; conversely, the directions parallel to the brane are associated with conserved momenta due to translational invariance. In addition, each compact direction has a winding number which is conserved since string endpoints are constrained to end on the brane.

Performing a $T$-duality in a direction transverse to the brane serves to restore the conservation of momentum along this direction due to the exchange of momentum and winding. Consequently, there is an additional direction in which the brane is space filling, and thus the dimension of the brane has increased. Hence performing a $T$-duality in a direction transverse to a $Dp$-brane produces a $D(p+1)$-brane.

Let us consider the case of the parallel directions. Exchanging the r\^ole of the momentum and winding we move to a theory with an additional non conserved momentum. Following our prior reasoning, directions perpendicular to the brane are characterised by non conserved momenta; hence, we have gained an additional perpendicular direction, which leads us to conclude that $T$-duality parallel to a $Dp$-brane produces a $D(p-1)$-brane.

It is interesting to note that one may interpret the original bosonic string theory as containing a twenty six dimensional $D25$-brane; such an object would be completely space filling, and thus string endpoints are free to move anywhere throughout the space. In this case, performing a $T$-duality along any direction is parallel to the $D25$-brane; accordingly, it takes one to a theory with a $D24$ dimensional brane.

If $T$-duality is indeed a duality, we should expect that it should give a description of the same physics and hence not change any predictions of the theory. The quantities which can be predicted are the conserved charges of particles, which arise from the zero mode spectrum, and scattering amplitudes. We have shown above that the zero mode spectrum is left invariant under the action of $T$-duality. In order to consider the issue of scattering amplitudes one should examine the interactions, described by vertex operator insertions. These are left invariant under the action of $T$-duality since they depend on the operator product expansion which is local and therefore not sensitive to ``global information'' such as the radius of the compact direction. 

We note in passing that, at $R=\sqrt{\alpha^{\prime}}$, the self dual radius, $T$-duality becomes an exact symmetry of the theory. The restoration of a symmetry is witnessed by the appearance of extra massless gauge particles, and thus we consider the zero modes, which may be written as

\eqn\zer{\eqalign{\tilde{\alpha}_0^{25}&= {1\over \sqrt{2}}(n-w) \cr \ \alpha_0^{25}&={1\over \sqrt{2}}(w+n)}}
\noindent
at $R=\sqrt{\alpha^{\prime}}$. It is clear that, for some values of $w$ and $n$, the zero modes vanish and thus one sees that new massless states appear in the spectrum. 

This concludes the discussion of $T$-duality in the bosonic string. In the following section we will discuss the fact that forms from the $R$-$R$ sectors of $IIA$ and $IIB$ supersymmetric string theories, which are perturbatively closed string theories, are sourced by Dirichlet branes, which naturally imply open string degrees of freedom\joep. 

\vfill\eject
\newsec{$R$-$R$ charge and $D$-branes in Type $IIA$ and $IIB$ string theory}

In the previous section, the basic concepts necessary to understand $T$-duality in the context of the open and closed bosonic string were developed. The action of $T$-duality in flat spacetime closed string theory appears to be rather trivial since it is not necessary to introduce the notion of a $D$-brane for the sake of consistency. In the case of Type $IIA$ and $IIB$ superstring theories, which are perturbatively closed string theories, the basic concepts of $T$-duality discussed for the bosonic string survive virtually unchanged but have some remarkable consequences.

In order to consider supersymmetric string theories, worldsheet supersymmetry requires that, for each scalar field $X^{\mu}$, there is a corresponding Majorana fermion $\psi^{\mu}$ {\foot{We note that this is in the R-NS formulation of string theory.}}. It is important to note that the requirement of worldsheet supersymmetry does not specify supersymmetry in spacetime. In order to achieve this one implements the GSO projection which requires that states with even fermion number are retained. This projects out the tachyon found in the bosonic theory, and leaves a theory supersymmetric in spacetime.

Just as in the bosonic case, the fermionic degrees of freedom separate into independent left and right moving sectors. Type $IIA$ is a theory which is non chiral in {\it{spacetime}}, that is to say, has both left and right moving fermionic degrees of freedom. By contrast, Type $IIB$ string theory is a chiral theory in spacetime \foot{This situation is reversed when considering worldsheet parity!}.
\noindent
There are two possible boundary conditions which may be applied to the fermions,

\eqn\fermbc{\psi^{\mu}(\sigma+2\pi)=\pm\psi^{\mu}(\sigma)}
\noindent
where the periodic boundary conditions are known as Neveu-Schwarz ($NS$), and the anti-periodic boundary conditions are referred to as Ramond ($R$).

As observed earlier in the context of the bosonic string, the action of $T$-duality acts like a parity transformation on one half of the theory. The action of $T$-duality on Type $IIB$ string performs a spacetime parity on one sector of the fermions, which transforms one from a chiral theory in spacetime to a non chiral theory--Type $IIA$ string theory.

In the beginning of this section we have had in the back of our minds the R-NS formalism, where one has supersymmetry on the worldsheet, but not in spacetime. For the following discussion, we will work in the GS formalism, where spacetime supersymmetry is manifest. This will be useful in what follows since our interest will be in the properties of the fields under spacetime transformations. 
In what follows, we will consider the fields $\lambda_1$, $\lambda_2$ from the Ramond sector as arising from the GS formalism, that is to say, they are spinors in spacetime. One constructs the ground states in the $R$-$R$ sector by tensoring left and right moving degrees of freedom. Hence we may infer the massless $R$-$R$ spectrum simply by requiring that we have objects which transform in a well defined way under Lorentz transformations, and noting the index structure $\Gamma _{\alpha \dot{\beta}}^\mu $ of the gamma matrices. In the Ramond sector of Type $IIB$ string we have chiral fermions, which  carry only either dotted or undotted indices, implying that an {\it{even}} number of gamma matrices are required to construct objects with well defined Lorentz transformation properties. Correspondingly, the forms we can construct in the $R$-$R$ sector of $IIB$ string are

\eqn\axion{\chi \sim \lambda _1^T\lambda _2}
\noindent
which is a scalar known as the axion. Similarly, we can have

\eqn\anti{B^{\mu \nu }\sim \lambda _1^T\Gamma ^\mu \Gamma ^\nu \lambda _2}

\noindent
and

\eqn\tbrane{A^{\mu \nu \rho \sigma }\sim \lambda _1^T\Gamma ^\mu \Gamma ^\nu \Gamma ^\rho \Gamma ^\sigma \lambda _2}

\noindent
In order to have the correct counting of degrees of freedom, we are compelled to require that $A^{\mu \nu \rho \sigma }$ is self dual.

Using the same reasoning in the case of $IIA$ string, which has non chiral spacetime fermions, an {\it{odd}} number of gamma matrices will be required. Accordingly, we obtain

\eqn\formone{A^\mu \sim \lambda _\alpha ^1\Gamma _{\alpha \dot{\beta}}^\mu \lambda _{\dot{\beta}}^2}

\noindent
and
\eqn\formtwo{A^{\mu \nu \rho }\sim \lambda _1^T\Gamma ^\mu \Gamma ^\nu \Gamma ^\rho \lambda _2}

\noindent
as the forms present in the $R$-$R$ sector of $IIA$ theory. The existence of these forms in the massless spectra of Type $IIA$ and $IIB$ string theory was known, however, it was not known how to switch them on. The first piece of tantalising evidence suggesting a connection to $D$-branes is given by $T$-duality. As discussed above, the massless $R$-$R$ sector of Type $IIA$ is populated by tensors of odd rank, while conversely $IIB$ has only even rank tensors in its spectrum. These forms naturally couple to higher dimensional objects, which we denote $p$-branes\foot{We use the term ``brane'' to emphasize that the forms couple to objects of higher dimensional volume. There is as yet no connection to Dirichlet branes.} where, as before, $p+1$ labels the spacetime dimensions of the object.  
\noindent
The four form $A^{\mu \nu \rho \sigma}$ couples naturally to the four dimensional worldvolume of a 3-brane, while the two form $B^{\mu \nu }$ couples to the worldvolume of a 1-brane. The axion is evaluated at a single time instant. The appearance of the 5-brane can be understood by acting with the exterior derivative on $B^{\mu \nu }$, which generates a three form field strength. The dual field strength associated with this three form field strength is a seven form in ten dimensions. The dual gauge potential associated with the dual field strength is a six form gauge potential, which couples naturally to the worldvolume of a six dimensional object--the 5-brane. Accordingly, the 5-brane is seen to be the magnetic dual of the 1-brane. Employing the same reasoning, one notes that the magnetic dual of the 3-brane is itself, while the magnetic dual of the -1 brane is the 7-brane. One concludes that $IIB$ contains -1-branes, 1-branes, 3-branes, 5-branes and 7-branes.

Similarly, in Type $IIA$ theory, the one form $A^\mu$ is sourced by a 0-brane, and the three form $A^{\mu \nu \rho }$ sources the 2-brane. Following our previous logic, one deduces their magnetic duals as 6-branes and 4-branes, respectively.

Thus the presence of the $R$-$R$ forms implies that Type $IIA$ contains $p$-branes of $p$ even, while Type $IIB$ contains branes of $p$ odd. In order to draw a link to $D$-branes, let us consider for the sake of example Type $IIA$ theory, which contains only $p$-branes of even dimension. It was argued earlier that performing a $T$-duality along any of the directions maps one into Type $IIB$ theory, a theory which contains $p$-branes of odd dimension. A link to the behaviour of $D$-branes under the action of $T$-duality becomes apparent, since performing a $T$-duality either parallel or transverse to a $D$-brane of even dimensionality must, of necessity, result in a brane of higher or lower {\it{odd}} dimension. One sees that the behaviour of $D$-branes under the action of $T$-duality is consistent with that of the $p$-branes in $IIA$ and $IIB$.

However, this is far from compelling. It was the work of Polchinski \joep, which provided strong evidence that Dirichlet branes carry $R$-$R$ charge and act as sources for the forms present in the $R$-$R$ sectors of Type $IIA$ or $IIB$ string theory. We briefly outline the argument here.
Let us imagine that we have a pair of parallel Dirichlet $p$-branes separated by some distance $L$, with the first brane at $X^{\mu}=0$, and the second at $X^{\mu}=Y^{\mu}$ with $\mu=p+1,...,9$.

\noindent
We would like to compute a one loop open string amplitude, which amounts to summing the zero point energies of the open string oscillators, and thus gives the vacuum energy of this configuration.
The amplitude has the following form,

\eqn\openamp{A=V_{p+1}\int {{d^{p+1}p}\over {(2\pi )^{p+1}}}\sum_i\int {{dt}\over t}e^{-t(p^2+m_i^2)/2}}
\noindent
which we can argue for as follows. Summing to one loop requires us to evaluate

\eqn\explan{\det (p^2+m^2)=e^{Tr{(\ln(p^2+m^2))}}}Inserting this between a complete set of string states allows us to evaluate the trace\foot{The trace over momentum is only over directions parallel to the brane; this is due to the fact that the boundary conditions of the configuration do not switch on all of the open string oscillators.},

\eqn\tra{\sum_i\int d^{p+1}p\ln (p^2+m_i^2)}
\noindent
We note that one may rewrite the logarithm as
\eqn\relog{\ln (p^2+m_i^2)=\int {dt \over t} e^{-t((p^2+m_i^2))}}
\noindent
This is the form of the amplitude \openamp\ up to numerical factors.

Following in Polchinski's footsteps, we are now going to do something rather bold: we will use the open string amplitude to infer something about the closed string by making use of the $s$-$t$ duality present in string theory. Swapping $\tau$ and $\sigma$ allows us to reinterpret the open string amplitude as arising from the exchange of a closed string mode between the branes.\foot{Note that, while we are computing a one loop quantum correction in the open string picture, the closed string exchange is tree level.} Sending the brane separation of the branes $L \to \infty $, which corresponds to taking $t \to 0$, ensures that we are studying the exchange of closed string {\it{massless}} modes.
\noindent
The amplitude, while vanishing overall, has two contributions, one from the closed string $NS$-$NS$ sector, and the other from the closed string $R$-$R$ sector. The contribution from the $R$-$R$ sector enters with a minus sign, and the $NS$-$NS$ sector with a plus sign. The fields in the $NS$-$NS$ sector are the dilaton and metric--hence the plus sign in the amplitude, since the exchange of even integer spin particles produces an attractive interaction.
\noindent
The amplitude is

\eqn\largesep{A=(1-1)V_{p+1}2\pi (4\pi ^2\alpha ^{\prime })^{3-p}G_{9-p}(Y^2)}
\noindent
where $G_{9-p}$ has the form
\eqn\green{G_{9-p}\sim (Y^2)^{(p-7)/2}}

\noindent
The Green's function for the propagation of a massless particle in $9-p$ dimensions must satisfy 
\eqn\gfe{\partial ^{2}\tilde{G}_{9-p}=\delta ^{(9-p)}}
\noindent
where $\partial ^{2}$ is the Laplacian in the $9-p$ dimensional space. Using dimensional analysis and exploiting the rotational symmetry of the problem, one sees that the Green's function must have the form

\eqn\greent{\tilde{G}_{9-p}\sim {1\over r^{7-p}}}

\noindent
Comparing \green\ with the result \greent, we identify the quantity $G_{9-p}$ as the Green's function describing the propagation of massless particles in the $9-p$ directions transverse to the Dirichlet $p$-branes. This is rather pretty, since the Green's function arose naturally out of the sum over the open string oscillators.

We may now compare the contribution from the $R$-$R$ sector of the open string calculation with that of a $p+1$ form potential $A_{p+1}$, which was argued to live in the massless $R$-$R$ sectors of the Type $IIA$ ($p$ even) and $IIB$ ($p$ odd) closed string theories. The field strength corresponding to this potential is obtained by taking the exterior derivative of the field strength, $F_{p+2}=dA_{p+1}$. The action describing the dynamics of this field is

\eqn\pact{S={{\alpha _p}\over 2}\int F_{p+2}F_{p+2}+i\mu _p\int_{branes}A_{p+1}}

\noindent
The constants $\alpha _p$ and $\mu _p$ have been left arbitrary. Remarkably, computing the amplitude for the exchange of a $p+1$ form between the branes gives a negative term equal to the contribution given by the $R$-$R$ amplitude from the open string calculation, and fixes the normalisation\joep

\eqn\norm{{\mu_{p}^{2}\over \alpha_{p}}=2\pi(4\pi^{2}\alpha^{\prime})^{3-p}}

\noindent
One can show that the Dirac quantization condition which the charges must satisfy is
\noindent
\eqn\diracq{{\mu_{p}\mu_{p-6}\over \alpha_p}=2\pi n}

\noindent
Examining the normalisation from the $p+1$ brane calculation, we see that the charges do indeed satisfy the Dirac quantization condition with minimum quantum $n=1$.

This calculation provides compelling evidence that Dirichlet branes source the forms present in the $R$-$R$ sectors of Type $IIA$ and $IIB$ theory. The implication of this calculation is remarkable, since Type $IIA$ and $IIB$ are theories which perturbatively only contain {\it {closed}}\ string excitations. However, by switching on forms present in the $R$-$R$ sectors of Type $IIA$ and $IIB$, we are describing configurations containing $D$-branes, which necessarily implies that Type $IIA$ and $IIB$ string theories contain {\it{open}} string degrees of freedom which  arise as non perturbative excitations.

\vfill\eject

\newsec{Open$\backslash$Closed String Duality}
\subsec{Dual Models}

The duality between open and closed strings is one of the most fundamental features of string theory. In the 60's, the experimental discovery of a proliferation of hadronic resonances with large spins spurred the search for a consistent theory of the strong interactions. A promising approach was put forward by Veneziano, who presented an amplitude which described many of the features of resonances\gsw. One of the most striking features of this amplitude was its invariance under the exchange of $s$ and $t$ channels, a feature known as $s$-$t$ duality. This duality implies that one needs only to sum over the $s$ channel when considering scattering amplitudes. This is at odds with traditional quantum field theory since studying scattering processes requires an amplitude with contributions from both the $s$ and $t$ channels.

In order to get a better insight into the appearance of this symmetry, we adopt the philosophy that the resonances cannot all be fundamental particles. The alternative, implying that one has a large number of adjustable parameters in the masses and couplings, is simply too unpleasant to contemplate. In accord with this philosophy, one sums only over $s$ channel poles. To make this more concrete, consider an $s$ channel process in the description of the scattering of two scalar particles. If we allow them to exchange states of arbitrary spin $J$, we should sum over these in the $s$ channel amplitude. This amplitude has the general form

\eqn\amp{A(s,t)\sim {\sum}_J{ {{(-t)^J} \over {M_{J}^2-s^2}}}}

\noindent
with explicit poles in the $s$ channel. Experimentally, however, it is known that there is also a $t$ channel contribution to the overall amplitude, which immediately implies that the sum over $s$ channel poles cannot be finite since it is quite plausible that the infinite sum may diverge for some finite values of $t$, generating poles in the $t$ channel. Therefore requiring that the sum over $s$ channel poles yields an amplitude which has both $s$ and $t$ channel contributions implies that the resonances increase indefinitely at higher and higher energies. It was Dolen, Horn and Schmid\gsw who argued, with the aid of some empirical evidence, that the duality between $s$ and $t$ channel amplitudes was approximately obeyed for some values of $s$ and $t$. However, it was left to Veneziano to propose an amplitude with an exact $s$-$t$ duality symmetry.

This proposal offers an additional advantage, since terms in Eq. \amp\ become increasingly divergent at high energies for $J > 1$. This is understood by realising that if one wishes to couple particles with spin to scalar fields, powers of momentum naturally enter in order to contract the tensor indices, which leads to the poor ultra violet behaviour of this amplitude for particles with spin greater than one. If the sum in Eq. \amp\ were finite, the high energy behaviour would be dominated by the exchange of the particle with the highest spin. This trend is in conflict with experimental evidence, which suggests hadron scattering amplitudes are very well behaved at high energies, in sharp contrast to the behaviour indicated by Eq. \amp\ for finite $J$. Assuming that the sum over $J$ is not finite offers the possibility that the sum may converge to something with a better high energy behaviour.

The amplitude suggested by Veneziano provided a brief hope for theorists, as it described many of the features of resonances with reasonable accuracy\foot{It is interesting to note that with the adoption of QCD as a description of the strong interactions, dual models fell out of favour. It is perhaps the ultimate irony that QCD may well be a string theory.}.
Subsequent analysis of this amplitude by Sakita, Gervais and Virasoro revealed that it had its origins in a relativistic theory of strings. The $s$-$t$ duality of the Veneziano amplitude could now be seen to reflect a duality between open and closed strings. To understand this duality, consider taking a long exposure photograph of a piece of glowing string propagating between two $D$-branes. As a closed string is emitted from one $D$-brane, it traces out a cylinder on the film until it is absorbed by the other. Next, consider an open string stretching between the branes. If the string endpoints move in a closed loop on the surface of the $D$-branes, a cylinder is traced out on the film. The relation between the two pictures is seen as an exchange of what one calls spacetime time. In the case of the closed string, time lies transverse to the branes, whereas for the open string, time runs like an angular coordinate parallel to the branes. The duality boils down to the fact that the amplitudes for these processes are equal up to an interchange of the time coordinate and the coordinate parametrizing the `length' of the string propagation. 

\subsec{The Maldacena Conjecture}

In recent times, the notion of open$\backslash$closed string duality has once again risen to prominence with the advent of the Maldacena conjecture. In the works \mal,\eddie,\gubbie\ strong evidence is provided for a conjecture of the duality of ${\cal N}=4$ super Yang Mills theory, which is the low energy limit of an open string theory, and $IIB$ supergravity, the low energy limit of a closed string theory. The notion that gauge theories are related to string theories is a fairly old one, originating with 't Hooft over twenty years ago\hooft. 't Hooft noticed that, quite generally, the large $N$ expansion of a gauge theory could be mapped into the sum over string worldsheets of arbitrary topology, provided that the string coupling $g_s$ is identified with $1/N$. The AdS$\backslash$CFT correspondence is a concrete realisation of this; however it is perhaps quite unexpected that it is a {\it{critical}} string theory which describes a gauge theory in four dimensions.

In the previous section, evidence was provided that Dirichlet branes source the $R$-$R$ forms present in the massless sector of $IIB$ string theory. Thus from the point of view of supergravity, one can study Dirichlet branes by looking for an appropriate supergravity solution which switches on these forms. In contrast, one can also study the Dirichlet branes from the point of view of a  worldvolume description. Considering a system of $N$ coincident $D$3-branes in flat ten dimensional Minkowski space and attempting to reconcile these two equivalent descriptions lead Maldacena to conjecture the duality of ${\cal N}=4$ super Yang Mills theory and $IIB$ supergravity.

Let us consider a configuration of $N$ $D$3-branes in $IIB$ string theory from a worldvolume perspective. We will consider this system at sufficiently low energies such that only massless modes may be excited, which corresponds to taking the limit ${\alpha^{\prime} \to 0}$. The threebranes have open strings stretching between them which can oscillate. The worldvolume theory of the threebranes is described by a Born-Infeld action which, after taking the low energy limit, reduces to the action for ${\cal{N}}=4$ super Yang-Mills theory. The strings themselves may begin and end on different branes, correspondingly one requires two sets of indices $i,j=1,...,N$ to label where the open strings begin and end, which implies that the super Yang-Mills theory has gauge group $U(N)$. In addition to the open string degrees of freedom, the threebranes are massive objects which gravitate, and consequently one expects the threebranes to interact with the bulk spacetime by the exchange of massless closed string modes; correspondingly we expect a term in the action which couples the brane action to the supergravity.

In taking the low energy limit, one takes ${\alpha^{\prime} \to 0}$ and at the same time switches off the string coupling $g_s \to 0$, which sends the interaction term between the brane and the bulk to zero, and consequently leaves two systems which have decoupled from one another; ten dimensional $IIB$ supergravity, and ${\cal{N}}=4$ super Yang-Mills theory. The Yang-Mills theory has a number of interesting features; it is maximally supersymmetric, that is to say, it has four spinor supercharges, which is the maximum allowed for a theory in four dimensions which does not contain gravity. In addition, the theory is conformally invariant and consequently the coupling constant does not depend on the energy scale. 

The next step is to consider the configuration of $N$ $D$3-branes from the point of view of supergravity.

\subsec{Closed string (Supergravity) description}

The solution which describes a set of $N$ coincident D3 branes inside $IIB$ supergravity switches on the metric and dilaton fields from the $NS$-$NS$ sector and a self dual $R$-$R$ four form. The threebranes are massive objects which deform the space around them. The metric describing an assembly of $N$ coincident threebranes is \strom

\eqn\metric
{ds^2 = f^{-1/2}dx_{\parallel}^2 + f^{1/2}(dr^2+r^2d\Omega_5^2)}
\noindent
where
\eqn\gf
{f=1+{4\pi g_{s} \alpha^{\prime 2} N \over r^4}}
with $g_s$ as the closed string coupling and $N$ the number of coincident threebranes.
\noindent
The coordinates $x_{\parallel}$ are parallel to the branes while the remaining coordinates lie transverse to them. The $r^{-4}$ dependence in the factor $f$ reflects the fact that the threebranes exchange massless modes in the six transverse directions. Note that as $r \to \infty$, $f \to 1$, and the metric reduces to the flat Minkowski metric in ten dimensions.

In general, the square root of the metric component $g_{00}$ is responsible for the gravitational red shift, and accordingly energies are redshifted by this metric as

\eqn\redshift
{E=f^{-1/4} E_r}
\noindent
where $E_r$ is the energy measured by an observer at position $r$, and $E$ is the energy of an observer at infinity. As $r \to 0$, $f^{-1/4}  \sim r \to 0 $, implying that particles near $r=0$ look like very low energy excitations to the observers at infinity.

To make contact with our earlier discussion in terms of the worldvolume description of this configuration, we now take the low energy limit. The redshift plays an important r\^ole, since we are permitted to keep excitations of arbitrarily high energy provided they are sufficiently close to $r=0$, since these look like low energy excitations at infinity. Consequently, states from the full string theory in the near horizon geometry survive the low energy limit. There are also the massless excitations which propagate in the bulk ten dimensional spacetime which have a very long wavelength; after taking the low energy limit the excitations in the bulk decouple from those in the near horizon region. This can be explained by realising that the excitations in the near horizon region are prevented from escaping to the asymptotically flat ten dimensional Minkowski spacetime by the gravitational potential, while heuristically the massless excitations in the bulk have wavelengths too large to `see' the near horizon region\klebo. Accordingly, taking the low energy limit leads to the decoupling of the system into two subsystems which do not interact with each other. In the worldvolume description, the low energy limit was taken by setting $\alpha^{\prime} \to 0$. In the supergravity setting, the near horizon limit is taken in such a way that masses of states in the field theory are held fixed, which corresponds to keeping the ratio $r/ \alpha^{\prime}$ fixed while taking $\alpha^{\prime} \to 0$.

Comparing the worldsheet and supergravity descriptions at low energy suggests that we should identify ${\cal {N}}=4$ super Yang-Mills and Type $IIB$ string theory living in the near horizon geometry. The near horizon region, which corresponds to small $r$, allows us to set $f = 4\pi g_{s} \alpha^{\prime 2} N/r^4$ in which case the metric becomes

\eqn\ADS
{ds^2 = {r^2 \over R^2}dx_{\parallel}^2 + {R^2 \over r^2}(dr^2+r^2d\Omega_5^2)}

\noindent
where we have set $R^4 \equiv 4\pi g_{s} \alpha^{\prime 2} N$. This may now be recognised as the metric of $AdS_{5} \times S^5$, where we interpret $R$ as the radius of the $AdS$ space. One of the interesting features of this space is that it has a boundary. Prior to the formulation of the Maldacena conjecture, it was argued that a theory of gravity must admit a holographic description \holo,\holt. This means that one is able to fully characterise a system which incorporates gravity simply by studying the degrees of freedom on a boundary which encloses the system.
The evidence for the holographic principle stems from the fact that the entropy of a black hole is well known to be characterised by the area of the event horizon rather than the volume\bht. Assuming this fact to be true, while contending that, with the sole exception of the black hole, the degrees of freedom of any configuration inside a theory of gravity are proportional to the {\it{volume}} of the system, can be shown to be inconsistent with the second law of thermodynamics. The Maldacena conjecture is a concrete realisation of the holographic principle, as the ten dimensional $IIB$ supergravity has a description in terms of a four dimensional super Yang-Mills theory.

One issue which needs to be addressed is the fact that, perturbatively at least, the super Yang-Mills theory bears no resemblance whatsoever to supergravity, and yet they are claimed to be equivalent. In order to understand this, we note that there are two possible sources of corrections to the supergravity; string loop corrections, controlled by the string coupling, and ${\alpha^{\prime}}$ corrections, which become important for large curvatures. One might ask what working in the uncorrected regime of the supergravity theory implies for the field theory. The identification

\eqn\ident{R^4=4\pi g_{s} \alpha^{\prime 2} N}
\noindent
connects quantities in the supergravity with those of the Yang-Mills theory. The relation between the open and closed string couplings has been known for some time, thanks to $s$-$t$ duality, which allows us to identify $g_s=g_{YM}^2$. The parameter $N$ labels the number of $D$3-branes in the supergravity, while on the field theory side it enters as the rank of the gauge group. In field theory, the product $g^2_{YM}N$ is known as the 't Hooft coupling, $\lambda$. One can compute the Ricci scalar as $\Re \sim \alpha^{\prime}/R^2$, and combining this result with \ident\ gives

\eqn\validity{\Re \sim {1 \over \sqrt{\lambda}}}

\noindent
One sees that suppression of the curvature corrections may be achieved by taking $\lambda$ large and fixed. However, simultaneously requiring that there are no corrections arising from string loop effects implies we should take $g_s \to 0$. In order to hold the t'Hooft coupling $\lambda$ large and fixed while simultaneously taking $g_s \to 0$ suggests one should take $N \to \infty$, as $g_s \sim 1/N$. One sees that strongly coupled field theory, ({\it{i.e.}} field theory at large $\lambda$) describes the supergravity. This explains why our knowledge of supergravity and perturbative field theory look very different while not threatening the conjecture. The Maldacena conjecture is often phrased as a ``duality" between supergravity and field theory. This is because, while both of these theories are simply different descriptions of the same physics, when the field theory is strongly coupled, the gravity is perturbative, and vice versa.

It was mentioned earlier that ${\cal{N}}=4$ super Yang-Mills has some rather special features, specifically it is maximally supersymmetric and conformally invariant. It is natural to ask how these features manifest themselves on the supergravity side. It turns out that $AdS_5$ has a large group of isometries $SO(2,4)$, which is the same as the group of conformal symmetries in four dimensions. Additionally, there is an $SO(6)$ symmetry associated with the $S_5$, which is recognised in the field theory as the $SU(4)$ $R$-symmetry which shuffles the supercharges amongst each other. Thus the superconformal symmetry of the field theory is seen to be encoded in the near horizon geometry of the supergravity. 

The thrust of this thesis will be to explore the gravity$\backslash$gauge theory correspondence, albeit from a slightly different point of view. The idea is to use the correspondence to extract quantities in a field theory which are not protected by supersymmetry; in particular, one would like to compute non-holomorphic corrections to the effective action of a field theory. One can argue for a link between the effective action of a gauge theory and a Born-Infeld action probing a supergravity background by a slight modification to the arguments presented earlier. Again, we consider a system of coincident $N$ $D$3-branes in flat ten dimensional Minkowski space; however, we introduce a $D$3-brane slightly displaced from the coincident $D$3-branes. Taking the low energy limit as before decouples the bulk and worldvolume descriptions from one another, leaving a supergravity theory in the bulk and an ${\cal{N}}=4$ super Yang-Mills theory in 3+1 dimensions. The separation of the single $D$3-brane from the clump is parametrised by a Higgs field of the super-Yang Mills theory and consequently the gauge group $SU(N+1)$ of the super Yang-Mills theory is broken to $SU(N) \times U(1)$. The $U(1)$ of the broken gauge group describes the massless modes living on the separated $D$3-brane.

The system described above has an alternative description in terms of supergravity. The geometry of the $N$ $D$3-branes is given by Eq. \metric, which interpolates between flat Minkowski space and $AdS_{5} \times S_5$ in the near horizon region.  In the limit of large $N$ the backreaction of the single brane on the geometry may be ignored, and hence the dynamics of the $D$3 is described by a Born-Infeld action probing the near-horizon supergravity solution. Taking the low energy limit on the supergravity side leads to the decoupling of the physics in the near horizon from the bulk. Reconciling the description of the separated brane in the supergravity$\backslash$worldvolume pictures forces one to identify the effective action of the $U(1)$ mode of ${\cal{N}}=4$ super Yang-Mills theory with gauge group $SU(N+1)$ broken to $SU(N)\times U(1)$ with the Born-Infeld action in the $AdS_5 \times S_5$ background. Thus the effective action of a field theory may be realised as the worldvolume theory of a threebrane probing a supergravity background.

Our interest will be the effective action of ${\cal{N}}=2$ super Yang-Mills theory with gauge group $SU(2)$. The low energy effective action of this model was determined exactly in \rSW. This was done by exploiting the fact that the low energy effective action is governed by a holomorphic prepotential. Sen \Sen\ has shown that the determination of a non-perturbative $IIB$ background consisting of $D$7 branes is mathematically identical to the problem of finding the exact effective gauge coupling for ${\cal{N}}=2$ super Yang-Mills theory with gauge group $SU(2)$ and $N_F=4$ flavors. 

In our analysis, a supergravity background consisting of $D$7 branes and $N$ $D$3-branes will be considered. The $D$7-brane background is chosen such that fields in the supergravity play the r\^ole of the effective coupling of ${\cal{N}}=2$ super Yang-Mills theory with gauge group $SU(2)$ and $N_F=4$ flavors. In order to ensure that the curvature of the background is small almost everywhere, and hence that the supergravity description can be trusted, $N$ $D$3-branes will be introduced. The introduction of the $D$3-branes does not disturb the dilaton-axion modulus set up by the $D$7-brane background. A $D$3-brane, described by a Born-Infeld action, will be introduced to probe the near horizon region of this geometry. The expansion of the Born-Infeld action about this background will be compared with the effective action of ${\cal{N}}=2$ super Yang-Mills theory with gauge group $SU(2)$ and $N_F=4$ flavors. One of the main goal of this work is to use a $D$3-brane probing the near horizon geometry of a supergravity background to obtain non-holomorphic corrections to the effective action of ${\cal{N}}=2$ super Yang-Mills theory. 

In the following section, the relevant details of ${\cal{N}}=2$ supersymmetric Yang-Mills theory will be reviewed, followed by a discussion of the link between the work of Seiberg and Witten\rSW\ and the $IIB$ backgrounds treated by Sen\Sen, with a view to introducing the probe analysis. 

\vfill \eject

\newsec{${\cal{N}}=2$ Supersymmetric Yang-Mills Theory}
Seiberg and Witten were able to compute the exact low energy effective action of ${\cal{N}}=2$ supersymmetric Yang-Mills theory with gauge group $SU(2)$ by exploiting the supersymmetry{\foot{For a review of supersymmetry, see \lyk.}} of the model\rSW. In this section we will give a brief summary of this analysis which will be relevant in what follows.

The ${\cal{N}}=2$ super Yang-Mills theory with gauge group $SU(2)$ contains a gauge field $A_\mu^i$ and a scalar $\phi^i$ in addition to their fermionic superpartners in the adjoint of $SU(2)$, and in addition may also have flavor hypermultiplets which transform in a different representation of the group. The potential for the scalar field $\phi$ is

\eqn\pot{V(\phi)\sim [\phi,\phi^{\dagger}]}
\noindent
One sees that the potential is minimised by taking $\phi=a\sigma^3$, which spontaneously breaks the $SU(2)$ symmetry down to $U(1)$. The symmetry breaking generates a mass for the gauge fields $A_\mu^1$, $A_\mu^2$ and their superpartners. One chooses a gauge invariant quantity $u=u(a)$ as a coordinate on the space of inequivalent vacua, or moduli space ${\cal{M}}$. In practice we will take the Higgs field $a$ as a coordinate (which is not gauge invariant), although it will become clear in what follows that such a coordinate does not provide a global description of the moduli space.

If one wished to obtain a description of the physics of the light degrees of freedom one could integrate out the massive particles to give an effective action for the light mode, the Wilsonian effective action. Schematically, 

\eqn\eff{\eqalign{S&=\int d\Phi_H \int d\phi_L e^{S(\Phi_H, \phi_L)} \cr &= \int \phi_L e^{\tilde S ( \phi_L)}}}
\noindent
where $\Phi_H$ and $\phi_L$ correspond to the massive and light fields, respectively, and $\tilde S$ is the effective action for the light degrees of freedom. The {\it{low energy}} effective action consists of the terms which are leading at small momenta, consisting of at most two derivatives or four fermions. It is important to note that the effective action has an infinite derivative expansion; these higher derivative terms are not in general holomorphic while, in contrast, it turns out that the two derivative/four fermion contributions are governed by a holomorphic function, the prepotential. Following \rHen\ one can make this statement more precise by introducing an `order in derivatives' $n$. An expansion of the effective action in powers of the momentum scale of the external particles divided by the characteristic scale of the theory is a profitable way to extract information from the low energy effective action. A spacetime derivative is associated with a power of the spacetime momentum; accordingly this procedure amounts to keeping terms with some maximum powers of the spacetime derivatives. We define the order in derivatives $n$ by requiring that the ${\cal{N}}=2$ vector superfield $\Phi$ have $n=0$ and the covariant derivative $D_{\alpha}^{i}$  on superspace $n=1/2$. The covariant derivatives on superspace satisfy the anticommutator

\eqn\comm{\{D_{\alpha}^A,{\bar{D}}_{\dot{\beta} B}\}=-2i\sigma^{\mu}_{\alpha \dot{\beta}} \delta^{A}_{B} \partial_{\mu}}
\noindent
which implies that $n=1$ for the spacetime derivative $\partial_\mu$. The form of the covariant derivatives then implies that that the Grassman coordinates $\theta^{A}_{\alpha}, A=1,2$ of ${\cal{N}}=2$ superspace have $n=-1/2$. The Grassman measure $d^4{\theta}$ has $n=2$, which is easily seen from the properties of Grassman coordinates under integration. 

In order to construct actions which are invariant under supersymmetry transformations, we note that a general superfield integrated over ${\cal{N}}=2$ superspace transforms as a total spacetime derivative under the action of the ${\cal{N}}=2$ supercharges $Q^A_{\alpha}, {\bar{Q}}^A_{\dot{\beta}}$. For a holomorphic superfield, which depends only on the $\theta^{A}_{\alpha}$ and not on the ${\bar{\theta}^{A}_{\dot{\beta}}}$, we integrate only over a subspace of superspace. Consequently, a candidate for a term in an action with ${\cal{N}}=2$ supersymmetry can be found by integrating a holomorphic function (the holomorphic prepotential) of the ${\cal{N}}=2$ vector superfield $A$ over a subspace of the full superspace,

\eqn\holo{\int d^4\theta {\cal{F}}(A)}
\noindent
The prepotential itself is of order $n=0$ and the Grassman measure $d^4{\theta}$ has $n=2$. Terms quadratic in the derivatives or quartic in the fermions are of order $n=2$, and consequently they must arise from the holomorphic prepotential. Another possible term in the ${\cal{N}}=2$ action is given by a superfield which depends both on $A$ and ${\bar{A}}$, and consequently is integrated over the whole of superspace,

\eqn\nonholo{\int d^4\theta d^4{\bar{\theta}}{\cal{H}}(A,{\bar{A}})}
\noindent 
Clearly, the Grassman measure is of order $n=4$, and the function ${\cal{H}}(A,{\bar{A}})$ is of order $n=0$, implying that terms which have four powers of the spacetime derivatives or eight fermion fields originate from this term in the action.

\noindent
The holomorphic prepotential has the form
 \eqn\ta
{{\cal F}(a)={1 \over 2} \tau_0 a^2+{i \over \pi}a^2 \log \left({a^2 \over \Lambda^2}\right)+{a^2 \over 2\pi i} \sum^\infty_{l=1} c_l \left ({\Lambda \over a}\right)^{4l}}

\noindent
The logarithmic term arises perturbatively from a one loop calculation. The fact that the prepotential is one loop exact may be attributed to supersymmetry. The last term corresponds to contributions from non-perturbative instanton corrections. The general form of the prepotential was known before the work of \rSW; it was the contribution of Seiberg and Witten to determine all of the coefficients $c_l$ and consequently fully determine the low energy effective action.

The prepotential is related to the coupling $\tau$ of the field theory as 

\eqn\rela{\tau (a) ={\partial^2 {\cal{F}}(a) \over \partial a^2}}

\noindent
We note that one must require $\Im\tau>0$ if unitarity is not to be violated. However, $\tau$ is a holomorphic function of $a$ and cannot have a minimum if it is globally defined. We will see in what follows how this inconsistency is to be resolved.

The key property which enables one to obtain the full low energy effective action is the holomorphic nature of the prepotential, which is ensured by supersymmetry. As is well known from complex analysis, the knowledge of the singularity structure of a holomorphic function is sufficient to determine it exactly. The singularity structure of the quantum moduli space ${\cal{M}}_q$ and the physical significance of the singularities was first motivated in \rSW. Essentially, the reasoning is that the appearance of singularities on the moduli space signifies that the effective action has not captured all of the relevant degrees of freedom. Accordingly, motivated by physical arguments, \rSW\ speculates that certain monopoles and dyons become massless at these points on ${\cal{M}}_q$. Since the effective action is obtained by integrating out the massive degrees of freedom, it is clear that if particles are to become massless at certain points on the moduli space, the effective action is no longer a complete description of the degrees of freedom at these points. Note that in the classical theory one would have a singularity on the classical moduli space at $a=0$ since the massive gauge bosons $A_\mu^1$, $A_\mu^2$ become massless. However, this picture is modified by quantum corrections, which split the classical singularity at the origin into two singularities at distinct points on the moduli space where monopoles and dyons become massless. In addition, when studying the gauge theory with hypermultiplets, the moduli space has singularities at the points where the hypermultiplets become massless.  

The fact that the masses of particles become small at the singularities suggests that one may formulate perturbation expansions about these points, and accordingly one adopts new local coordinates in patches about the singularities which are dual to $a$ in the sense that, for $a \to \infty$, perturbation expansions phrased in terms of the dual coordinate $a_D$ are strongly coupled, while they are weakly coupled in terms of the coordinate $a$. Remarkably, a globally consistent solution can be found by requiring the mutual consistency of the perturbation expansions formulated in the neighbourhood of the singularities. 

We note that the prepotential and the complex coupling $\tau$ as derived from Eq. {\ta} are multiple valued functions due to the presence of the logarithm, in particular we see that circling about the singularity $a=\infty$ induces non-trivial monodromies. In general, circling each singularity on the moduli space induces the complex coupling $\tau$ to transform with a particular monodromy. The monodromies generate a subgroup of the modular group $SL(2,{\cal{Z}})$ which suggests that one might reconcile the monodromies about these points consistently by identifying the complex coupling $\tau$ with the modular parameter of a torus. Let us introduce the curve\rSW  

\eqn\swcurve{y^2=x^3+f(a)x+g(a)}
\noindent
whose moduli space is identical to ${\cal{M}}_q$. The coordinates $x$, $y$ are complex. For fixed $a$, we can interpret Eq. \swcurve\ as defining a $T^2$. This is seen by realising that for each value of $x$, one has two corresponding $y$ values; accordingly, varying $x$ describes two spheres which are connected via the two branch cuts which run between the four square root branch points{\foot{We have taken one of the branchpoints to be at infinity in Eq. \swcurve.}}. The modular parameter $\tau$ of the $T^2$ depends on the functions $f$ and $g$, which in turn depend on the moduli space coordinates. As a consequence the modular parameter typically varies as we move about the moduli space, implying that we have associated a $T^2$ with each point on ${\cal{M}}_q$. In the context of the Seiberg-Witten curve, the monodromies around the singularities on ${\cal{M}}_q$ correspond to the monodromies of the basis homology cycles of the torus, while at the singularities the cycles of the torus degenerate, which is reflected in Eq. \swcurve\ as a coalescence of some of the branchpoints.
The modular parameter $\tau$ is computed as the ratio of the integrals of the holomorphic differential over the two basis cycles of the torus, {\it{i.e.}}

\eqn\period{\tau ={{\oint_\beta \omega} \over {\oint_\alpha \omega}}}

\noindent 
where
\eqn\diff{\omega = {dx \over y^2(x,u)}}

\noindent
is the holomorphic differential, and $\alpha$ and $\beta$ are the basis cycles of the torus. The $\tau$ recovered from the Seiberg-Witten curve satisfies the requirement of positivity. 

In the next section, we discuss the remarkable link between the Seiberg-Witten monodromy problem and supergravity.

\newsec{F-theory and the Seiberg-Witten Moduli space}

In this section, we will discuss the remarkable fact that the solutions of of the Seiberg-Witten monodromy problem of ${\cal{N}}=2$ super Yang-Mills theory with gauge group $SU(2)$ and $N_F=4$ flavors correspond to exact $D$7-brane backgrounds in compactifications of Type $IIB$ string theory which, in turn, correspond to F-theory compactifications. 

F-theory on an elliptically fibered manifold ${\cal{M}}$ with base ${\cal{B}}$ is equivalent to type $IIB$ string theory compactified on ${\cal{B}}$ with the axion-dilaton modulus set equal to the modular parameter of the fiber. The axion-dilaton modulus is formed by assembling the scalar from the $R$-$R$ sector, the axion, and the dilaton from the $NS$-$NS$ sector into a single complex coupling $\tau=\chi+ie^{-\phi}$. From the previous section, it is clear that $R$-$R$ fields are sourced by $D$-branes. It is easily verified that the magnetic source which excites the axion is the $D$7-brane, and consequently the $IIB$ backgrounds corresponding to an F-theory compactification may contain $D$7-branes.

We study a particular $IIB$ background \Sen, which perturbatively consists of $D$7-branes and an $O$7 plane{\foot{Consider a symmetry operation $G$ of a string theory which is the product of orientation reversal on the worldsheet and a target spacetime symmetry. Gauging this symmetry gives rise to an orientifold plane. See \tasi\ for more details.}}. The geometry set up by the $D$7-branes is such that the transverse space is compact. Such a background does not satisfy the vacuum Einstein equations $R_{\mu \nu}=0$, since the presence of the axion and dilaton ensures that the energy-momentum tensor is nonvanishing. The orientifold plane carries an $R$-$R$ charge, which requires that four $D$7-branes are added lying parallel and coincident to the $O$7 plane in order to ensure that the charge of the $O$7 plane is cancelled locally. Thus we have a trivial $R$-$R$ background where the axion-dilaton modulus does not vary over the base. 

If one wishes to deform this configuration by moving the $D$7 branes away from the orientifold plane, the complex coupling $\tau$ varies over the base since the $R$-$R$ charge is no longer cancelled locally; however one finds an interesting inconsistency in that the imaginary part of the dilaton-axion modulus $\Im \tau=e^{-\phi}$ becomes negative, suggesting that strong coupling physics begins to play a r\^ole. In addition one notes that circling the $D$7 branes in the transverse space produces non-trivial monodromies since the coupling behaves as $\tau \sim \log (z-z_i)$ in the neighbourhood of a $D$7-brane located at $z_i$, where $z$ is a complex coordinate parametrizing the two directions transverse to the $D$7-branes. These facts suggest that the $IIB$ background has some features reminiscent of the Seiberg-Witten monodromy problem. In fact Sen\Sen, guided by the F-theory description of the $IIB$ background, showed that understanding the non perturbative $IIB$ physics is mathematically identical to solving the Seiberg-Witten monodromy problem for the gauge theory with $N_F=4$ flavor hypermultiplets. Type $IIB$ string theory on this background is equivalent to F-theory compactified on an elliptically fibered manifold described by 
\eqn\fcurve{y^2=x^3+f(z)x+g(z)}
\noindent
where the complex coupling $\tau$ of the fiber is to be identified with the axion-dilaton modulus of the $IIB$ background. The directions $y$ and $x$ are accessible only in F-theory, and are not visible in the $IIB$ description. At a fixed point on the base, described by the coordinate $z$, Eq. \fcurve\ describes a torus, which is the Seiberg-Witten curve. As in the context of the field theory, the complex coupling $\tau$ recovered from Eq. \fcurve\ is sensible{\foot{That is, it has a positive imaginary part.} at all points on the base, which leads one to expect that the F-theory background describes the correct non perturbative $IIB$ physics. 
One may now draw a correspondence between the field theory moduli space and the $IIB$ background. The classical singularity at $a=0$ on the field theory side, which was interpreted as massive gauge fields becoming massless, corresponds to the presence of the $O7$ plane situated at $z=0$ in the supergravity. In the case that the $D$7 branes are parallel and coincident with the $O$7 plane, this corresponds to the situation where one is describing Seiberg-Witten theory with $N_F=4$ {\it{massless}} flavors.
The perturbative beta function for ${\cal N}=2$ super Yang-Mills theory with gauge group $SU(N_{c})$ and $N_{f}$ flavor hypermultiplets is proportional to $(2N_{c}-N_{f})$, and consequently vanishes in this case. Moreover, since the flavor hypermultiplets are massless, the theory is conformally invariant. The complex coupling of the field theory $\tau$ is a constant, which is reflected in the supergravity as the local cancellation of $R$-$R$ charge of the $O$7 plane by the coincident $D$7-branes. As the $D$7 branes are moved away from the $O$7 plane, one finds that the correct non perturbative description of the background splits the orientifold plane in two, in direct analogy with the fate of the classical singularity on the Seiberg-Witten moduli space. Moving the $D$7-branes away from the $O$7 plane corresponds to giving a mass to the flavor hypermultiplets in the field theory. The $IIB$ background is also capable of describing the pure gauge ${\cal{N}}=2$ super Yang-Mills theory. This is accomplished by moving the four $D$7-branes infinitely far away from $z=0$, which corresponds to giving the four flavor hypermultiplets in the field theory infinite masses. These states can no longer be excited in the field theory, and one is left with a description of the pure gauge theory.

In the following sections, some results from field theory will be introduced to be compared with the probe analysis. We begin by collecting relevant results for ${\cal{N}}=2$ super Yang-Mills theory with $N_f=4$ massless multiplets and gauge group $SU(2)$, which will be compared with results obtained for a threebrane probing the appropriate supergravity background. 

\newsec{${\cal{N}}=2$ super Yang-Mills theory with $N_f=4$ massless multiplets}

The perturbative beta function for ${\cal N}=2$ super Yang-Mills theory with gauge group $SU(N_{c})$ and $N_{f}$ flavor hypermultiplets is proportional to $(2N_{c}-N_{f})$. Thus, for $N_{c}=2$ and $N_{f}=4,$ the perturbative beta function vanishes. If in addition all of the flavors of matter are massless, we obtain a finite conformally invariant theory\rSW. The exact effective coupling of the theory has the form 

$$ \tau=\tau_{1}+i\tau_{2}=\tau_{cl}
+{i\over\pi}\sum_{n=0,2,4,...}c_{n}e^{in\pi\tau_{cl}}
=\tau_{cl}+{i\over\pi}\sum_{n=0,2,4,...}c_{n}q^{n},$$

\noindent
where $\tau_{cl}$ is the classical coupling of the theory. The coefficient $c_{0}$ is a one loop perturbative correction, which has the value $c_{0}=4\log(2)$ in the Pauli-Villars scheme\rdorey. The coefficients $c_{n}$ with $n>0$ and even come from nonperturbative (instanton) effects. The two instanton coefficient has been computed and has the value $c_{2}=-7/(2^{6}3^{5})$. As motivated earlier, the leading contribution to the low energy effective action comprises all terms with the equivalent of two derivatives or four fermions and is determined in terms of the effective coupling. The next-to-leading contribution to the low energy effective action contains all terms with four derivatives or eight fermions. Dine and Seiberg, using the scale invariance and $U(1)_{\cal R}$ symmetry of the model, were able to argue that the four 
derivative term is one loop exact\rDS. In \rKD\ the vanishing of instanton corrections to the four derivative terms was explicitly verified and a rigorous proof of this non-renormalization theorem has recently been given in \rOvrut. The one loop contribution to the four derivative terms has been considered in \rOLoop. The result for the low energy effective action, up to and including four derivative terms, in ${\cal N}=2$ superspace, is given by 

\eqn\LEEA
{8\pi\big(S_{eff}^{(2)}+S_{eff}^{(4)}\big)=
{\cal I}m\int d^{4}xd^{4}\theta\big({1\over 2}A^{2}\big)+
{3\over 128\pi^{2}}\int d^{4}xd^{4}\theta 
d^{4}\bar{\theta}\log A\log\bar{A},}

\noindent
where $A$ is an ${\cal N}=2$ Abelian chiral superfield. The number of terms that contribute to the low energy effective action at each order, for six derivative terms or higher, increases rapidly and a direct approach to these
terms is not feasible. An elegant approach to study these terms has been developed in \rocek\ for ${\cal N}=4$ super Yang-Mills\rbuch, based on the conjectured $SL(2,Z)$ duality of the theory. This duality was used to fix the 
form of the effective action up to six derivatives. The theory that we are studying is also believed to have
an exact $SL(2,Z)$ duality\rSW, and under this assumption the analysis of \rocek applies. 

\noindent
The unique $SL(2,Z)$ invariant form for the six
derivative terms is

\eqn\SixDTer
{\eqalign{8\pi S_{eff}^{6}=
\Big({3\over 128 \pi^{2}}\Big)^{2}
\lambda^{(6)}\int &d^{4}x d^{4}\theta d^{4}\bar{\theta}
\Big({1\over\sqrt{\tau\bar{\tau}}}
{\bar{D}^{\dot{\alpha}}{}_{a}
\bar{D}_{\dot{\alpha} b}\log (\bar{A})\over A}
{D_{\alpha}{}^{a}D^{\alpha b}\log (A)\over\bar{A} }\Big)\cr
+{i\over 2}\Big({3\over 128\pi^{2}}\Big)^{2}
\int &d^{4}x d^{4}\theta d^{4}\bar{\theta}
\Big(\log(\bar{A})
{\bar{D}_{\dot{\alpha}1}\bar{D}^{\dot{\alpha}}{}_{1}\bar{D}_{\dot{\beta} 2}
\bar{D}^{\dot{\beta}}{}_{2}\log\bar{A}\over \tau A^{2}}\cr
&\qquad-
\log (A){D_{\alpha}{}^{1}D^{\alpha 1}D_{\beta}{}^{2}
D^{\beta 2}\log (A)\over \bar{\tau}\bar{A}^{2}}\Big),}}

\noindent
Under duality, the second term above mixes with the two and four derivative terms and consequently its coefficient is fixed. The requirement of self duality does not fix $\lambda^{(6)}$, since duality maps this term into itself
at lowest order. These are the field theory results that we wish to compare to gravity.

On the gravity side, we will consider a probe moving in a background to be specified below. The probe worldvolume dynamics is captured by a Born-Infeld action. The Born-Infeld action itself is self-dual, but the duality does not
act on the separation of the branes. This separation is parametrized by the Higgs fields which belong to the same supermultiplet as the gauge fields. This implies, as pointed out in \rocek,  that the Higgs fields that realize
${\cal N}=2$ supersymmetry linearly must be related by a nonlinear gauge field dependent redefinition to the separation. It is interesting to note that a similar field redefinition is needed to map the linear realization of
conformal symmetry in super Yang-Mills theory into the isometry of the Anti de-Sitter spacetime of the supergravity description\rJev. We refer the reader to \rocek\ for the detailed form of the field redefinitions.
The result after performing the field redefinitions, in terms of component fields, reads

\eqn\Reslt
{\eqalign{S_{eff}=&\int d^{4}x
\Big(-{1\over 4g^{2}}\partial_{m}\bar{\varphi}
\partial^{m}\varphi-{1\over 8g^{2}}(F_{\alpha\beta}F^{\alpha\beta}
+\bar{F}_{\dot{\alpha}\dot{\beta}}\bar{F}^{\dot{\alpha}\dot{\beta}})
+\Big({3\over 128\pi^{2}}\Big)^{2}\times\cr
&\times {1\over 32\pi}{F_{\alpha\beta}F^{\alpha\beta}
\bar{F}_{\dot{\alpha}\dot{\beta}}\bar{F}^{\dot{\alpha}\dot{\beta}}
+(\partial_{m}\varphi\partial^{m}\varphi)(\partial_{n}\bar{\varphi}
\partial^{n}\bar{\varphi})-F^{\beta\alpha}\partial_{m}\varphi
\sigma^{m}{}_{\alpha\dot{\beta}}\bar{F}^{\dot{\beta}\dot{\alpha}}
\partial_{n}\bar{\varphi}\sigma^{n}{}_{\beta\dot{\alpha}}\over
\varphi^{2}\bar{\varphi}^{2}}\cr
&-{g^{2}\over 256\pi^{2}}
\Big({3\over 128\pi^{2}}\Big)^{2}
{F_{\alpha\beta}F^{\alpha\beta}
\bar{F}_{\dot{\alpha}\dot{\beta}}\bar{F}^{\dot{\alpha}\dot{\beta}}
(F^{\rho\tau}F_{\rho\tau}+\bar{F}^{\dot{\rho}\dot{\tau}}
\bar{F}_{\dot{\rho}\dot{\tau}})
\over\varphi^{4}\bar{\varphi}^{4}}
\Big),}}

\noindent
where we have set $\tau=i{4\pi\over g^{2}}.$ The six derivative terms for the scalars are not displayed since they depend on the arbitrary constant $\lambda^{(6)}$. The value of $\lambda^{(6)}$ as well as the structure of the effective action given above can be checked by explicitly computing instanton corrections to the six derivative terms. Notice that all acceleration terms were eliminated by the field redefinition, something first noted in\rHo.

\subsec{Supergravity Results}

As motivated earlier, the supergravity background relevant for the study of ${\cal N}=2$ supersymmetric field theory is generated by sevenbranes and a large number of threebranes\rOfer, i.e. threebranes in F-theory\foot{Supergravity backgrounds corresponding to ${\cal N}=1$ field theories have been considered in \rAhn.}. To construct this background it is convenient to start with a solution for the sevenbranes by themselves\rOfer. The sevenbrane solution is described in terms of non-zero metric, dilaton and axion fields. The dilaton and axion are assembled into a single complex coupling $\tau=\chi+ie^{-\phi}=\tau_{1}+i\tau_{2}$. The coupling $\tau$ is identified with the modular parameter of the elliptic fiber of the F-theory compactification. The $(8,9)$ plane is taken to be orthogonal to the sevenbranes. In terms of the complex coordinate $z=x^{8}+ix^{9}$ we make the following ansatz for the metric

\eqn\frst
{ds^{2}=e^{\varphi(z,\bar{z})}dzd\bar{z}+(dx^{7})^{2}+...
+(dx^{1})^{2}-(dx^{0})^{2}.}

\noindent
The parameter $z$ is to be identified with the Higgs field appearing in the low energy effective action of the ${\cal N}=2$ field theory. With this ansatz, the type IIB supergravity equations of motion reduce to\rGSVY\

\eqn\scnd
{\eqalign{\partial\bar{\partial}\tau &=
{2\partial\tau\bar{\partial}\bar{\tau}\over\bar{\tau}-\tau}\cr
\partial\bar{\partial}\varphi &=
{\partial\tau\bar{\partial}\bar{\tau}\over(\bar{\tau}-\tau)^{2}}.}}

\noindent
The complex coupling $\tau$ is identified with the low energy effective coupling of the ${\cal N}=2$ field theory. Supersymmetry constrains the effective coupling of the field theory to be a function of $z$, so that
the first equation in \scnd\ is automatically satisfied. The general solution to the second equation in \scnd\ is 

\eqn\thrd
{\varphi (z,\bar{z})=\log(\tau_{2}) +F(z)+\bar{F}(\bar{z}).}

\noindent
The functions $F(z)$ and $\bar{F}(\bar{z})$ should be chosen in order that \frst\ yields a sensible metric. For the case that we are considering, the explicit form for the metric transverse to the sevenbranes is

\eqn\htgdgf
{ds^{2}=e^{\varphi (z,\bar{z})}dzd\bar{z}=\tau_{2}|da|^{2},}

\noindent
where $a$ is the quantity that appears in the Seiberg-Witten solution\rOfer. This specifies the solution for the sevenbranes by themselves.

Next following \rOfer, we introduce threebranes into the problem\foot{See also \rKeh\ where this solution was independently discovered.}. The presence of the threebranes modifies the metric and switches on a non-zero flux for the self dual $R$-$R$ five-form field strength. The world volume coordinates of the threebranes are  $x^{0},x^{1},x^{2},x^{3}$. One obtains a valid 
solution\rOfer\ by making the following ansatz for the metric

\eqn\frth
{ds^2 = f^{-1/2}dx_{\parallel}^2 + f^{1/2}{g}_{ij}dx^{i}dx^{j}}

\noindent
and the following ansatz for the self-dual 5-form field strength

\eqn\ffth
{F_{0123i}  =  -{{1}\over{4}}\partial_{i}f^{-1}~. }

\noindent
The complex field $\tau$ is unchanged by the introduction of the threebranes. Inserting the above ansatz into the IIB supergravity equations of motion, one finds that $f$ satisfies the following equation of motion

\eqn\sxh
{ {1\over\sqrt{g}} 
\partial_{i}(\sqrt{ g}{ g}^{ij}\partial_j f)=
- (2 \pi)^4 N { \delta^6(x-x^0) \over \sqrt{\ g}. }}

\noindent
This last equation corresponds to the case in which all of the three branes are located at the same point. In the limit that $N\to\infty$ the curvature becomes small almost everywhere and the supergravity solution can be used to reliably compute quantities in the field theory limit as explained 
in\rOfer. 

In the case of $N_f=4$ massless hypermultiplets, we have a constant $\tau$. Explicitly, $\tau_2=\sqrt{g}$ and $g^{ij}=\partial_y^{2}+{\tau_2}^{-1} \partial_a \partial_{\bar{a}}$ and consequently \sxh\ becomes

\eqn\explicit
{\big[\tau_{2}\partial_{y}^{2}+4\partial_{a}\partial_{\bar{a}}\big]
f=-(2\pi )^{4}N\delta^{(4)}(y)\delta^{(2)}(a).}

\noindent
The solution is given by \foot{Threebranes in a IIB orientifold background were first considered in \rSpa.} 

$$ f={4N\pi\over\big[y^{2}+\tau_{2}|a|^{2}\big]^{2}}. $$

To reproduce the low energy effective action of the field theory, we now consider the dynamics of a threebrane probe moving in this geometry. It is well known that the probe has a low energy effective action which matches that of the corresponding low energy field theories\rDoug,\rsennet. Here we are interested in checking the form predicted by the probe for the higher order
corrections. The leading low energy effective action plus corrections for the bosons in the background described above, is obtained by expanding the self-dual action \rDaction,
\eqn\svnth
{S={T_{3}\over 2}
\int d^{4}x\Big[\sqrt{det(G_{mn}+e^{-{1\over 2}\phi}F_{mn})}+\chi
F\wedge F\Big].}

\noindent
where $G_{mn}$ is the induced metric, $G_{mn}=g_{\mu \nu}\partial_m X^{\mu} \partial_n X^{\nu}$. Note that we choose to identify $x^a$ as $X^a=x^a$, for $a=0,...,3$, where $X^{\mu}$ are the spacetime coordinates and $x^a$ the worldvolume coordinates of the $D$3-brane probe.
\noindent
$T_3$ has no dependence on the string coupling constant. We obtain for the scalar terms obtained from the expansion of \svnth\ after setting $y=0$:

\eqn\svnthexp
{\eqalign{S&\sim {1\over 2}\int d^{4}x
\Big(\tau_{2}\partial_{m}a\partial^{m}\bar{a}
-{f\over 2}\tau_{2}^{2}(\partial_{m}a\partial^{m}a)
(\partial_{n}\bar{a}\partial^{n}\bar{a})\cr
&+{f^{2}\over 2}\tau_{2}^{3}(\partial_{p}a\partial^{p}\bar{a})
(\partial_{m}a\partial^{m}a)
(\partial_{n}\bar{a}\partial^{n}\bar{a})+ ...\Big)\cr
&={1\over 2}\int d^{4}x \Big(
\tau_{2}\partial_{m}a\partial^{m}\bar{a}
-{2N\pi\over (a\bar{a})^{2}}(\partial^{m}a\partial_{m}a) 
(\partial^{n}\bar{a}\partial_{n}\bar{a})\cr
&\qquad +{8\pi^{2}N^{2}\over \tau_{2}(a\bar{a})^{4}}
(\partial_{p}a\partial^{p}\bar{a})
(\partial^{m}a\partial_{m}a) 
(\partial^{n}\bar{a}\partial_{n}\bar{a})+ ...\Big).}}

\noindent
Notice that each term in this action comes multiplied by a different power of $N$. As things stand, the $2n$ derivative term will come with a coefficient of $\tau_{2}^{n}f^{n-1}\sim N^{n-1}$. The full effective action for the probe interacting with $N$ coincident source threebranes, should come with an overall factor of $N$\rTseyt. This is achieved by noting that the coupling of the background, $\tau_{2}$ should be identified with

\eqn\Rel
{\tau_{2}\equiv {N\tau_{2,SW}} = {1 \over g_s},}

\noindent
where $\tau_{SW}$ is the Seiberg-Witten effective coupling for the field theory of interest. 
With this identification, the string coupling is $O({1\over N})$ and explicitly goes to zero as $N\to\infty$. After making the $N$ dependence of $\tau_{2}$ explicit, we find that the probe action is indeed proportional to $N$.
We will not always show this dependence explicitly in what follows. Following \rocek, we find the Taylor expansion of \svnth\ exactly matches the super Yang-Mills effective action \Reslt\ after identifying

$$  a\bar{a}={1\over T_{3}}\varphi\bar{\varphi},\qquad
    F_{s,\alpha\beta}F_{s}{}^{\alpha\beta}=
    {1\over 4T_{3}}
    F_{f,\alpha\beta}F_{f}{}^{\alpha\beta}  $$

\noindent
where $F_{s,\alpha\beta}$ is the field strength appearing on the probe worldvolume and $F_{f,\alpha\beta}$ is the field strength of the field theory. Note that $\tau_{2}$ appearing in \svnthexp\ is the classical coupling plus all instanton corrections. The fact that the four derivative terms are independent of $\tau_{2}$ shows that the supergravity result
explicitly reproduces the nonrenormalization theorem for these terms\rBob.

\newsec{${\cal N}=2$ Super Yang-Mills Theory with gauge group $SU(2)$ and 
$N_{f}=4$ Massive Multiplets}

In this section we consider the supergravity background corresponding to 
the case where all flavor multiplets of the field theory on the probe world
volume have a mass. In this case, both the
effective coupling and the four derivative terms get contributions from 
instantons. We are able to show that the supergravity solution is capable
of producing what is expected for the one instanton correction. We are not
however able to fix the coefficient of this correction. The dilaton
of the supergravity solution is no longer a constant and there are corrections
to the $AdS_{5}$ geometry reflecting the fact that the field theory is no 
longer conformally invariant. We compute the quark-antiquark potential and
show that its form is remarkably similar to that for a quark-antiquark pair 
in the ${\cal N}=4$ theory at finite temperature.

\subsec{Field Theory Results}

The masses of the quark flavors breaks the conformal invariance that is
present in massless theory. In this case, the effective coupling does 
pick up a dependence on the energy scale as a result of instanton corrections.
At high enough energies we expect these corrections can be neglected and
the theory flows to the conformal field theory corresponding to the case of
massless flavors. Indeed, the perturbative beta function still vanishes and
the coupling goes to a constant at high energies. We will focus attention
on the two and four derivative terms appearing in the low energy effective
action. These terms are completely specified by a holomorphic prepotential
${\cal F}$ and a real function ${\cal H}$

$${S_{eff}={1\over 2i}\int d^{4}x\Big(\int d^{4}\theta{\cal F}(A)
-\int d^{4}\bar{\theta}\bar{{\cal F}}(\bar{A})\Big)+\int d^{4}x
\int d^{4}\theta d^{4}\bar{\theta}{\cal H}(A,\bar{A}).}$$

\noindent
In what follows, we will only account for the one instanton corrections to 
both the prepotential and the four derivative terms. The prepotential does
not receive any loop corrections for $N_f =4$. The one instanton 
correction to the prepotential was computed in \rKh. The one instanton 
corrected prepotential is

$$ {\cal F}={1\over 2}\tau_{cl}A^{2}-{i\tau_{cl}\over 2\pi}
   {q\over A^{2}}m_{1}m_{2}m_{3}m_{4}. $$

\noindent
This corresponds to a low energy effective coupling

\eqn\LEC
{\tau=\tau_{cl}-{3iq\tau_{cl}\over\pi \varphi^{4}}m_{1}m_{2}m_{3}m_{4}}

\noindent
The one loop correction to the real function ${\cal H}$ is\rOLoop\

$$ {\cal H}= {3\over 256\pi^{2}}\log^{2}\Big(
   {A\bar{A}\over\langle A\rangle \langle \bar{A}\rangle}\Big), $$

\noindent 
and the one instanton correction is given by

$$ {\cal H}(\varphi,\bar{\varphi})={-qm_{1}m_{2}m_{3}m_{4}\over
   8\pi^{2}\varphi^{4}} \log\bar{\varphi}.$$

\noindent
The one anti-instanton contribution is given by the complex conjugate
of the one instanton correction. The pure scalar two and four derivative 
terms appearing in the low energy effective action, after performing the
field redefinition needed to compare to the brane result, are easily
obtained by using the formulas quoted in \rHo,\rBob. The results are

\eqn\LowEn
{S=\int d^{4}x\Big( K_{\varphi\bar{\varphi}}
\partial_{\mu}\varphi\partial^{\mu}\bar{\varphi}+
\tilde{\cal H}_{\varphi\varphi\bar{\varphi}\bar{\varphi}}
(\partial^{m}\varphi)
(\partial_{m}\varphi)(\partial^{n}\bar{\varphi})
(\partial_{n}\bar{\varphi})\Big),}

\noindent
where

$$  K_{\varphi\bar{\varphi}}\equiv Im\Big(
    {\partial^{2}{\cal F}\over\partial\varphi^{2}}\Big) 
    =\tau_{2}={4\pi^{2}\over g_{cl}^{2}}-
{6\pi\over g_{cl}^{2}}m_{1}m_{2}m_{3}m_{4}\Big[
{q\over a^4}+{\bar{q}\over \bar{a}^4}\Big]$$

\noindent
and

$$ {\eqalign{\tilde{\cal H}_{\varphi\varphi\bar{\varphi}\bar{\varphi}}&=
  16\Big({\partial^{4}{\cal H}\over \partial\varphi\partial\varphi
  \partial\bar{\varphi}\partial\bar{\varphi}}-
  {\partial^{3}{\cal H}\over \partial\varphi\partial\varphi
  \partial\bar{\varphi}}
  (K_{\bar{\varphi}\varphi})^{-1}{\partial 
  K_{\varphi\bar{\varphi}}\over\partial\bar{\varphi}}-
  {\partial K_{\varphi\bar{\varphi}}\over \partial\varphi}
  (K_{\bar{\varphi}\varphi})^{-1}
  {\partial^{3}{\cal H}\over \partial\varphi
  \partial\bar{\varphi}\partial\bar{\varphi}}\cr
  &+2{\partial K_{\varphi\bar{\varphi}}\over\partial\varphi}
  (K_{\bar{\varphi}\varphi})^{-1}
  {\partial^{2}{\cal H}\over \partial\varphi\partial\bar{\varphi}}
  (K_{\bar{\varphi}\varphi})^{-1}{\partial K_{\varphi\bar{\varphi}}
  \over\partial\bar{\varphi}}\Big)\cr
  &={3\over 8\pi^{2}}
  {1\over\varphi^{2}\bar{\varphi}^{2}}
  +{40 m_{1}m_{2}m_{3}m_{4}\over\pi^{2}}
  \Big[{q\over\bar{\varphi}^{2}\varphi^{6}}
  +{\bar{q}\over\varphi^{2}\bar{\varphi}^{6}}\Big].}} $$

\noindent
Notice that for large $\varphi$ the fall off of the four derivative terms
is like $|\varphi|^{-4}$. This has an interesting supergravity interpretation.

\subsec{Supergravity Results}

The first step in the supergravity analysis entails solving \sxh\ for the
background geometry, with the complex coupling $\tau$ given in \LEC. The
coupling $\tau$ is only valid for large $|\varphi|$. For small $|\varphi|$
higher instanton corrections can not be neglected. For this reason, we will
construct a solution to \sxh\ which is valid for large $|\varphi|$. Towards
this end, split $\tau_{2}$ into two pieces as follows

$$ \tau_{2}=V_{1}-V_{2},\quad V_{1}={4\pi^{2}\over g_{cl}^{2}}
   \equiv\tau_{2cl},\quad V_{2}=
   {3\tau_{2cl}\over 2\pi}m_{1}m_{2}m_{3}m_{4}\Big[
   {q\over a^4}+{\bar{q}\over \bar{a}^4}\Big]. $$

\noindent
We can now solve \sxh\ perturbatively by writing $f=f_{0}+f_{1}+...$ where

\eqn\FrtPbn
{\Big[V_{1}{\partial^{2}\over\partial y^{2}}+{\partial^{2}\over\partial a\partial\bar{a}}\Big]
f_{0}=-N(2\pi )^{4}\delta^{(4)}(y)\delta^{(2)}(a),}

\eqn\SndPbn
{\Big[V_{1}{\partial^{2}\over\partial y^{2}}+{\partial^{2}\over\partial a\partial\bar{a}}\Big]
f_{n}=V_{2}{\partial^{2}\over\partial y^{2}}f_{n-1}.}

\noindent
To find the leading corrections to the four derivative terms, it is
sufficient to focus attention on $f_{0}$ and $f_{1}$. The solution for
$f_{0}$ is 

$$ f_{0}={4N\pi\over\big[y^{2}+\tau_{2cl}|a|^{2}\big]^{2}}. $$

\noindent
The function $f_{1}$ satisfies

$$  \Big[\tau_{2cl}{\partial^{2}\over\partial y^{2}}+
    {\partial^{2}\over\partial a\partial \bar{a}}\Big]f_{1}=
    {3m_{1}m_{2}m_{3}m_{4}\tau_{2cl}\over 2\pi}
    \Big({q\over a^{4}}+{\bar{q}\over\bar{a}^{4}}\Big)
    \Big(-{16N\pi\over \big[y^{2}+\tau_{2cl}a\bar{a}\big]^{3}}
    +{96N\pi y^{2}\over \big[y^{2}+\tau_{2cl}a\bar{a}\big]^{4}}
    \Big). $$

\noindent
We will look for solutions to this equation that preserve rotational symmetry 
in the $y_{i}$ variables. To do this it is useful to move into radial 
coordinates. Denoting the angular variable in the $a,\bar{a}$ plane by 
$\theta$ and the radial coordinate in the $a,\bar{a}$ plane by $r$ and 
in the $y_{i}$ plane by $\rho$, we find\foot{The factors $q$ and $\bar{q}$
appearing in $V_{2}$ are pure phases and can be absorbed into a convenient
choice for $\theta=0$.}

$$ {\eqalign{\Big[\tau_{2cl}&
   {\partial^{2}\over\partial \rho^{2}}+\tau_{2cl}
   {3\over\rho}{\partial\over\partial\rho}+
   {\partial^{2}\over\partial r^{2}}+
   {1\over r}{\partial\over\partial r}+{1\over r^{2}}
   {\partial^{2}\over\partial\theta^{2}}\Big]f_{1}\cr
   &\qquad=
   {3m_{1}m_{2}m_{3}m_{4}\tau_{2cl}\over \pi}
   {cos(4\theta)\over r^{4}}
   \Big(-{16N\pi\over \big[\rho^{2}+\tau_{2cl}r^{2}\big]^{3}}
   +{96N\pi \rho^{2}\over \big[\rho^{2}+\tau_{2cl}r^{2}\big]^{4}.}
   \Big).}} $$

\noindent
By inspection, it is clear that the angular dependence of $f_{1}$ is given
by $f_{1}=cos(4\theta)g(r,\rho).$ The function $g$ satisfies

\eqn\Forg
{\eqalign{\Big[\tau_{2}^{(0)}&
{\partial^{2}\over\partial \rho^{2}}+\tau^{(0)}_{2}
{3\over\rho}{\partial\over\partial\rho}+
{\partial^{2}\over\partial r^{2}}+
{1\over r}{\partial\over\partial r}-{16\over r^{2}}\Big]g\cr
&\qquad=
{3m_{1}m_{2}m_{3}m_{4}\tau_{0}^{(2)}\over \pi}
{1\over r^{4}}
\Big(-{16N\pi\over \big[\rho^{2}+\tau_{2}^{(0)}r^{2}\big]^{3}}
+{96N\pi \rho^{2}\over \big[\rho^{2}+\tau_{2}^{(0)}r^{2}\big]^{4}.}
\Big).}}

\noindent
This equation admits a power series solution. To set up the solution, 
notice 
that both $r$ and $\rho$ have the dimensions of length ($L$). It is not 
difficult to see that $g$ has dimension $L^{-8}$. Thus, by dimensional
analysis, it must have an expansion of the form

\eqn\Expnsn
{g=\sum_{m}{c_{m}\over r^{m}\rho^{8-m}}.}

\noindent
For consistency, we require that $g\to 0$ at least as $r^{-4}$ as $ r\to\infty$.
If this is not the case, $f_{1}$ is not a small correction to $f_{0}$. 
Thus, we restrict $m\ge 4$ in \Expnsn. With this restriction, after inserting 
\Expnsn\ into \Forg, one finds for the first few $c_{m}$:

$${\eqalign{c_{m}&=0,\qquad m<8,\qquad
c_{8}=\alpha m_{1}m_{2}m_{3}m_{4}N,\cr
c_{9}&=0\qquad
c_{10}=-6m_{1}m_{2}m_{3}m_{4}N\Big[
{1+\alpha\over (\tau_{2cl})^{3}}
\Big].}}$$

\noindent
The full solution is not needed, since only $f_{1}$ at $y=0$ enters the
probe action. Notice that this solution for $f_{1}$ is labeled by an 
arbitrary parameter $\alpha$ which cannot be fixed by the above iterative calculation. It is an interesting
open question to see if $\alpha$ can be fixed by a more sophisticated
analysis. The correction to the leading term in $f$

$$ f(y=0,a,\bar{a})={4N\pi\over (\tau_{2cl}a\bar{a})^{2}}+
\alpha m_{1}m_{2}m_{3}m_{4}N{1\over 2(\bar{a}a)^{2}}
\Big({q\over a^{4}}+{\bar{q}\over\bar{a}^{4}}\Big)
+O\Big({1\over |a|^{12}}\Big)$$

\noindent
represents a correction to the $AdS_{5}$ geometry. This correction is
expected because we are no longer dealing with a conformal field theory.
Notice that the $AdS_{5}$ geometry is recovered in the limit of large energies
($|a|\to\infty$) and in the limit of massless matter $m_{i}\to 0$.
Expanding the probe action in this background, we find that the pure scalar
terms read

$$ {\eqalign{S={T_{3}\over 2}\int\Big(&\Big[\tau_{2cl}-
{3\tau_{2cl} \over 2\pi}m_{1}m_{2}m_{3}m_{4}\Big(
{q\over a^4}+{\bar{q}\over \bar{a}^4}\Big)\Big]
\partial_{m}a\partial^{m}\bar{a}\cr
+&\Big[{2N\pi\over (a\bar{a})^{2}}+
{\alpha Nm_{1}m_{2}m_{3}m_{4}(\tau_{2cl})^{2}\over4(a\bar{a})^{2}}
\Big({q\over a^{4}}+{\bar{q}\over\bar{a}^{4}}\Big)\Big]
\partial_{n}a\partial^{n}a
\partial_{m}\bar{a}\partial^{m}\bar{a}
\Big).}} $$

\noindent
Notice that the correction to the four derivative terms has the structure of
the one instanton corrections computed using field theory. We have not been
able to fix $\alpha$ with our asymptotic analysis, so that the coefficient
of this correction could not be checked. As reviewed above, instanton 
effects explicitly break the conformal symmetry of the field theory. The
breaking of the $SO(2,4)$ conformal symmetry in the field theory is reflected 
in the corrections to the $AdS_{5}$ geometry, which break the $SO(2,4)$
isometry of the $AdS_{5}$ space. Note that the coupling runs with a power law.
Solutions of type-0 string theories with a power law running for the coupling
have been studied in \rAlv. Power law running of the coupling has also
played a prominent r\^ole in gauge-coupling unification in theories with
large internal dimensions\rDie. The presence of the large internal dimensions
is reflected in the fact that massive Kaluza-Klein modes run in loops of the
four dimensional theory. This gives rise to a power law running of the 
couplings. In \rGuby\ it was suggested that this effect may be responsible 
for the power law running of couplings in the IIB background discussed in 
that study. In our case, there is no need for effects due to large internal 
dimensions and the power law running of the coupling is simply explained
by instanton effects in the four dimensional field theory.

Before leaving this section we would like to make some comments on the 
supergravity interpretation of the leading $|a|^{-4}$ behaviour of the
four derivative terms. The operator on the worldvolume which couples to the
dilaton is given by\rDas\

$$ {\cal O}_{\phi}=-{1\over 4}F^{\mu\nu}F_{\mu\nu}. $$

\noindent
By expanding the Born-Infeld action one finds a four derivative term in 
the effective potential that has the form

$$ {\cal O}_{\phi}{\cal O}_{\phi}\Big[{2N\pi\over (a\bar{a})^{2}}+
{\alpha Nm_{1}m_{2}m_{3}m_{4}\tau_{2cl}^{2}\over 4(a\bar{a})^{2}}
\Big({q\over a^{4}}+{\bar{q}\over\bar{a}^{4}}\Big)\Big]. $$

\noindent
The leading term of $|a|^{-4}$ comes from the static massless propagator in 
the six dimensional transverse space. This term is due to exchange of a dilaton
and appears because the supergravity modes which couple to constant gauge fields
on the brane have zero momentum along the brane \rDas. 

\subsec{Instanton Effects and the static Quark-Antiquark Potential from
Supergravity}

In this section we study the static quark-antiquark potential, in 
the large $|\varphi|$ region where the background geometry described above
is valid. This corresponds to studying the effects of instantons on the 
static quark-antiquark potential in the large $N$ field theory. There is a large body
of evidence from lattice calculations that indicate that instanton effects
play a major r\^ole in the physics of light hadrons\rNeg. Below we will argue 
that the supergravity description provides a powerful new approach to these 
questions.

The energy of a quark-antiquark pair can be read off of the expectation value
of a Wilson loop. This Wilson loop is identified with a fundamental string
ending on the boundary of the asymptotically $AdS_{5}$ space\rLoop. The Wilson
loop configuration is thus obtained by minimizing the Nambu-Goto 
action\foot{In this expression $\alpha,\beta=\tau,\sigma,$ 
and $M,N=0,1,...,9$}

$$S={1\over 2\pi}\int d\tau d\sigma\int\sqrt{det(g_{MN}
\partial_{\alpha}x^{M}\partial_{\beta}x^{N})} $$

\noindent
The metric $g_{MN}$ felt by the strings is not the Einstein metric \frth, 
but rather the string frame metric. We are interested in a static string
configuration and take $\sigma=x^{1}$ and $\tau=x^{0}$. The string is at
a fixed $x^{2},x^{3},x^{4},x^{5},x^{6},x^{7}$ and $\theta$ where $\theta$
is the angular variable in the $(8,9)$ plane. In terms of the variable
$r^{2}\equiv a\bar{a},$ the Nambu-Goto action takes the
form

$$ S={T\over 2\pi}\int d\sigma
\sqrt{(\partial_{\sigma}r)^{2}+{1\over f\tau_{2}}}
={T\over 2\pi}\int d\sigma
\sqrt{(\partial_{\sigma}r)^{2}+{r^{4}\over a}-{b\over a^{2}}} $$

$$a={4N\pi\over \tau_{2cl}},\quad
  b=\cos (4\theta)m_{1}m_{2}m_{3}m_{4}N\Big(
{\alpha\tau_{2cl}\over 2}-{6\over \tau_{2cl}}\Big) $$

\noindent
where $T=\int d\tau$. We have dropped terms of $O(r^{-4})$ in the square 
root above. The solution to the Euler-Lagrange equations of motion following 
from this action is obtained in the usual way: The action does not
depend explicitly on $\sigma$ so that the Hamiltonian in the $\sigma$
direction is a constant of the motion

$$
{{r^{4}\over a}-{b\over a^{2}}\over
\sqrt{(\partial_{\sigma}r)^{2}+{r^{4}\over a}-{b\over a^{2}}}}=
const=\sqrt{{r_{0}^{4}\over a}-{b\over a^{2}}}
$$

\noindent
where $r_{0}$ is the minimal value of $r$. By symmetry we have
$r(\sigma=0)=r_{0}$. It is now straight forward to obtain

$$
\sigma=\sqrt{{r_{0}^{4}\over a}-{b\over a^{2}}}
\int_{r_{0}}^{r} {dy\over
\sqrt{({y^{4}\over a}-{b\over a^{2}})
      ({y^{4}\over a}+{r_{0}^{4}\over a})} }.$$

\noindent
The string endpoints are at the boundary of the asymptotically $AdS_{5}$
space ($r\to\infty$) so that we can trade the integration constant $r_{0}$
for the distance $L$ between the quark and anti-quark

$$
L=2\sqrt{{r_{0}^{4}\over a}-{b\over a^{2}}}
\int_{r_{0}}^{\infty} {dy\over
\sqrt{({y^{4}\over a}-{b\over a^{2}})
      ({y^{4}\over a}+{r_{0}^{4}\over a})} }.$$

\noindent
The energy is now computed by evaluating our action at this classical 
solution. After subtracting twice the self-energy of a quark, we obtain 
the following result for the quark-antiquark potential

$$
E={1\over\pi\sqrt{({r_{0}^{4}\over a}-{b\over a^{2}})}}
\int_{r_{0}}^{\infty}dy\left(
{\sqrt{(y^{4}-{b\over a})}
\over\sqrt{y^{4}+r_{0}^{4}}}-1\right).
$$ 

\noindent
To extract the dependence of this energy on the quark-antiquark separation $L$
we need to determine $r_{0}$ as a function of $L$. The expression for the 
energy given above is identical to the static potential for the 
quark-antiquark pair in the ${\cal N}=4$ theory at finite temperature\rSon.
>From the results of \rSon\ we know that $E(L)$ has the form

$$
E=-{c_{1}\over L}-c_{2}L^{3}.
$$

\noindent
The constant $c_{1}$ is positive. In the limit that $m_{i}\to 0$, $c_{2}\to 0$
and we regain the ${1\over L}$ dependence, a fact which is determined by
conformal invariance. The sign of the constant $c_{2}$ is dependent on 
$\theta.$ For $c_{2}$ positive (negative) we have a screening (antiscreening)
of the quark-antiquark pair due to the instantons.
This expression can't be trusted for very large $L$: for larger
and larger $L$ the Wilson loop is able to move further and further into the
bulk. Our solution is however only valid for large $r$, so that the Wilson loop
begins to explore regions in the bulk for which our solution is not valid. The
long distance behaviour of the quark-antiquark potential could be extracted 
from the exact supergravity background.

\newsec{Pure Gauge ${\cal N}=2$ Super Yang-Mills Theory with Gauge Group 
$SU(2)$}

In this section we obtain the leading correction to the four derivative terms
in both the field theory and the supergravity descriptions in the probe approach.

\subsec{Field Theory Results}

In the case where there are no flavor multiplets, the perturbative beta 
function does not vanish and the field theory is asymptotically free. We 
will focus attention on the perturbative contributions to the two and four 
derivative terms appearing in the low energy effective action. The one loop
results for the K\"ahler metric and real function ${\cal H}$ are

$$ K_{\varphi\bar{\varphi}}\sim 
{\log (\varphi\bar{\varphi}/\Lambda^{2})\over\varphi\bar{\varphi}}, $$

$$ {\cal H}(A,\bar{A})\sim \log \Big({A\over\Lambda}\Big)
\log\Big({\bar{A}\over\Lambda}\Big). $$

\noindent
This leads to the following four derivative term for the scalars, after
performing the field redefinition\rHo\

$$ \eqalign{S&=\int d^{4}x 
   (\partial^{m}\varphi\partial_{m}\varphi)
   (\partial^{n}\bar{\varphi}\partial_{n}\bar{\varphi})
   {8+4\log\Big({\varphi\bar{\varphi}\over\Lambda^{2}}\Big)
   +\Big[\log\Big({\varphi\bar{\varphi}\over\Lambda^{2}}\Big)\Big]^{2}\over
   \varphi^{2}\bar{\varphi}^{2}
   \Big[\log\Big({\varphi\bar{\varphi}\over\Lambda^{2}}\Big)\Big]^{2}}\Big]\cr
  &=\int d^{4}x (\partial^{m}\varphi\partial_{m}\varphi)
  (\partial^{n}\bar{\varphi}\partial_{n}\bar{\varphi})
  \Big[{1\over (\varphi\bar{\varphi})^{2}}+
  O\Big({1\over (\varphi\bar{\varphi})^{2}log|\varphi|}\Big)\Big].}$$

\noindent
The $|\varphi|^{-4}$ fall off at large separations (large $|\varphi|$) 
suggests that the dominant interaction between the branes is again due to 
the exchange of massless supergravity modes propagating in the six 
dimensional space transverse to the three branes.

\subsec{Supergravity Results}

The problem of finding the relevant background geometry corresponding to
the asymptotically free gauge theory is considerably more complicated. The
Laplace equation \sxh\ becomes (we will show all $N$ dependence in this 
section)

$$ \Big[{N\over\lambda }
\Big({8\pi\over g_{cl}^{2}}+{6\over\pi}+{2\over\pi}
\log \Big({a\bar{a}\over\Lambda^{2}}\Big)
\Big){\partial^{2}\over\partial y^{2}}
+{\partial\over\partial a}{\partial\over\partial \bar{a}}\Big]f=
-(2\pi )^{4}N\delta^{(4)}(y)\delta^{(2)}(a). $$

\noindent
Performing a Fourier transform on the $y$ variables and working in the large 
$|a|$ region (which corresponds to the semi-classical regime of the field 
theory) we find

$$ \Big[{\partial\over\partial a}{\partial\over\partial \bar{a}}
-k^{2}{N\over\lambda}
\Big({8\pi\over g_{cl}^{2}}+{6\over\pi}+{2\over\pi}
\log \Big({a\bar{a}\over\Lambda^{2}}\Big)\Big)
\Big]f=0. $$

\noindent
We have not been able to solve this equation exactly. However an approximate
solution in the large $|a|$ region is given by

$$ f\approx Ne^{-{2k\sqrt{N}|a|\over\sqrt{\lambda\pi}}
\sqrt{{2\pi^{2}\over g_{cl}^{2}}+{1\over 2}+\log(|a|)}},$$

\noindent
Using this approximate solution we obtain

$$ f(y=0,a,\bar{a})=\int d^{4}k f(k,a,\bar{a})=
{12\pi^{2}\lambda^{2}\over 16N(a\bar{a})^{2}
\Big(\log\Big[{\sqrt{a\bar{a}}\over\Lambda}exp({2\pi^{2}\over g_{cl}^{2}}+{1\over 2})\Big]
\Big)^{2}}. $$

\noindent
This result determines the coefficient of the four derivative terms

$$
\tau_{2}^{2}f(y=0,a,\bar{a})\sim {N\over (a\bar{a})^{2}}+
O\Big({1\over (a\bar{a})^{2}log|a|}\Big). $$

\noindent
This is exactly the same behaviour as obtained from the field theory
analysis.

The complex coupling $\tau$ in this supergravity background has a logarithmic
dependence on $a\bar{a}$ corresponding to the logarithmic dependence of the 
field theory coupling on the energy scale. Gravity solutions that have
couplings with this logarithmic dependence have been constructed in 
type-0 theories\rIgK. 

We should now address the validity of this computation. There are two potential
sources of corrections to the supergravity background - string loop effects
and curvature corrections. At large $N$ and large 't Hooft coupling both of
these types of corrections are small and supergravity is a reliable
description of the background. As the 't Hooft coupling decreases, curvature 
corrections become
important and the uncorrected supergravity can no longer be trusted\rWK. 
We would like to determine whether the uncorrected supergravity is a valid description for the 
perturbative field theory living on the probe. The simplest way to assess the 
validity of the supergravity description is simply to compute the square of
the Ricci tensor. This calculation should be carried 
out in string frame because
we are interested in the region in which the dilaton is going to zero. We will
show both the Einstein and string frame results in what follows. In the large
$|a|$ region the Einstein frame metric that we have computed above takes the
form\foot{In what follows $\mu,\nu=0,1,2,3$; $i,j=4,5,6,7$ and
$M,N=0,1,...,9$.}

$$ g_{MN}^{(e)}dx^{M}dx^{N}={\sqrt{N}}\Big(a\bar{a}\log|a|
\eta_{\mu\nu}dx^{\mu}dx^{\nu}
+{1\over a\bar{a}\log |a|}\Big(\delta_{ij}dx^{\prime i}dx^{\prime j}
+\log|a| dad\bar{a}\Big)\Big),$$

\noindent
where we have rescaled $x^{i}\to x^{\prime i}$ where $x'=x/\sqrt{N}$. 
The corresponding string frame metric is

$$ g_{MN}^{(s)}dx^{M}dx^{N}={}
\Big(a\bar{a}(\log|a|)^{1/2}
\eta_{\mu\nu}dx^{\mu}dx^{\nu}
+{1\over a\bar{a}(\log |a|)^{3/2}}\Big(\delta_{ij}dx^{\prime i}dx^{\prime j}
+\log|a| dad\bar{a}\Big)\Big).$$

\noindent
The leading contribution to the square of the Ricci tensor in the Einstein
frame is a constant

$$R_{MN}R^{MN}={32\over N}.$$

\noindent
In the string frame, the leading contribution to the square of the Ricci tensor
diverges logarithmically for large $|a|$

$$R_{MN}R^{MN}={32}\log |a|.$$

\noindent
To interpret these results, note that the Yang-Mills coupling squared is
$g^{2}=\lambda(N\log|a|)^{-1}$, so that the 't Hooft coupling is 
$\lambda\equiv g^{2}N=1/\log|a|.$ We see that the square of the Ricci 
tensor in the string frame is inversely proportional to the 't Hooft coupling,
so that we recover the well known result that curvature effects in the background 
are small at large $\lambda$. The perturbative probe field theory is valid for $|a|>>1$. It is clear that in the $N\to\infty$ limit, large $\lambda$ and large $|a|$ are incompatible, since the 't Hooft coupling is large when the probe worldvolume field theory is non-perturbative, {\it{i.e.}} $|a| \to 0$, and vice versa. Consequently, the fact that the probe analysis captures the {\it{perturbative}} behaviour of the field theory is unexpected. It is not unreasonable to conclude that this correspondence comes about due to the constraints of ${\cal{N}}=2$ supersymmetry, which prevents certain quantities from being renormalised beyond the one loop level.


We note that the effects that we have computed in this section are 
linear in both the number of source branes and the number of probe branes. The
supergravity will not capture effects which do not have this linear dependence.

In the approach followed in this thesis, ${\cal N}=2$ super Yang-Mills theory is realised as the worldvolume theory of a threebrane probe in an F-theory background\rDoug,\rOfer,\Sen. However, an alternative approach to Super Yang-Mills theory with gauge group $SU(N)$ and ${\cal N}=2$ supersymmetry is to realise it at low energy and weak string coupling on the worldvolume of $N$ Dirichlet fourbranes stretched between two Neveu-Schwarz fivebranes\rWitten. 
The singularities where the fourbranes meet the Neveu-Schwarz fivebranes are
resolved in the strong string coupling limit where this brane set up is
replaced by a single M-theory fivebrane wrapping the Seiberg-Witten curve.
In this M-theory description, the fourbranes become tubes, wrapping the
eleventh dimension and stretching between the two flat asymptotic sheets of
the M-fivebrane. The point at which they meet the fivebrane is resolved -
the fourbrane (tube) ending on a fivebrane creates a dimple in the fivebrane. 
It is a fascinating result that the non-trivial bending of the branes due
to these dimples encode the perturbative plus all instanton corrections to the
low energy effective action describing the original $SU(N)$ ${\cal N}=2$ super
Yang-Mills theory\foot{The Seiberg-Witten effective action can also be 
recovered from a Dirichlet fivebrane of IIB string theory wrapping the 
Seiberg-Witten curve\rGukov.}\rOz. The encoding of quantum effects in the field theory in a bending of the brane geometry is a general feature of gauge theories realised on brane worldvolumes, and will make an appearance in the latter part of this thesis. 

The M-theory description of the fivebrane is only valid at large string coupling. Field theory however, is only recovered
at weak string coupling\rOoguri. Quantities that are protected by
supersymmetry are not sensitive to whether they are computed at weak or strong
coupling. Consequently, computing these quantities we find the fivebrane and
field theory results agree. However, the fivebrane does not reproduce the field theory
result for quantities that are not protected by supersymmetry. In particular,
higher derivative corrections to the low energy effective action computed
using the fivebrane do not agree with the field theory results\rOz.


\vfill \eject

\newsec{Field theory BPS states from the Probe Worldvolume}

As is well known, ${\cal N}=2$ super Yang-Mills theories have a rich spectrum of BPS{\foot{For a general discussion of BPS states see, for example, \rHarv.} states\rSW,\rFerrari.
The latter part of this thesis will be devoted to exploring field theory BPS states from the point of view of the threebrane worldvolume. Sen \rSen\ has identified these states with strings that stretch along geodesics from the probe to a sevenbrane. This description of BPS states provides an elegant description of the selection rules\rSel\ determining the allowed BPS states for the $N_f\le 4$ $SU(2)$ Seiberg-Witten theory\rFayy. However, open string junctions can fail to be smooth\rGaber. In this case, one needs to consider geodesic string junctions to describe the BPS states\rBerg.
An assumption implicit in this approach is that the threebrane probe is infinitely heavy as compared to the string. In this approximation, the three brane remains flat and its geometry is unaffected by the string ending on it. Recalling the
r\^ole that bending plays in the M-theory fivebrane description of field theory, it is natural to identify this approximation with a classical description of the BPS states. 

In the following, we will construct a description of BPS states of the Born-Infeld
probe world-volume action, which accounts for the bending of the threebrane into
a dimple due to the string attached to it\foot{An analogous computation for
the M-theory fivebrane has been carried out in\rWst. However, in this case the
BPS states correspond to self-dual strings stretched along the Riemann surface
associated with the Seiberg-Witten curve, instead of along the moduli space.}.

The analogy to the M-fivebrane description of field theory discussed earlier suggests that we are accounting for quantum
effects in the background of these BPS states. We are able to provide a supergravity
solution corresponding to a string stretching over a finite interval. On the 
world-volume of the D3 probe this translates into a cut-off allowing for the existence
of {\it finite energy} classical solutions of the abelian Born-Infeld action. We show that
the long distance properties of the monopole solution to non-abelian gauge theories 
are reproduced by the abelian low energy effective action. This caculation again
requires precise knowledge of the near horizon background geometry. This background 
geometry pinches the dimple on the D3 brane into an endpoint. By appealing to the holographic 
principle we are able to interpret the long distance ``cut-off" on the dimple in terms of the
correct UV cut-off entering in the definition of the Wilsonian effective action. 

In section (12) we will review the problem of solving for the background set up by the sevenbranes plus $N$ threebranes. This reduces to the problem of solving the Laplace equation on the background generated by the 
sevenbranes\rOfer. We show that the solution to this Laplace equation is
duality invariant. By making an explicitly duality invariant ansatz, we are
able to accurately construct the background geometry in the large
$|a|$ region and close to the sevenbranes. A computation of the square of the
Ricci tensor shows that this background has curvature singularities. The
interpretation of these singularities in the field theory are as the
points in moduli space where the effective action breaks down due to the appearance of new 
massless particles. 

In section (13) we construct solutions corresponding to Dirichlet strings stretching from the probe to the
(0,1) sevenbrane{\foot{The sevenbranes carry R and NS charge. We may define a notation for the sevenbranes by saying that a $(p,q)$ string can end on a $(p,q)$ sevenbrane. Thus, for example, a (1,0) string is an elementary type IIB string, a $(1,0)$ sevenbrane is a Dirichlet sevenbrane. In this notation, the F-theory background contains one $(0,1)$ sevenbrane, one $(2,1)$ sevenbrane and $N_f$
$(1,0)$ sevenbranes.}. We do this explicitly for the $N_f=0$ case, but the
extension to $N_f\le 4$ is trivial.\foot{F-theory backgrounds which correspond to 
the superconformal limit of the field theories we consider have been constructed in \rSpa.} 
These ``magnetic dimple" solutions have
the interesting property that they capture the long distance behaviour of
the Prasad-Sommerfeld monopole solution\rPS. 
The Prasad-Sommerfeld solution is an
exact solution for an $SU(2)$ gauge theory. An important feature of this
monopole solution, intimately connected to the non-Abelian structure of the
theory, is the way in which both the gauge field and Higgs field need to be
excited in order to obtain finite energy solutions. It is interesting that
the Abelian worldvolume theory of the probe correctly captures this structure.
This is not unexpected since the probe worldvolume theory is describing the 
low energy limit of a non-Abelian theory. The energy of these solutions is 
finite due to a cut off which must be imposed, as indicated earlier. The UV cut off in the field
theory maps into an IR cut off in the supergravity, which is a manifestation
of the IR/UV correspondence\rES.

In section (14) we consider the moduli space of the solutions we have 
constructed. We are able to compute the metric on the one monopole moduli
space exactly. For two monopoles and higher, it does not seem to be possible to do things exactly. However, under the assumption that the monopoles are very widely separated, we are able to construct the asymptotic form of the metric.
In this case, we see that the metric receives both perturbative and instanton corrections. The hyper-K\"ahler structure of these moduli spaces in unaffected by these corrections.

\vfill\eject

\newsec{Gravitational Interpretation of Singularities in the Field Theory Moduli Space}

As motivated earlier, the model that we consider comprises a large number $N$ of 
coincident threebranes and a group of separated
but parallel sevenbranes. We will briefly review the construction of the background setup by the $N$ threebranes and the sevenbranes (first discussed earlier in section (8)) with a view to obtaining a duality invariant solution to the Laplace equation on the sevenbrane background. The sevenbranes have
worldvolume coordinates $(x^0,x^1,x^2,x^3,x^4,x^5,x^6,x^7)$; the 
threebranes have worldvolume coordinates $(x^0,x^1,x^2,x^3).$ The presence 
of the sevenbranes gives rise to nontrivial monodromies for the complex 
coupling\foot{The $N$ dependence of $\tau$ has been discussed in\rUs. We
will not indicate this dependence explicitly.}
$\tau=\tau_1 +i\tau_2$ as it is moved in the $(x^8,x^9)$ space transverse to
both the sevenbranes and the threebranes. The spacetime
coordinates $(x^8,x^9)$ play the dual r\^ole of coordinates in the
supergravity description, and of the (complex) Higgs field $a$ in the field
theory\foot{As we have already noted, the precise identification between $(x^8,x^9)$ and $a$ involves a field redefinition as explained in \rJe.}. The metric due to the sevenbranes by
themselves is given by\rGSVY\

\eqn\Met
{ds^2=-(dx^0)^2+(dx^i)^2+\tau_2\big[(dx^8)^2+(dx^9)^2\big],}

\noindent
where $i$ runs over coordinates parallel to the sevenbrane.
The addition of $N$ threebranes to this pure sevenbrane 
background deforms the metric and leads to a non-zero self-dual $R$-$R$ fiveform
flux. This flux and the deformed metric are given in terms of $f$

\eqn\Sln
{\eqalign{ds^2 =f^{-1/2}dx_{\parallel}^{2}+f^{1/2}\big[ (dx^{i})^{2}
+&\tau_{2}\big( (dx^{8})^{2}+(dx^{9})^{2}\big)\big]=
f^{-1/2}dx_{\parallel}^2 + f^{1/2}{g}_{jk}dx^{j}dx^{k},\cr
F_{0123j}&= -{1\over 4}\partial_{j}f^{-1},}}

\noindent
where $f$ is a solution to the Laplace equation in the sevenbrane 
background\rOfer\

\eqn\LaplacEqn
{{1\over\sqrt{g}} 
\partial_{i}(\sqrt{ g}{ g}^{ij}\partial_j f)=
- (2 \pi)^4 N { \delta^6(x-x^0) \over \sqrt{\ g}. }}

\noindent
The position of the $N$ source threebranes is $x^0$.
The index $i$ in \Sln\ runs over coordinates transverse to the threebrane
but parallel to the sevenbrane, $j,k$ runs over coordinates transverse to
the threebrane and $x_{\parallel}$ denotes the coordinates
parallel to the threebrane. For the backgrounds under consideration, 
\LaplacEqn\ can be written as

\eqn\LapEqn
{\big[\tau_{2}\partial_{y}^{2}+4l_s^{-4}\partial_{a}\partial_{\bar{a}}\big]
f=-(2\pi )^{4}N\delta^{(4)}(y-y^0)\delta^{(2)}(a-a^0)}

\noindent
where $a,\bar{a}$ and $\tau_2$ are the quantities appearing in the 
Seiberg-Witten solution\rSW\ and $y$ denotes the coordinates transverse
to the threebranes but parallel to the sevenbranes.
In sections (8)--(11) of the thesis this equation was solved in the large $|a|$ limit, for $\tau$ 
corresponding to ${\cal N}=2$ field theories with gauge group $SU(2)$ and
$N_f=0,4$ flavors of matter. It is possible to go beyond this approximation,
by noting that the form of Eq. \LapEqn\ implies that $f$ is invariant under
a duality transformation. To see this, change from the electric variables
$a$ to the dual magnetic variables $a_D$. After a little rewriting, we find
that Eq. \LapEqn\ becomes

\eqn\Trnfrmd
{\big[{1\over 2i}(\partial_{\bar{a}} \bar{a}_D-
\partial_a a_D )\partial_{y}^{2}
+4l_s^{-4}\partial_{a_D}\partial_{\bar{a}_D}\big]
f=-(2\pi )^{4}N\delta^{(4)}(y-y^0)\delta^{(2)}(a_D-a_D^0)}

\noindent
If we now identify $\tau_D=-{1\over \tau}$ we see that $f$ is also a solution 
of the Laplace equation written in terms of the dual variables. This result is
a consequence of the fact that the Einstein metric is invariant under the
classical $SL(2,R)$ symmetry of IIB supergravity. It is possible to construct 
an approximate solution $f$ which is valid in the region corresponding to
large $|a|$ and in the region corresponding to small $|a_D|$ 

\eqn\fsoln
{f(y,a,\bar{a})={l_s^4\over\big[y^{i}y^{i}-
{i\over 2}l_s^4(a_{D}(\bar{a}-\bar{a}_0)-\bar{a}_{D}(a-a_0))\big]^{2}}.}

\noindent
In this last formula, $a_0$ is the constant value of $a$ at $a_D=0$.
For concreteness, we will focus on the pure gauge theory in the discussion
which follows, but our results are valid for all $N_f\le 4$.
In the large $a$ limit, from the work of Seiberg and Witten\rSW, we know
that $a_D$ can be expressed as a function of $a$ as (see for example
\rLerche)

\eqn\AdA
{a_D=\tau_0 a+{2ia\over\pi}\log\Big[{a^2\over\Lambda^2}\Big]
+{2ia\over\pi}+{a\over 2\pi i}\sum_{l=1}^\infty c_l (2-4l)\Big(
{\Lambda\over a}\Big)^{4l},}

\noindent
where $\tau_0$ is the classical coupling and $\Lambda$ is the dynamically
generated scale at which the coupling becomes strong.
Inserting this into \fsoln\ and evaluating 
the expression at $y=0$ we find the
following asymptotic behaviour for $f$

\eqn\AsymptBeh
{f\sim {1\over (a\bar{a}\log|a|)^2}.}

\noindent
This reproduces the large $|a|$ solution of \rUs. In the small $|a_D|$ limit,
$a$ can be expressed as\rLerche\

\eqn\Asymptotic
{a=\tau_{0D}a_D-{2ia_D\over 4\pi}\log\Big[{a_D\over\Lambda}\Big]
-{ia_D\over 4\pi}-{1\over 2\pi ia_D}\Lambda^2\sum_{l=1}^\infty
c_l^D l\Big({ia_D\over \Lambda}\Big)^{l}.}

\noindent
Thus, in this limit and at $y=0$ we find

\eqn\LimitForF
{f\sim {1\over (\bar{a}_D a_D\log|a_D|)^2}.}

\noindent
It is easy to again check that this is the correct leading behaviour for $f$
close to the monopole singularity at $|a_D|=0$. The corrections to $f$, at
$y=0$ are of
order $|a_D|^{-4}(\log |a_D|)^{-4}$. In the large $|a|$ limit, $f\to 0$ and 
in the small $|a_D|$ limit, $f\to\infty$ signaling 
potential singularities in both limits. From
the field theory point of view, both of these limits correspond to weakly
coupled limits of the field theory (or its dual) which leads us to suspect
that curvature corrections may not be small in these regions\rWit. 
This is easily confirmed by computing the square of the Ricci tensor in 
string frame, which for large $|a|$ behaves as $R^{MN}R_{MN}\sim \log|a|$
and for small $|a_D|$ behaves as 
$R^{MN}R_{MN}\sim \log|a_D|.$
It is clear that the point $|a_D|=0$ corresponds to a
curvature singularity in the supergravity background. If we circle this
point in the field theory moduli space, the coupling transforms with a
nontrivial monodromy, so that this point is to be identified with the
position of a sevenbranes. The appearance of these naked singularities in
the supergravity background could have been anticipated from no-hair
theorems\rnh.

Recently a number of interesting type IIB supergravity backgrounds were 
constructed\rGubser. These backgrounds exhibit confinement and a running
coupling. In addition, a naked singularity appears in spacetime. It is
extremely interesting to determine whether this naked singularity can be
attributed to a weakly coupled dual description of the theory, since this
would provide an example of duality in a non-supersymmetric confining gauge
theory.

\newsec{Magnetic Dimples in the Probe Worldvolume}

The BPS states of the field theory living on a probe which explores the
sevenbrane background, have been identified with strings that stretch along
geodesics, from the probe to a sevenbrane\rSen.\foot{As mentioned in the introduction
a large number of BPS states of the field theory do not correspond to strings stretched
between the threebrane and a sevenbrane, but rather to string junctions (or webs) which 
connect the threebrane to more than one sevenbrane \rGaber,\rBerg.}  In this section
we will look for a description of these 
BPS states of the probe worldvolume, which
accounts for the bending of the threebrane into a dimple, due to the string
attached to it. We will focus on the case when a Dirichlet string ends on the 
probe, corresponding to a ``magnetic dimple".  

The dynamics of the threebrane probe is given by the Born-Infeld action for a
threebrane in the background geometry of the sevenbranes and $N$ threebranes.
This background has non-zero $R$-$R$ five-form flux, axion, dilaton and metric.
The worldvolume action is computed using the induced (worldspace) metric 

\eqn\IndMet
{g_{mn}=G_{MN}\partial_m X^M\partial_n X^N+l_s^2 F_{mn}}

\noindent 
where $F_{mn}$ is the worldvolume field strength tensor.
We will use capital letters to denote spacetime coordinates ($X^M$) and lower
case letters for worldvolume coordinates ($x^n$). We are using the static
gauge $(X^0,X^1,X^2,X^3)=(x^0,x^1,x^2,x^3)$. In addition to the Born-Infeld
term the action includes a Wess-Zumino-Witten coupling to the 
background $R$-$R$ five-form field strength. The probe is taken parallel to the
stack of $N$ threebranes so that no further supersymmetry is broken when the
probe is introduced. The worldvolume soliton solutions that we will construct
have vanishing instanton number density so that there is no coupling to the
axion. The explicit action that we will use is\rDaction\

\eqn\explctact
{S=T_3\int d^4 x\sqrt{-\det (g_{mn}+e^{-{1\over 2}\phi}l_s^2F_{mn})}
+T_3\int d^4 x\partial_{n_1}X^{N_1}
\wedge ...\wedge\partial_{n_4}X^{N_4} A_{N_1 N_2 N_3 N_4},} 

\noindent
where $A_{N_1 N_2 N_3 N_4}$ is the potential for the self-dual fiveform and
we have omitted the fermions. We work in the Einstein frame so that the
threebrane tension $T_3=l_s^{-4}$ is independent of the string coupling.
In what follows, we will use $a_D$ to denote the magnetic variable that 
provides the correct description of the field theory in the small $|a_D|$ 
regime. The corresponding low energy effective coupling is denoted $\tau_D$.
The electric variable $a$ is the correct variable to use in the large $|a|$
limit and the associated coupling is denoted $\tau$.

\subsec{Spherically Symmetric Solution}

In this section we will construct a worldvolume soliton that 
can be interpreted as a single Dirichlet string 
stretching from the probe to the ``magnetic" (0,1) sevenbrane in the case of the pure gauge theory. The extension to $0<N_f\le 4$ is 
straightforward and amounts to the same computation with a different $\tau$. From
the point of view of the worldvolume, the Dirichlet string endpoint behaves as
a magnetic monopole\rDia. If we consider the situation in which the probe and the
magnetic sevenbrane have the same $x^9$ coordinate and are separated in the
$x^8$ direction, then the only Higgs field which is excited is $x^8$. In 
addition, because we expect our worldvolume soliton is a magnetic monopole
we make the spherically symmetric ansatz $x^8=x^8(r)$ and assume that the
only non-zero component of the worldvolume field strength tensor in 
$F_{\theta\phi}$. With this ansatz, the threebrane action takes the form

\eqn\tbact
{S=T_3\int dtdrd\theta d\phi
\Big(\sqrt{f^{-1}+\tau_{2D}(\partial_{r}x^{8})^2}
\sqrt{f^{-1}r^4\sin^2\theta
-\tau_{2D}l_s^4 F_{\theta\phi}F_{\phi\theta}}
-{1\over f}r^{2}\sin\theta\Big).}

\noindent
We will consider the situation in which the probe is close to the (0,1)
sevenbrane corresponding to a region where $a_D$ is small. For this reason
we use the dual coupling in \tbact\ and we will identify $x^8$ with the
dual Higgs $a_D$.  
Recall the fact that the Born-Infeld action has the property that a BPS
configuration of its Maxwellian truncation satisfies the equation of motion
of the full Born-Infeld action\rMal. 
In our case, the Maxwellian truncation of
the Born-Infeld action is just Seiberg-Witten theory. 
The supersymmetric
variation of the gaugino of the Maxwell theory is

\eqn\MT
{\delta \psi=(\Gamma_{\theta\phi}F^{\theta\phi}
+\Gamma_{8r}\partial_{r}x^8)\epsilon.}

\noindent
A BPS background is one for which this variation vanishes. Now, note that if
we identify ($\eta^{mn}$ is the flat four dimensional Minkowski metric)

\eqn\BPSCond
{\eta^{\theta\theta}\eta^{\phi\phi}l_s^4(F_{\theta\phi})^2=
{F_{\theta\phi}^2l_s^4\over r^4\sin^2\theta}=(\partial_r x^8)^2=
\eta^{rr}(\partial_r x^8)^2,}

\noindent
we obtain a background invariant under supersymmetries \MT, where $\epsilon$
satisfies $(\hat{e}^{\theta}\Gamma_{\theta}\hat{e}^{\phi}\Gamma_{\phi}$
$\pm\hat{e}^8\Gamma_{8}\hat{e}^r\Gamma_{r})\epsilon =0.$ The sign depends on
whether we consider a monopole or an anti-monopole background. This
condition can be rewritten as
$\Gamma_{1}\Gamma_{2}\Gamma_{3}\Gamma_{8}\epsilon=\pm\epsilon$.
Thus, this is a BPS background of Seiberg-Witten theory. We will assume 
that \BPSCond\ provides a valid BPS condition for the full Born-Infeld action.
Upon making this ansatz, we find that the determinant factor in the Born-Infeld
action can be written as a perfect square, as 
expected\rHash. The full Born-Infeld
action plus Wess-Zumino-Witten term then simplifies to the Seiberg-Witten
low energy effective action \foot{The study of the quantum corrections to solitons by 
studying the minima of the low energy effective action has been considered in \rChalm.}

\eqn\SWAct
{S=T_3\int d^4 x\tau_{2D}(\partial_{r}x^{8})^2.}

\noindent
We have checked that the arguments above can also be made at the level of the 
equations of motion. The preserved supersymmetries have a natural 
interpretation. The probe preserves supersymmetries of the form
$\epsilon_L Q_L+\epsilon_R Q_R$ where
$\Gamma_0\Gamma_1\Gamma_2\Gamma_3\epsilon_L=\epsilon_R$. The Dirichlet
string preserves supersymmetries for which
$\Gamma_0\Gamma_8\epsilon_L=\pm\epsilon_R$, with the sign depending on whether
the string is parallel or anti-parallel to the $x^8$ axis. The two
conditions taken together imply that 
$\Gamma_1\Gamma_2\Gamma_3\Gamma_8\epsilon_i=\pm\epsilon_i$, $i=L,R$. Thus,
the supersymmetries preserved by the ansatz \BPSCond\ are exactly the 
supersymmetries that one would expect to be preserved by a Dirichlet string
stretched along the $x^8$ axis. With a suitable choice of the phase of $a_D$
we can choose \foot{This identification is only correct to leading 
order at low energy. One needs to perform a field redefinition of the field theory Higgs fields before they can be identified with supergravity 
coordinates\rJe.} $a_Dl_s^2=-ix^8$. With this choice, 
using the formulas for the dual prepotential
quoted in \rLerche, we find that the dual coupling 
can be expressed in terms of
$x^8$ as 

\eqn\DulCoup
{\tau_{2D}={4\pi\over g_{cl,D}^2}
-{3\over 4\pi}-{1\over 2\pi}\log\Big[{x^8\over l_s^2\Lambda}\Big]
-{1\over 2\pi}\Lambda^2l_s^4\sum_{l=1}^\infty {c_l^D l(l-1) (x^8)^{l-2}\over
(l_s^2\Lambda)^l}.}

\noindent
Note also that $a$ is real and can be expressed in terms of $x^8$ and 
$\tau_{2D}$
as

\eqn\ExpFrA
{a=\int dx^8 {\tau_{2D}\over l_s^2}={4\pi\over g_{cl,D}^2}{x^8\over l_s^2}
-{1\over 4\pi}{x^8\over l_s^2}
-{1\over 2\pi l_s^2}x^8\log\Big[{x^8\over l_s^2\Lambda}\Big]
-{1\over 2\pi}\Lambda^2l_s^2\sum_{l=1}^\infty {c_l^D l (x^8)^{l-1}\over
(l_s^2\Lambda)^l},}

\noindent
The equation of motion which follows from \SWAct\

\eqn\SWEom
{{d\over d r}\Big(\tau_{2D} r^2{dx^8\over dr}\Big)=0,}

\noindent
is easily solved to give

\eqn\Dimple
{\gamma-{\alpha\over r}=a={4\pi\over g_{cl,D}^2 l_s^2}x^8
-{1\over 4\pi l_s^2}x^8
-{1\over 2\pi l_s^2}x^8\log\Big[{x^8\over l_s^2\Lambda}\Big]
-{1\over 2\pi}\Lambda^2l_s^2 \sum_{l=1}^\infty 
{c_l^D l (x^8)^{l-1}\over (l_s^2\Lambda)^l}.}

\noindent
In principle, this last equation determines the exact profile of our magnetic
dimple as a function of $r$. At large $r$ our fields have the following
behaviour

\eqn\Asympt
{{x^8\over l_s^2}=\nu -{\beta\over r},\quad
l_s^4 B^{r}B_r=l_s^4 F^{\theta\phi}F_{\theta\phi}=(\partial_r x^8)^2=
{\beta^2 l_s^4\over r^4},}

\noindent
where $\nu$ and $\beta$ are constants.
To interpret these results, it is useful to recall the Prasad-Sommerfeld
magnetic monopole solution\rPS. This monopole is a solution to the following
$SU(2)$ non-Abelian gauge theory

\eqn\Nalbi
{\eqalign{S&=-{1\over e^2}\int d^4 x\Big({1\over 4}(F^a_{mn})^2+
{1\over 2}(D^n\phi^a)^2\Big),\cr
F^a_{mn}=\partial_m A_n^a&-\partial_n A_m^a +\epsilon^{abc}
A_m^bA_n^c,\quad D_n\phi^a=\partial_n\phi^a+\epsilon^{abc}
A_n^b\phi^c.}}

\noindent
An important feature of the monopole solution, intimately connected to
the non-Abelian gauge structure of the theory, is the fact that in order
to get a finite energy solution, the monopole
solution excites both the gauge field and the Higgs field\rHarv. The Higgs
field component of the Prasad-Sommerfeld monopole is

\eqn\PSMonopole
{\eqalign{\phi^a
&=\hat{r}^a\Big(\gamma \coth ({\gamma r\over\alpha})
-{\alpha\over r}\Big),\cr
&=\gamma -{\alpha\over r}+2\gamma e^{-2\gamma r\over\alpha}+2\gamma
e^{-4\gamma r\over\alpha}...,}}

\noindent
where on the second line we have performed a large $r$ expansion. Clearly,
the electric variable\foot{It is the electric variable $a$ of
Seiberg-Witten theory that is related to the Higgs field appearing in the
original microscopic $SU(2)$ theory.} $a$ reproduces the large $r$ behaviour
of the Higgs field of the Prasad-Sommerfeld solution. Thus, we see that
the Abelian worldvolume theory of the threebrane probe catches some
of the structure of the non-Abelian field theory whose low energy limit
it describes. Comparing $a$ to the asymptotic form of the Higgs field in the
Prasad-Sommerfeld solution allows us to interpret the constants of integration
$\alpha$ and $\gamma$. The constant $\alpha$ is related to the inverse of
the electric charge $1/e$. As $r\to\infty$ $a\to\gamma$ so that $\gamma$ 
determines the asymptotic Higgs expectation value; i.e. it determines the 
moduli parameters $a,\bar{a}$ of the field theory or, equivalently, the 
position of the threebrane in the $(8,9)$ plane. The mass of the $W^\pm$
bosons are given by the ratio $m_W=\gamma/\alpha.$ In a similar way, $\nu$
fixes the asymptotic expectation value of the dual Higgs field and $\beta$
is related to the inverse magnetic charge $1/g$. Thus, the mass of the BPS
monopole is given by $m_g=\nu/\beta.$ We are working in a region of moduli
space where $|a_D|$ is small, so that the monopoles are lighter that the
$W^\pm$ bosons. We will return to the exponential corrections in \PSMonopole\
below.

Consider next the small $r$ limit. It is clear that in this limit 
$a\to\infty$. To correctly interpret this divergence in $a$, recall that
the Seiberg-Witten effective action is a Wilsonian effective action, obtained
by integrating out all fluctuations above the scale set by the mass of
the lightest BPS state in the theory. In our case, the lightest BPS states 
are the monopoles and this scale is $m_g$. 
The largest fluctuations left in the effective low energy
theory all have energies less than $m_g$. By Heisenberg's uncertainty
relation, the effective theory must be cut off at a smallest distance of
$1/m_g$. Clearly then, the divergence in the $r\to 0$ limit is unphysical and
it occurs at length scales below which the effective theory is valid. What is
the interpretation of this short distance (UV) field theory cut off in the
supergravity description? To answer this question, we will need a better 
understanding of the small $r$ behaviour of $x^8$. Towards this end, note that
with our choice $a_D$, $\tau$ is pure imaginary so that

\eqn\Chad
{\tau=-{1\over\tau_D}={i\over\tau_{2D}}=i\tau_2.}
 
\noindent
Thus, the action \SWAct\ can be written as

\eqn\NwAct
{S=T_3\int d^4 x\tau_2 (\partial_r a)^2=
T_3\int d^4 x {(\partial_r a)^2\over\tau_{2D}}.}

\noindent
Noting that $\tau_{2D}=\partial a/\partial x^8$, the equation of motion 
for $a$ implies 

\eqn\Crnch
{r^2\tau_2{\partial a\over\partial r}=r^2{\partial x^8\over \partial r}}

\noindent
is a constant. Thus, the expressions in \Asympt\ are exact.
At $r=1/m_g=\nu/\beta$ we find that

$$ a_D=-i\Big(\nu-{\beta\over r}\Big)=0.$$

\noindent
Thus, the magnetic dimple is cut off at $a_D=0$. Intuitively this is pleasing:
the magnetic dimple (Dirichlet string) should end on the 
magnetic (0,1) sevenbrane
which is indeed located at $a_D=0$. To understand the geometry of the dimple
close to $a_D=0$, note that the induced metric is

\eqn\WorldMet
{\eqalign{ds^2&=f^{-1/2}(-dt^2+d\Omega_2)
+(f^{-1/2}+f^{1/2}\tau_2\partial_r x^8\partial_r x^8)dr^2\cr
&=l_s^2 a_D\bar{a}_D\log|a_Dl_s| (-dt^2+d\Omega_2)+\cr
&(l_s^2 a_D\bar{a}_D\log|a_Dl_s|+
{\tau_{2D} \over l_s^2 a_D\bar{a}_D\log|a_Dl_s|}
\partial_r x^8\partial_r x^8)dr^2.}}

\noindent
In the above expression we have used the approximate solution \fsoln\
for the metric, obtained in the last section. This is a valid
approximation since we are interested in the geometry close to $a_D=0$
where our expression for $f$ becomes exact. The proper length to $a_D=0$ is
clearly infinite. It is also clear from \WorldMet\ that the dimple ends at
a single point at $a_D=0$, which is again 
satisfying. Thus, although we motivated
the need to introduce a cut off from the point of view of the low energy
effective action realised on the probe, one could equally well 
argue for {\it exactly
the same} cut off from the dual gravity description. 
In the super Yang-Mills theory 
one has a short distance (UV) cut off; on the gravity side one has a long
distance (IR) cut off. The connection between these cut offs is expected
as a consequence of the UV/IR correspondence.

A direct consequence of the short distance cut off in field theory, is that
the monopole appears as a sphere of radius $1/m_g$. From the point of view of 
the effective field theory, it is not possible to localize the monopole any
further. The more a particle in quantum field theory is localized, the higher
the energy of the cloud of virtual particle fluctuations surrounding it will 
be. If one localizes the monopole in the effective field theory any further,
the energy of the fluctuations becomes high enough to excite virtual 
monopole-antimonopole 
pairs, and hence we would leave the domain of validity of the effective
field theory. The origin of the exponential corrections in \PSMonopole\ can
be traced back to virtual $W^\pm$ bosons by noting that the factors multiplying
$r$ in the exponent are proportional to the boson mass 
$m_W$. The Higgs field of the 
Prasad-Sommerfeld monopole goes smoothly to zero as $r\to 0$ so that these 
bosons resolve the singular monopole core. The fact that these fluctuations have
been integrated out of the effective theory naturally explains why $a$ does
not receive any exponential corrections. Although these corrections are crucial
for the description of the monopole core, they are not needed by the effective 
field theory which describes only low energy (large distance) phenomena. In a
similar way, virtual monopole-antimonopole pairs will resolve the singular
behaviour of $x^8$ as $r\to 0$.

The cut-off that we employ in our work has already been anticipated in a completely 
different context in \rDenef. In this work the effective field theory realisation of BPS states
was studied in the field theory limit of IIB string theory compactified on a Calabi-Yau
manifold. In this description, the BPS states arise as limits of the attractor ``black holes" 
in ${\cal N}=2$ supergravity. To analyse the BPS equations around $a_D=0$ (a singular point 
in the field theory moduli space), it is safer to employ the ``attractor-like" formulation of 
the BPS equations. The solution with a finite $a_D=0$ core then arises as a solution of the BPS
equations of motion. Our results are in agreement with those of \rDenef. The work of this section was published in \rUst. Related work was pulished in the reference \rTown. In this paper finite energy BPS states corresponding to open strings that start and end on threebranes were constructed. These correspond to BPS states of the ${\cal N}=4$ super Yang-Mills theory. These authors argue for the same cut off employed in our work.

We will now comment on the relation between our study and the results presented in \rWst.
Above we argued that there was a need for a cut-off because the world-volume theory of 
the probe is a Wilsonian effective action, obtained by integrating all fluctuations 
above some energy scale out of the theory. However, this argument could also be made 
directly in the supergravity description by examining the induced metric given above. 

The form of this induced metric relies crucially on the fact that the probe is moving 
in the background of sevenbranes and $N$ threebranes. In particular, by taking $N\to\infty$
the authors of \rOfer\ argued that the background geometry can be trusted in the field theory 
limit. By accounting for the deformation of the background due to these $N$ threebranes 
our analysis goes beyond the low energy approximation and, in particular, can be trusted when
computing quantities not protected by supersymmetry. Note that the induced metric is non-holomorphic.
In the case of the M-theory fivebrane, one does not expect to reproduce quantities that are non-holomorphic \rOz.
Indeed, the background geometry for the M-theory fivebrane analysis is flat, so it is difficult
to see how an analogue of the induced metric could be realised. It is an open question as to whether 
a cut-off can consistently be used in this case.  

We now compute the energy of the magnetic dimple. Since the dimple is at rest,
this energy should be proportional to the mass of the monopole. As discussed
above, $1/\alpha$ is playing the r\^ole of the electric charge $e$. The Dirac
quantization condition is $eg=4\pi$, so that we can identify the magnetic
charge $g=4\pi\alpha.$ The magnetic dimple is an excitation of the flat
threebrane located at $a=\gamma$. We will denote the corresponding value
of the dual variable $a_D$ by $a_D^0$. Thus, the mass of a magnetic monopole
of this theory is $m=g|a_D^0|=4\pi\alpha |a_D^0|$. To compute the mass of the
magnetic dimple note that (we have set the $\theta$-angle to zero) 

$$\tau_{2D}=-i{\partial a\over\partial a_D}=-i
{\partial a\over\partial r}{\partial r\over\partial a_D}
=i{\alpha\over r^2}{\partial r\over\partial a_D}.$$

\noindent
Since the kinetic energy vanishes, the energy of the soliton is proportional
to the Lagrangian

$$\eqalign{E&=-L=-l_s^{-4}\int_{0}^{\infty} dr \int_{0}^{\pi}d\theta 
\int_{0}^{2\pi}d\phi \sin\theta r^2\tau_{2D} 
(\partial_r x^8)^2 \cr 
&=+i4\pi\alpha\int^{a_D^0}_{0} da_D=4\pi\alpha|a_D^0|.}$$

\noindent
The mass of the dimple lends further support to its interpretation as a
magnetic monopole.

\subsec{Multimonopole Solutions}

In this section we will describe solutions containing an arbitrary number of
separated static dimples. We will look for static solutions, that have
$x^8$ and $F_{ij}$ $i,j=1,2,3$ excited. It is useful 
to again consider the Maxwellian
truncation of the full Born-Infeld action to motivate an ansatz. The variation
of the gaugino of the Maxwellian theory, assuming the most general static
magnetic field, is given by

\eqn\Vart
{\delta \psi=(\Gamma_{ij}F^{ij}
+\eta^{kk}\Gamma_{8k}\partial_{k}x^8)\epsilon.}

\noindent
Thus, a solution which satisfies

\eqn\ScndAnz
{{1\over 2}\eta_{il}\epsilon^{ljk}F_{jk}=
\pm \partial_{i}x^8,}

\noindent
will be invariant under supersymmetries \Vart\ as long as $\epsilon$ satisfies
$\Gamma_1\Gamma_2\Gamma_3\Gamma_8\epsilon=\pm\epsilon.$ The choice of sign
again depends on whether one is describing a monopole or an anti-monopole
background. As explained above, these are exactly the supersymmetries that one
would expect to be preserved by a Dirichlet string stretched along the $x^8$
axis. Upon making this ansatz, we again find that the determinant appearing 
in the Born-Infeld action can be written as a perfect square and that once again
the dynamics for the field theory of the threebrane probe worldvolume is
described by the Seiberg-Witten low energy effective action. After noting that
the identification $a_Dl_s^2=-ix^8$ 
implies that $\tau_{2D}=l_s^2 \partial a/\partial x^8$,
the equation of motion following from the Seiberg-Witten effective action 
can be written as

\eqn\SWEqn
{{\partial\over\partial x^i}
{\partial\over\partial x^i} a=0.}

\noindent
This is just the free Laplace equation which is easily solved

\eqn\MultiMonSol
{a=\gamma +\sum_{i=1}^{n}{\alpha\eta\over \big[
(x-x_i)^2+(y-y_i)^2+(z-z_i)^2\big]^{1/2}},}

\noindent
where $\eta=\pm 1$.
The number of dimples $n$ and their location $(x_i,y_i,z_i)$ is completely
arbitrary. This is to be expected - BPS states do not experience a static force.
In the above, we have set each of the numerators equal to $\alpha$, because as
we have seen above this factor is related to the electric charge $e$. The 
classical solution would of course allow arbitrary coefficients. The
coefficients of these terms can be positive or negative. All strings must
have the same orientation for this to be a BPS state. Thus the interpretation
of terms with a positive coefficient is that they correspond to
strings that end on the threebrane; terms
with a negative coefficient correspond to strings that
start from the threebrane. The finite energy solutions
that we consider only allow for strings that start from the probe and end
on the sevenbrane, and consequently we fix the sign of all terms to be negative.
Each of the dimples above ends in a point at the sevenbranes at $a_D=0$. The
electric Higgs field $a$ again reproduces the known asymptotic behaviour of the
Higgs field component of multi-monopole 
solutions in non-Abelian ($SU(2)$) gauge theories\rSols.

We will now consider the mass of this multi-monopole solution.
In the limit that the dimples are very widely separated, the mass
of each dimple can be computed separately. 
In this widely separated dimple limit, the
total energy is simply the sum of the energy for each dimple, so that the
energy of the $n$-dimple solution is indeed consistent with its interpretation as an $n$-monopole state. Of course the total energy of the $n$-dimple state
is independent of the locations of each dimple so that once the energy is
known for widely separated dimples, it is known for all possible positions of
the dimples.
 
\newsec{Metric of the Monopole Moduli Space}

In the previous section we constructed smooth finite energy solutions to the
equations of motion. The solution describing $n$ BPS monopoles is known to be 
a function of $4n$ moduli parameters\rMSD. 
The position of each monopole accounts 
for $3n$ of these parameters. The remaining $n$ parameters correspond to
phases of the particles. By allowing these moduli to become time dependent
it is possible to construct the low energy equations of motion for the 
solitons\rManton. The resulting low energy 
monopole dynamics is given by finding
the geodesics on the monopole moduli space. The low energy dynamics of
${\cal N}=2$ supersymmetric monopoles has been studied in \rGaunt.
In this section we would like
to see if the metric on monopole moduli space can be extracted from the
solutions we constructed above.

\subsec{One Monopole Moduli Space}

A single monopole has four moduli parameters. Three of these parameters may
be identified with the center of mass position of the monopole, $x^i$, whilst
the fourth moduli parameter corresponds to the phase of the monopole, $\chi$. 
We will denote these moduli as $z^{\alpha}=(\chi,x^1,x^2,x^3)$. After allowing
the moduli to pick up a time dependence, the action picks up the following
additional kinetic term

\eqn\AddKin
{S=\int d^4 x\tau_{2D} \dot{a}_D\dot{\bar{a}}_D=\int dt 
{\cal G}_{\alpha\beta}\dot{z}^{\alpha}
\dot{z}^{\beta},}

\noindent
where we have introduced the metric on the one monopole moduli space

\eqn\OneMonMet
{{\cal G}_{\alpha\beta}\equiv \int d^3 x\tau_{2D}
{\partial a_D\over\partial z^{\alpha}} 
{\partial \bar{a}_D\over\partial z^{\beta}}.}

\noindent
Consider first the center of mass of the monopole. To construct the moduli space
dependence on these coordinates, replace $x^i_0\to x^i_0+x^i(t)$, where $x^i_0$
denotes the initial (time independent) monopole position. The action
picks up the following kinetic terms

\eqn\AdKn
{S=\int d^4 x \tau_{2D}  \dot{a}_D\dot{\bar{a}}_D=
\int d^4 x \tau_{2D} {\partial a_D\over \partial x^{i}(t)}
{\partial \bar{a}_D\over \partial x^{j}(t)}\dot{x}^{i}(t)
\dot{x}^{j}(t)=\int dt \dot{x}^{i}(t)
\dot{x}^{j}(t){\cal G}_{ij}.}

\noindent
Since the velocity of the monopole is small, we keep only terms which
are quadratic in the monopole velocity. Thus, when we compute

\eqn\OneMonMet
{{\cal G}_{ij}=-\int d^3 x\tau_{2D} {\partial a_D\over\partial x^i}
{\partial a_D\over\partial x^j}=-\int drd\theta d\phi r^2\sin \theta\tau_{2D}
\Big({\partial a_D\over\partial r}\Big)^2{\partial r\over\partial x^i}
{\partial r\over\partial x^j}}

\noindent
we evaluate the integrand at the static monopole solution. Using the
result 

$$ \int_0^{\pi} d\theta\int_{0}^{2\pi} d\phi 
\sin (\theta) {\partial r\over\partial x^i}
{\partial r\over\partial x^j}={1\over 2}4\pi\delta_{ij},$$

\noindent
we find

\eqn\ReltsFrMet
{{\cal G}_{ij}=-{1\over 2}\delta_{ij}\int dr r^2
\tau_{2D}\Big({\partial a_D\over\partial r}\Big)^2={1\over 2}
4\pi\alpha|a_D^0|
\delta_{ij}.}

\noindent
Consider now the fourth moduli parameter. Under the large gauge transformation
$g=e^{\chi(t) a_D}$ one finds that

\eqn\LGT
{\delta A_i =\partial_{i}(\chi (t) a_D),\quad 
   \delta A_0= \partial_{0}(\chi (t) a_D),\quad
   \delta a_D=0. }

\noindent
Note that since the gauge group $U(1)$ is compact, the parameter $\chi (t)$ is
a periodic coordinate. It is possible to modify this transformation so that
the potential energy remains constant and a small electric field is turned on. 
The modified transformation is\rHarv\

\eqn\Modfd
{\delta A_i =\partial_{i}(\chi (t) a_D),\quad 
   \delta A_0= \partial_{0}(\chi (t) a_D)-\dot{\chi}a_D,\quad
   \delta a_D=0.}

\noindent
After the transformation, the electric field is 
$E_i=F_{0i}=i\dot{\chi}\partial_i a_D=\dot{\chi}B_i.$ The fourth parameter is
$\chi$ and its velocity controls the magnitude of the electric field which is
switched on. The extra kinetic contribution to the action is

\eqn\LstKin
{S={1\over 2}\int d^4 x F_{0r}^2=
{1\over 2}\int dt \dot{\chi}^2\int d^3 x B_{r}B^{r}=
{1\over 2}\int dt \dot{\chi}^2 4\pi\alpha |a_D^0|. }

\noindent
Thus, we find that the quantum mechanics for the collective coordinates on 
the one monopole moduli space is described by the action

\eqn\FnlRslt
{S=\int dt {1\over 2}4\pi\alpha
|a_{D}^0|\delta_{\alpha\beta}z^{\alpha}z^{\beta}.}

\noindent
This is the correct result\rGM. Thus, 
the monopole moduli space is topologically
$R^3\times S^1$. The induced metric on the moduli space is simply the flat
metric. It is well known that this metric is hyper-K\"ahler. This fits nicely
with the structure of the moduli space quantum mechanics: the dimple
preserves ${\cal N}=1$ supersymmetry in the four dimensional field theory.
As a result, we would expect an action with ${\cal N}=4$ worldline 
supersymmetry. In one dimension there are two types of multiplets with
four supercharges. The dimensional reduction of two dimensional $(2,2)$
supersymmetry leads to ${\cal N}=4A$ supersymmetry. The presence of this
supersymmetry requires that the moduli space be a 
K\"ahler manifold\rSusyO. The 
second supersymmetry, ${\cal N}=4B$ is obtained by reducing two dimensional
$(4,0)$ supersymmetry. The presence of this supersymmetry requires that the
moduli space is a hyper-K\"ahler manifold\rSusyO.
The only fermion zero modes in
the monopole background in ${\cal N}=2$ super Yang-Mills theory is chiral  
in the sense $\Gamma_1\Gamma_2\Gamma_3\Gamma_8\epsilon=\pm\epsilon$ as
explained above. Thus, the supersymmetry on the worldline is 
${\cal N}=4B$ supersymmetry. The fact that the moduli space is 
hyper-K\"ahler is a direct consequence of supersymmetry. 

\subsec{Multi Monopole Moduli Space}

The asymptotic metric on the moduli space of 
two widely separated BPS monopoles
has been constructed by Manton\rManton. In this approach, one considers
the dynamics of two dyons. After constructing the Lagrangian that describes
the dyons motion in $R^3$, with constant electric charges as parameters, one
identifies the electric charges as arising from the motion on circles
associated with the fourth moduli parameter of the monopoles. The explicit form
of the metric for the relative collective coordinates is\rManton

\eqn\GMMet
{ds^2=U(r)dr^idr^i+{g^2\over 2\pi MU(r)}(d\chi+\omega^idr^i)^2,\quad
U(r)=1-{g^2\over 2\pi M(r^jr^j)^{1/2}},}

\noindent
where $r^i$ is the relative coordinate of the two monopoles, $\chi$ is the
relative phase, $M$ the monopole mass, $g$ the magnetic charge and $\omega^i$
the Dirac monopole potential which satisfies 
$\epsilon^{ijk}\partial^j\omega^k=r^i/(r^jr^j)^{3/2}.$
This is just the Taub-NUT metric with negative mass.
In this section, we will show that it is possible to reproduce the first term
in this metric from the dimple solutions. The second term could presumably be
reproduced by studying dyonic dimples.

We will consider the following two-dimple solution

\eqn\MDSol
{a=\gamma -\sum_{i=1}^{2}{\alpha\over |\vec{x}-\vec{x}_{i}|}.} 

\noindent
In the case of the two dimple solution, it is no longer possible to compute 
the moduli space metric exactly, and we have to resort to approximate
techniques. To reproduce the first term in \GMMet\ we need to evaluate

\eqn\IntToEval
{{\cal G}_{\alpha\beta}=-\int d^3 x\tau_{2D}
{\partial a_D\over\partial x^\alpha}
{\partial a_D\over\partial x^\beta}=
i\int d^3 x
{\partial a_D\over\partial x^\alpha}
{\partial a\over\partial x^\beta},}

\noindent
where $\alpha,\beta$ can take any one of six values corresponding to any of
the three spatial coordinates of either monopole. We do not know $a_D$ as a
function of the $x^\alpha$; we will write \MDSol\ as $a=\gamma-\Delta $. 
Setting $a_D= a_D^0+\beta_1\Delta+O(\Delta^2)$ in the expression
for $a(a_D)$ we find

\eqn\Expnd
{a(a_D)=a(a_D^0)+\beta_1\Delta{\partial a\over\partial a_D}|_{a_D=a_D^0}
=\gamma-\Delta.}

\noindent
This approximation is excellent at large $r$ where $\Delta<<1$.
Thus, we find that 

\eqn\Expnded
{a_D=a_D^0-\Delta\left({\partial a\over\partial a_D}|_{a_D=a_D^0}\right)^{-1}
=a_D^0+i{\Delta\over\tau_{2D}(a_D^0)}.}

\noindent
The dual theory is weakly coupled, so that $\tau_D(a_D^0)$ is large. Thus,
the correction to $a_D^0$ in the expression for $a_D$ is indeed small and
the approximation that we are using is valid. Using this expression for
$a_D$ in \IntToEval\ we find

\eqn\NextStep
{{\cal G}_{ij}=\int d^3 x{\alpha^2\over\tau_{2D}(a_D^0)}
\left({\partial\over\partial x^i_1}{1\over |\vec{x}-\vec{x}_1|}\right)
\left({\partial\over\partial x^j_1}{1\over |\vec{x}-\vec{x}_1|}\right).}

\noindent
To evaluate this integral it is useful to change coordinates to a spherical
coordinate system centered about monopole 1. In these coordinates

\eqn\NStep
{{\cal G}_{ij}=\delta_{ij}{1\over 2}{4\pi\alpha^2\over\tau_{2D}(a_D^0)}
\int dr{1\over r^2}.}

\noindent
The integral must again be cut off at the lower limit\foot{Strictly speaking
we should also exclude a circular region centered around
$\vec{x}=\vec{x}_1-\vec{x}_2$. However, the integrand is of order 
$|\vec{x}_1-\vec{x}_2|^{-2}$ in this region and in addition the area of the
region to be excluded is $\pi/m_g^2$, so that this is a negligible effect.}
where $a_D=0$. The value of the cut off is given by

\eqn\CtOff
{\eqalign{0=&a_D^0+{i\alpha\over \tau_{2D}(a_D^0)r}+
{i\alpha\over \tau_{2D}(a_D^0)|\vec{x}+\vec{x}_1-\vec{x}_2|}
=a_D^0+{i\alpha\over \tau_{2D}(a_D^0)r}+
{i\alpha\over \tau_{2D}(a_D^0)r_{12}}
+O({r\over r_{12}^2})\cr
{1\over r}&={\tau_{2D}(a_D^0)|a_D^0|\over\alpha}
-{1\over r_{12}},}}

\noindent
where $r_{12}$, the magnitude of the relative coordinate, is assumed to be
large. Thus, we find that

\eqn\Resltt
{{\cal G}_{ij}={1\over 2}\delta_{ij}4\pi {\alpha^2\over\tau_{2D}(a_D^0)}
\Big({\tau_{2D}(a_D^0)|a_D^0|\over\alpha}
-{1\over r_{12}}\Big)={1\over 2}\delta_{ij}
\left(4\pi\alpha|a_D^0|-
{g^2\over 4\pi\tau_{2D}(a_D^0)r_{12}}\right).}

\noindent
In a similar way, we compute

\eqn\NextSep
{\eqalign{{\cal G}_{i+3,j+3}&=\int d^3 x{\alpha^2\over\tau_{2D}(a_D^0)}
\left({\partial\over\partial x^i_2}{1\over |\vec{x}-\vec{x}_2}|\right)
\left({\partial\over\partial x^j_2}{1\over |\vec{x}-\vec{x}_2}|\right)\cr
&={1\over 2}\delta_{ij}
\left(4\pi\alpha|a_D^0|-
{g^2\over 4\pi\tau_{2D}(a_D^0)r_{12}}\right).}}

\noindent
To finish the calculation of the metric on the two monopole moduli space,
we need to compute

\eqn\NtStep
{{\cal G}_{i+3,j}=\int d^3 x{\alpha^2\over\tau_{2D}(a_D^0)}
\left({\partial\over\partial x^i_2}{1\over |\vec{x}-\vec{x}_2|}\right)
\left({\partial\over\partial x^j_1}{1\over |\vec{x}-\vec{x}_1|}\right).}

\noindent
We do not need to evaluate \NtStep\ directly. If we introduce center of mass
and relative coordinates as

\eqn\Comrel
{\vec{r}_{cm}={1\over 2}(\vec{x}_1+\vec{x}_2),\quad
\vec{r}_{12}=\vec{x}_1-\vec{x}_2,}

\noindent
it is a simple exercise to compute the center of mass contribution
to the action

\eqn\COMcont
{S=\int dt\dot{r}_{cm}^i\dot{r}_{cm}^i 4\pi\alpha |a_D^0|.}

\noindent
The integral that had to be performed to obtain this result was
proportional to the action itself. This result fixes

\eqn\LstRlt
{{\cal G}_{i+3,j}={\cal G}_{i,j+3}=
{1\over 2}\delta_{ij} 
{g^2\over 4\pi\tau_{2D}(a_D^0)r_{12}}.}

\noindent
Putting the above results together, we find the action which governs
the relative motion of the two monopoles is

\eqn\RelMot
{S_{rel}=\int dt\Big( {4\pi\alpha |a_D^0|\over 4}-
{g^2\over 8\pi\tau_{2D}(a_D^0) r_{12}}\Big)
{dr_{12}^i\over dt}{dr_{12}^i\over dt}.} 

\noindent
Thus, the low energy relative motion is geodesic motion for a metric on $R^{3}$
given by $ ds^2= U(r)dr^i dr^i,$ with 
$U(r)=1-g^2/(8\pi^2\alpha |a_D^0|\tau_{2D}(a_D^0) r)$. This
reproduces the first term in \GMMet. Notice however that the magnetic coupling
comes with a factor of $1/\tau_{2D}(a_D^0)$. This 
factor has its origin in the loop plus
instanton corrections that were summed to obtain the low energy effective
action. These corrections do not change the fact that the two monopole moduli
space is hyper-K\"ahler.
Thus, the dimple solutions on the probe reproduces the quantum corrected
metric. This non-trivial metric has its origin in the fact that the forces due
to dilaton and photon exchange no longer cancel at non-zero velocity due to the
different retardation effects for spin zero and spin one exchange\rHarv.

The exact two monopole metric has been determined by Atiyah and Hitchin\rAH. 
The fact that it has an $SO(3)$ isometry arising from rotational invariance, 
that in four dimensions hyper-K\"ahler implies self-dual curvature and the fact
that the metric is known to be complete determines it exactly. Expanding the
Atiyah-Hitchin metric and neglecting exponential corrections, one recovers the
Taub-NUT metric\rGM. If we again 
interpret the origin of the exponential corrections
as having to do with virtual $W^\pm$ boson effects, it is natural to expect
that the exact treatment of the dimples in the probe worldvolume theory will
recover the (quantum corrected) Taub-NUT metric and not the Atiyah-Hitchin 
metric. The metric on the moduli space of $n$ well separated monopoles has been
computed by Gibbons and Manton\rGM\ by studying 
the dynamics of $n$ well separated
dyons. The exact monopole metric computed for a tetrahedrally symmetric charge
4 monopole was found to be exponentially 
close to the Gibbons-Manton metric\rSut.
This result was extended in \rBiel\ where it was shown that the Gibbons-Manton
metric is exponentially close to the exact metric for the general $n$
monopole solution. In view of these results it is natural to conjecture that
an exact treatment of the $n$-dimple solution in the probe worldvolume theory
will recover the (quantum corrected) Gibbons-Manton metric.

\vfill\eject

\newsec{Summary and Conclusion}

In this thesis we have studied the effective action of ${\cal N}=2$ super Yang-Mills theory with $N_F=4$ hypermultiplets and gauge group $SU(2)$ by realizing it as the worldvolume theory of a $D$3-brane probing a supergravity background. The motivation for the study was to obtain non-holomorphic corrections to the effective action, which are not protected by supersymmetry. The expansion of the Born-Infeld action was shown to reproduce the correct structure for the non-holomorphic terms in the effective action of the field theory.
In particular, three specific cases were studied: ${\cal N}=2$ super Yang-Mills theory with four massless flavors, ${\cal N}=2$ super Yang-Mills theory with four massive flavors, and the pure gauge theory. 

In the first scenario, the effective gauge coupling is a constant due to the fact that the theory is conformally invariant. This is reflected in the supergravity by the fact that the sevenbranes are parallel and coincident with the $O7$ plane, ensuring that the $R$-$R$ charge is cancelled locally. In this case, a non-renormalization theorem is known to protect the four derivative terms of the field theory, which receive no instanton corrections. As noted earlier, $\tau_2$ receives instanton corrections. Remarkably, the expansion of the Born-Infeld action about the supergravity background is such that the $\tau_2$ dependence of the four derivative terms cancels, ensuring consistency with the field theory result, while the six derivative terms, which are not protected, have a $\tau_2$ dependence. 

In the second scenario, the masses of the hypermultiplets (arising since the Higgs field develops a non vanishing vacuum expectation value) break the conformal invariance of the massless theory. The effective coupling is no longer a constant, and receives instanton corrections. In the supergravity picture, giving the hypermultiplets a mass corresponds to pulling the sevenbranes away from the $O7$ plane. In doing so, quantum corrections to the supergravity background split the orientifold plane in two. The $R$-$R$ charge is no longer cancelled locally, and thus we expect the dilaton-axion modulus (and thus the effective coupling) to vary over the transverse space (moduli space). 

In this case there are no corresponding non-renormalization theorems for the four derivative terms; indeed, the four derivative terms receive instanton corrections along with the coupling. In the supergravity, the $AdS_5$ geometry is corrected when switching on a mass for the hypermultiplets reflecting the broken conformal invariance of the field theory. Expanding the Born-Infeld about the corrected $AdS_5$ geometry, the structure of the one instanton correction to the four derivative term so obtained matched that predicted by field theory computations; however, the coefficient of this term could not be fixed by our analysis. The manner in which the instanton corrections to the four derivative terms are reflected in the background geometry was studied by computing the quark-antiquark potential.

In the last scenario, we considered the pure gauge theory which is asymptotically free. In the supergravity picture, the $D$7-branes are moved infinitely far away which corresponds to giving the four flavor hypermultiplets in the field theory infinite masses. These states can no longer be excited in the field theory, and one is left with a description of the pure gauge theory. In this case, we were able to obtain an approximate solution to the background valid in the perturbative regime of the field theory. Expanding the Born-Infeld action about this background yielded the same structure of the four derivative terms as that predicted by a field theory analysis. This result is interesting, since the uncorrected supergravity is not expected to reproduce perturbative field theory. As mentioned earlier, we comment that this may be a consequence of the fact that theories with ${\cal{N}}=2$ supersymmetry are rather special since many quantities of interest are one loop exact. 

We conclude these results by noting that while the background does not affect the leading Seiberg-Witten low energy effective action it is essential in determining the correct structure for the higher derivative terms, which are not protected by supersymmetry.

Next, the BPS states of the ${\cal{N}}=2$ supersymmetric field theory were constructed as finite energy solutions of the worldvolume theory of a threebrane probe in F-theory. We began by constructing an approximate solution for the background, valid for large $|a|$ and small $|a_D|$. By computing the Ricci tensor in the string frame it was shown that the points $|a_D|=0$ and $|a| \to \infty$ correspond to curvature singularities in the supergravity background. These points where the supergravity breaks down correspond to regions on the moduli space where the field theory becomes weakly coupled in its magnetic and electric descriptions, respectively. 

A worldvolume soliton was constructed in the pure gauge theory, which is interpreted as a single Dirichlet string stretching from the probe to a ``magnetic" (0,1) sevenbrane. The place where the Dirichlet string meets the surface of the threebrane is pulled into a dimple. The solution breaks one half of the supersymmetries. From the worldvolume point of view, the endpoint of the Dirichlet string behaves like a magnetic monopole. The solution in terms of the electric variable $a$ reproduces the leading long distance behaviour of the Higgs field in the Prasad-Sommerfeld solution, implying that the abelian worldvolume theory captures the long distance  monopole physics of the non-abelian field theory it describes. One is able to argue for a short distance (ultra-violet) cut-off in the field theory. The same cut-off is motivated from the supergravity point of view by making use of the supergravity background constructed in section (12), which is able to describe the geometry in the vicinity of the monopole point $|a_D|=0$. This cut-off appears as an infra-red cut-off in the supergravity, in accord with the UV$\backslash$IR correspondence. The mass of the dimple solution is found to be consistent with its interpretation as a magnetic monopole. Next, solutions were constructed containing an arbitrary number of separated static dimples. Finally, some features of the moduli spaces of these solutions were investigated. We were able to compute the metric on the one monopole moduli
space exactly. For two monopoles and higher, this does not seem to be possible. Assuming that the monopoles are very widely separated allowed us to construct the asymptotic form of the metric. In this case, the metric receives both perturbative and instanton corrections. The hyper-K\"ahler structure of these moduli spaces is not disturbed by these corrections.

A comment needs to be made regarding the fact that, as our argument for the link between the Born-Infeld action and the effective action of a field theory implies, one would expect the expansion of Born-Infeld to be compared with the $U(1)$ effective action of a spontaneously broken $SU(N+1)$ theory, rather than $SU(2)$ as has been the case in this thesis. One may motivate this heuristically by noting that, although we have $N+1$ objects, we are really studying the two body interaction between the threebrane probe and the clump of $N$ threebranes. The lack of a Dirac quantization condition in the supergravity suggests that one is unable to distinguish between one object of charge $N$, and $N$ objects of unit charge. Consequently, one may reinterpret the system of interest as a two body system, comprised of the probe threebrane and a threebrane of charge $N$, which implies a gauge group $SU(2)$ broken down to $U(1)$. 

In conclusion, the central motivation behind this thesis has been to exploit the new insights into the intimate relationship between string theory and gauge theories in order to move closer to one of the Holy Grails of physics, namely, a knowledge of the strong coupling physics of QCD. Naturally, to come anywhere near realising this ambition, one has to begin to come to terms with theories which are not blessed with as much supersymmetry as ${\cal{N}}=4$ super Yang-Mills theory. In this thesis a step was taken in this direction by studying a theory with a reduced number of supersymmetries, although in some instances it was apparent that supersymmetry continued to work its magic behind the scenes.

It now seems likely that QCD is dual to a string theory, although it is by no means clear what form this string theory will take. For the time being, at least, the strong coupling physics of QCD seems safe from prying eyes. However, it appears that the faith theorists have placed in string theory may bear fruit sooner than expected, with powerful conceptual tools promising new approaches to gauge theories. However, whether or not these new insights will aid in shattering the inscrutability of QCD remains an open question.
\listrefs

\bye